\documentstyle[latexsym,amssymb,epsfig,aps,floats,eqsecnum,preprint,amsfonts]{revtex}

\def\laq{\raise 0.4ex\hbox{$<$}\kern -0.8em\lower 0.62
ex\hbox{$\sim$}}
\def\gaq{\raise 0.4ex\hbox{$>$}\kern -0.7em\lower 0.62ex\hbox{$\sim$}}

\font\tenbb=msbm10
\font\sevenbb=msbm7
\font\fivebb=msbm5
\newfam\bbfam
\textfont\bbfam=\tenbb \scriptfont\bbfam=\sevenbb
\scriptscriptfont\bbfam=\fivebb

\newcommand{\beq}{\begin{equation}} 
\newcommand{\eeq}{\end{equation}}
\newcommand{\bea}{\begin{eqnarray}} 
\newcommand{\eea}{\end{eqnarray}}
\newcommand{\beam}{\begin{mathletters}} 
\newcommand{\eeam}{\end{mathletters}}

\begin{document}
\draft
\preprint{\vbox{\baselineskip=12pt
\rightline{IHES/P/98/74} 
\vskip 0.2truecm
\rightline{gr-qc/9811091}}}

\title{\Large\bf Effective one-body approach to general 
relativistic two-body dynamics}
\author{A. Buonanno${}^{a}$ and T. Damour${}^{a,b}$}
\address{$^a$ {\it Institut des Hautes Etudes Scientifiques, 91440
Bures-sur-Yvette, France} \\ 
{$^b$ {\it DARC, CNRS-Observatoire de Paris, 92195 Meudon, France}}}
\maketitle
\begin{abstract}
We map the general relativistic two-body problem onto that of a test 
particle moving in an effective external metric. This effective-one-body 
approach defines, in a non-perturbative manner, the late dynamical evolution 
of a coalescing binary system of compact objects.
The transition from the adiabatic inspiral, driven by gravitational 
radiation damping, to an unstable plunge, induced by strong spacetime 
curvature, is predicted to occur for orbits more tightly bound than the 
innermost stable circular orbit in a Schwarzschild metric of mass $M=m_1 + 
m_2$. The binding energy, angular momentum and orbital frequency of the 
innermost stable circular orbit for the time-symmetric two-body problem are 
determined as a function of the mass ratio.
\end{abstract}
\maketitle
\newpage
\section{Introduction}
\label{sec1}
Binary systems made of compact objects (neutron stars or black holes), and 
driven toward coalescence by gravitational radiation damping, are among the 
most promising candidate sources for interferometric gravitational-wave 
detectors such as LIGO and VIRGO. It is therefore important to study the late 
dynamical evolution of a coalescing binary system of compact objects and, in 
particular, to estimate when occurs the transition from an adiabatic inspiral, 
driven by gravitational radiation damping, to an unstable plunge, induced by 
strong spacetime curvature. The global structure of the gravitational wave 
signal emitted by a coalescing binary depends sensitively on the location of 
the transition from inspiral to plunge. For instance, in the case of a system 
of two equal-mass neutron stars, if this transition occurs for relatively 
loosely bound orbits, the inspiral phase will evolve into a plunge phase 
before tidal disruption takes place. On the other hand, if the transition 
occurs for tightly bound orbits, tidal effects will dominate the late 
dynamical evolution.

In this paper we introduce a novel approach to the general relativistic 
two-body problem. The basic idea is to map (by a canonical transformation) the 
two-body problem onto an effective one-body problem, i.e. the motion of a test 
particle in some effective external metric. When turning off radiation 
damping, the effective metric will be a static and spherically symmetric 
deformation of the Schwarzschild 
metric. [The deformation parameter is the symmetric mass ratio 
$\nu \equiv m_1\,m_2/(m_1 + m_2)^2$.]
Solving exactly the effective problem of a test particle in this 
deformed Schwarzschild metric amounts to introducing a particular 
{\it non-perturbative} method for re-summing the post-Newtonian expansion of the 
equations of motion.

Our effective one-body approach is inspired by (though different from) an 
approach to electromagnetically interacting quantum two-body problems 
developed in the works of Br\'ezin, Itzykson and Zinn-Justin \cite{BIZ} (see 
also \cite{IZ}) and of Todorov and coworkers \cite{todorov}, \cite{MNT}. 
Ref.~\cite{BIZ} has shown that an approximate summation (corresponding to the 
eikonal approximation) of the ``crossed-ladder'' Feynman diagrams for the 
quantum scattering of two charged particles led to a ``relativistic Balmer 
formula'' for the squared mass of bound states which correctly included recoil 
effects (i.e. effects linked to the finite symmetric mass ratio $\nu = 
m_1 m_2 / (m_1 + m_2)^2$). However, the eikonal approximation does not capture 
some of the centrifugal barrier shifts which have to be added by hand through 
a shift $n \rightarrow n-\epsilon_j$ of the principal quantum number 
\cite{BIZ}, \cite{IZ}. The approach of Ref.~\cite{todorov} is more 
systematic, being based on a (Lippmann-Schwinger-type) quasi-potential equation 
whose solution is fitted to the Feynman expansion of the (on-shell) 
scattering amplitudes $\langle p'_1 \, p'_2 \, \vert S \vert \, p_1 \, p_2 
\rangle$. However, several arbitrary choices have to be made to define the 
(off-shell) quasi-potential equation and the nice form of the relativistic 
Balmer formula proposed in Ref.~\cite{BIZ} is recovered only at the end, 
after two seemingly accidental simplifications: (i) the ratio of some 
complicated energy-dependent quantities simplifies \cite{nbtod}, and (ii) the 
second-order contribution to the quasi-potential contributes only to third 
order. We note also that the extension of Todorov's quasi-potential approach 
(initially developed for quantum two-body electrodynamics) to the 
gravitational two-body problem \cite{MNT} leads to much more complicated 
expressions than the approach developed here.

Before entering the technical details of the effective one-body approach, let 
us outline the main features of our work. We use as input the explicit, 
post-Newtonian (PN) expanded classical equations of motion of a 
gravitationally interacting system of two compact objects. In harmonic 
coordinates (which are convenient to start with because they are standardly 
used for computing the generation of gravitational radiation), these equations 
of motion are explicitly known up to the 2.5PN level ($(v/c)^5$-accuracy) 
\cite{DD81a}, \cite{D82}. They have the form ($a,b=1,2$)
\begin{equation}
\mbox{\boldmath$a$}_a = \mbox{\boldmath${\cal A}$}_a^{2{\rm PN}} 
(\mbox{\boldmath$z$}_b , \mbox{\boldmath$v$}_b) + \mbox{\boldmath$A$}_a^{\rm 
reac} (\mbox{\boldmath$z$}_b , \mbox{\boldmath$v$}_b) + {\cal O} (c^{-6}) \, ,
\label{eq1.1}
\end{equation}
where $\mbox{\boldmath${\cal A}$}^{2{\rm PN}} = \mbox{\boldmath$A$}_0 + c^{-2} 
\, \mbox{\boldmath$A$}_2 + c^{-4} \, \mbox{\boldmath$A$}_4$ denotes the 
time-symmetric part of the equations of motion, and $\mbox{\boldmath$A$}^{\rm 
reac} = c^{-5} \, \mbox{\boldmath$A$}_5$ their time-antisymmetric part. Here, 
$\mbox{\boldmath$z$}_a$, $\mbox{\boldmath$v$}_a$, $\mbox{\boldmath$a$}_a$, 
denote the positions, velocities and accelerations, in harmonic coordinates, 
of the two bodies. [In this work we consider only non-spinning objects.] 
Throughout this paper, we shall use the following notation for the 
quantities related to the masses $m_1$ and $m_2$ of the two bodies:
\begin{equation}
M \equiv m_1 + m_2 \,, \quad \quad \mu \equiv \frac{m_1\,m_2}{M} \,, 
\quad \quad  \nu \equiv \frac{\mu}{M} 
\equiv \frac{m_1\, m_2}{(m_1 + m_2)^2} \, . 
\label{eq1.2}
\end{equation}
Note that the ``symmetric mass ratio'' $\nu$ varies between 0 (test mass 
limit) and $\frac{1}{4}$ (equal mass case).

We first focus on the time-symmetric, 2PN dynamics defined by 
$\mbox{\boldmath${\cal A}$}_a^{2{\rm PN}} (\mbox{\boldmath$z$}_b , 
\mbox{\boldmath$v$}_b)$. After going to the center of mass frame (uniquely 
defined by the Poincar\'e symmetries of the 2PN dynamics), and after a 
suitable coordinate transformation (from harmonic coordinates to ADM 
coordinates $\mbox{\boldmath$z$}_a \rightarrow \mbox{\boldmath$q$}_a$), the 
dynamics of the relative coordinates $\mbox{\boldmath$q$} \equiv 
\mbox{\boldmath$q$}_1 - \mbox{\boldmath$q$}_2$ is defined by a 2PN 
Hamiltonian $H (\mbox{\boldmath$q$},\mbox{\boldmath$p$})$. Starting from $H 
(\mbox{\boldmath$q$},\mbox{\boldmath$p$})$, we shall uniquely introduce a 
2PN-accurate static and spherically symmetric ``effective metric''
\begin{equation}
ds_{\rm eff}^2 = -A (R_{\rm eff}) \, c^2 \, dt_{\rm eff}^2 + \frac{D (R_{\rm 
eff})}{A (R_{\rm eff})} \ d R_{\rm eff}^2 + R_{\rm eff}^2 (d\theta_{\rm eff}^2 
+ \sin^2 \theta_{\rm eff} \, d\varphi_{\rm eff}^2) \, , \label{eq1.3}
\end{equation}
where
\begin{equation}
A(R) = 1 + \frac{a_1}{c^2 R} + \frac{a_2}{c^4 R^2} + \frac{a_3}{c^6 R^3} \,, 
\quad \quad  
D(R) = 1 + \frac{d_1}{c^2 R} + \frac{d_2}{c^4 R^2} \, , \label{eq1.4}
\end{equation}
such that the ``linearized'' effective metric (defined by $a_1$ and $d_1$) is 
the linearized Schwarzschild metric defined by the total mass $M=m_1 + m_2$, 
and such that the effective Hamiltonian $H_{\rm eff} (\mbox{\boldmath$q$}_{\rm 
eff} , \mbox{\boldmath$p$}_{\rm eff})$ defined by the geodesic action $-\int 
\mu \, c \, ds_{\rm eff}$, where $\mu = m_1 m_2 / M$ is the reduced mass, can 
be mapped onto the relative-motion 2PN Hamiltonian $H (\mbox{\boldmath$q$}, 
\mbox{\boldmath$p$})$ by the combination of a canonical transformation 
$(\mbox{\boldmath$q$}_{\rm eff} , \mbox{\boldmath$p$}_{\rm eff}) \rightarrow 
(\mbox{\boldmath$q$}, \mbox{\boldmath$p$})$ and of an energy transformation $H 
= f (H_{\rm eff})$, corresponding to an energy-dependent ``canonical'' 
rescaling of the time coordinate $dt_{\rm eff} = dt\,(dH / dH_{\rm eff})$.

The effective metric so constructed is a deformation of the Schwarzschild 
metric, with deformation parameter the symmetric mass ratio $\nu = \mu / M$. 
Considering this deformed Schwarzschild metric as an exact external metric 
then defines (in the effective coordinates) a $\nu$-deformed 
Schwarzschild-like dynamics, which can be mapped back onto the original 
coordinates $\mbox{\boldmath$q$}_a$ or $\mbox{\boldmath$z$}_a$. Our 
construction can be seen as a non-perturbative way of re-summing the 
post-Newtonian expansion in the relativistic regime where $GM / (c^2 
\vert \mbox{\boldmath$q$}_1 - \mbox{\boldmath$q$}_2 \vert)$ becomes of order 
unity. In particular, our construction defines a specific $\nu$-deformed 
innermost stable circular orbit (ISCO). Superposing the gravitational reaction 
force $\mbox{\boldmath$A$}^{\rm reac}$ onto the ``exact'' 
deformed-Schwarzschild dynamics (defined by mapping back the effective problem 
onto the real one) finally defines, in a non-perturbative manner, a dynamical 
system which is a good candidate for describing the late stages of evolution 
of a coalescing compact binary.

\section{Second post-Newtonian dynamics of the relative motion of a two-body 
system}
\label{sec2}
Let us recall some of the basic properties of the dynamics defined by 
neglecting the time-odd reaction force in the Damour-Deruelle equations of 
motion (\ref{eq1.1}). The 2PN (i.e. $(v/c)^4$-accurate) truncation of these 
equations of motion defines a time-symmetric dynamics which is derivable from 
a {\it generalized} Lagrangian $L(\mbox{\boldmath$z$}_1 , 
\mbox{\boldmath$z$}_2 ,\mbox{\boldmath$v$}_1 ,\mbox{\boldmath$v$}_2 
,\mbox{\boldmath$a$}_1 ,\mbox{\boldmath$a$}_2)$ \cite{DD81b}, \cite{D82} (a 
function of the harmonic positions, $\mbox{\boldmath$z$}_1 , 
\mbox{\boldmath$z$}_2$, velocities $\mbox{\boldmath$v$}_1 
,\mbox{\boldmath$v$}_2$ and accelerations $\mbox{\boldmath$a$}_1 
,\mbox{\boldmath$a$}_2$). The generalized Lagrangian $L(\mbox{\boldmath$z$}_1 
, \mbox{\boldmath$z$}_2 ,\mbox{\boldmath$v$}_1 ,\mbox{\boldmath$v$}_2 
,\mbox{\boldmath$a$}_1 ,\mbox{\boldmath$a$}_2)$ is (approximately) invariant 
under the Poincar\'e group \cite{DD81c}.
This invariance leads (via Noether's theorem) to the explicit construction of 
the usual ten relativistic conserved quantities for a dynamical system: energy 
${\cal E}$, linear momentum $\mbox{\boldmath${\cal P}$}$, angular momentum 
$\mbox{\boldmath${\cal J}$}$, and center-of-mass constant 
$\mbox{\boldmath${\cal K}$} = \mbox{\boldmath${\cal G}$} - 
\mbox{\boldmath${\cal P}$} \, t$. Because of the freedom to perform a 
Poincar\'e transformation (in harmonic coordinates), we can go to the (2PN) 
center-of-mass frame, defined such as
\begin{equation}
\mbox{\boldmath${\cal P}$} = \mbox{\boldmath${\cal K}$} = 
\mbox{\boldmath${\cal G}$} = \mbox{\boldmath$0$} \, . \label{eq2.1}
\end{equation}

Refs.~\cite{DS85}, \cite{DS88} explicitly constructed the coordinate 
transformation between the harmonic (or De Donder) coordinates, say $z^{\mu}$, 
used in the Damour-Deruelle equations of motion, and the coordinates, say 
$q^{\mu}$, introduced by Arnowitt, Deser and Misner \cite{ADM60} in the 
framework of their canonical approach to the dynamics of the gravitational 
field. The Lagrangian giving the 2PN motion in ADM coordinates has the 
advantage of being an ordinary Lagrangian $L(\mbox{\boldmath$q$}_1 
,\mbox{\boldmath$q$}_2 ,\dot{\mbox{\boldmath$q$}}_1 
,\dot{\mbox{\boldmath$q$}}_2)$ (depending only on positions and velocities), 
which is equivalent to an ordinary Hamiltonian $H(\mbox{\boldmath$q$}_1 
,\mbox{\boldmath$q$}_2 ,\mbox{\boldmath$p$}_1 ,\mbox{\boldmath$p$}_2)$ 
\cite{OOKH74}, \cite{S86}. The explicit expression of the 2PN Hamiltonian 
in ADM coordinates, $H (\mbox{\boldmath$q$}_1 ,\mbox{\boldmath$q$}_2 
,\mbox{\boldmath$p$}_1 ,\mbox{\boldmath$p$}_2)$, has been derived in 
Ref.~\cite{DS88} by applying a contact transformation
\begin{equation}
\mbox{\boldmath$q$}_a (t) = \mbox{\boldmath$z$}_a (t) - \delta^*
\mbox{\boldmath$z$}_a (z,v) \, , 
\label{eq2.2}
\end{equation}
to the generalized Lagrangian $L(\mbox{\boldmath$z$}_a ,\mbox{\boldmath$v$}_a 
,\mbox{\boldmath$a$}_a)$. The shift $\delta^* \, \mbox{\boldmath$z$}_a$ is of 
order ${\cal O} (c^{-4})$ and is defined in equation (35) of \cite{DS85} or 
equations (2.4) of \cite{DS88}.
The contact transformation (\ref{eq2.2}) removes the acceleration dependence 
of the harmonic-coordinates Lagrangian $L^{\rm harm} (z,v,a)$ and transforms it 
into the ADM-coordinates ordinary Lagrangian $L^{\rm ADM} (q, \dot q )$. A further 
Legendre-transform turns $L^{\rm ADM} (\mbox{\boldmath$q$}_1 
,\mbox{\boldmath$q$}_2 ,\dot{\mbox{\boldmath$q$}}_1 
,\dot{\mbox{\boldmath$q$}}_2)$ into the needed 2PN Hamiltonian 
$H(\mbox{\boldmath$q$}_1 ,\mbox{\boldmath$q$}_2 ,\mbox{\boldmath$p$}_1 
,\mbox{\boldmath$p$}_2)$ in ADM coordinates. The explicit expression of this 
Hamiltonian is given in equation (2.5) of Ref.~\cite{DS88}. It has also been 
shown in Ref.~\cite{DS85} that the Hamiltonian $H(\mbox{\boldmath$q$}_1 
,\mbox{\boldmath$q$}_2 ,\mbox{\boldmath$p$}_1 ,\mbox{\boldmath$p$}_2)$ can be 
directly derived in ADM coordinates from the (not fully explicit) $N$-body 
results of Ref.~\cite{OOKH74} by computing a certain integral entering the 
two-body interaction potential. [For further references on the general 
relativistic problem of motion, see the review \cite{D87}; for recent work on 
the gravitational Hamiltonian see \cite{OK89}, \cite{SW}, \cite{SGauss}, \cite{JS}.]

The ADM expression of the total Noether linear momentum 
$\mbox{\boldmath${\cal P}$}$ associated to the translational invariance of $L 
(\mbox{\boldmath$z$}, \mbox{\boldmath$v$}, \mbox{\boldmath$a$})$ is simply 
$\mbox{\boldmath${\cal P}$} = \mbox{\boldmath$p$}_1 + \mbox{\boldmath$p$}_2$. 
Therefore it is easily checked that, in the center-of-mass frame 
(\ref{eq2.1}), the relative motion is obtained by substituting in the two-body 
Hamiltonian $H(\mbox{\boldmath$q$}_1 , \mbox{\boldmath$q$}_2 , 
\mbox{\boldmath$p$}_1 , \mbox{\boldmath$p$}_2)$,
\begin{equation}
\mbox{\boldmath$p$}_1 \rightarrow \mbox{\boldmath$P$} \,, \quad \quad  
\mbox{\boldmath$p$}_2 \rightarrow - \mbox{\boldmath$P$} \, , \label{eq2.3}
\end{equation}
where $\mbox{\boldmath$P$} = \partial S / \partial \mbox{\boldmath$Q$}$ is the 
canonical momentum associated with the relative ADM position vector 
$\mbox{\boldmath$Q$} \equiv \mbox{\boldmath$q$}_1 - \mbox{\boldmath$q$}_2$. 
[For clarity, we modify the notation of Ref.~\cite{DS88} by using 
$\mbox{\boldmath$q$}_1$, $\mbox{\boldmath$q$}_2$, $\mbox{\boldmath$Q$}$ and 
$\mbox{\boldmath$q$}$ for the ADM position coordinates which are denoted 
$\mbox{\boldmath$r$}_1$, $\mbox{\boldmath$r$}_2$, $\mbox{\boldmath$R$}$ and 
$\mbox{\boldmath$r$}$ , respectively, in Ref.~\cite{DS88}.]

Our technical starting point in this work will be the {\it reduced 
center-of-mass} 2PN Hamiltonian (in reduced ADM coordinates). We introduce the 
following reduced variables (all defined in ADM coordinates, and in the 
center-of-mass frame):
\begin{eqnarray}
&&\mbox{\boldmath$q$} \equiv \frac{\mbox{\boldmath$Q$}}{GM} \equiv 
\frac{\mbox{\boldmath$q$}_1 - \mbox{\boldmath$q$}_2}{GM}\,, \quad  \quad 
\mbox{\boldmath$p$} \equiv \frac{\mbox{\boldmath$P$}}{\mu} \ , \nonumber \\
&&\widehat t \equiv \frac{t}{GM}\,, \quad  \quad \widehat H \equiv \frac{H^{{\rm 
NR}}}{\mu} \equiv \frac{H^{\rm R} - Mc^2}{\mu} \, . \label{eq2.4}
\end{eqnarray}
In the last equation, the superscript ``NR'' means ``non-relativistic'' (i.e. 
after subtraction of the appropriate rest-mass contribution), while ``R'' means
``relativistic'' (i.e. including the appropriate rest-mass contribution). From 
equation (3.1) of \cite{DS88} the reduced 2PN relative-motion Hamiltonian 
(without the rest-mass contribution) reads
\begin{equation}
\widehat H (\mbox{\boldmath$q$},\mbox{\boldmath$p$}) = \widehat{H}_0 
(\mbox{\boldmath$q$},\mbox{\boldmath$p$}) + \frac{1}{c^2} \, \widehat{H}_2 
(\mbox{\boldmath$q$},\mbox{\boldmath$p$}) + \frac{1}{c^4} \, \widehat{H}_4 
(\mbox{\boldmath$q$},\mbox{\boldmath$p$}) \, , \label{eq2.5}
\end{equation}
where
\begin{mathletters}
\label{eq2.6}
\begin{eqnarray}
\widehat{H}_0 (\mbox{\boldmath$q$},\mbox{\boldmath$p$}) &=& \frac{1}{2} \, 
\mbox{\boldmath$p$}^2 - \frac{1}{q} \, , \label{eq2.6a} \\
\widehat{H}_2 (\mbox{\boldmath$q$},\mbox{\boldmath$p$}) &=& - \frac{1}{8} \, 
(1-3\nu) \, \mbox{\boldmath$p$}^4 - \frac{1}{2q} \, [(3+\nu) \, 
\mbox{\boldmath$p$}^2 + \nu (\mbox{\boldmath$n$} \cdot \mbox{\boldmath$p$})^2] 
+ \frac{1}{2q^2} \, , \label{eq2.6b} \\
\widehat{H}_4 (\mbox{\boldmath$q$},\mbox{\boldmath$p$}) &=& \frac{1}{16} \, 
(1-5\nu + 5\nu^2) \, \mbox{\boldmath$p$}^6 \nonumber \\
&+& \frac{1}{8q} \, [(5-20\nu - 3\nu^2) \, \mbox{\boldmath$p$}^4 - 2\nu^2 \, 
\mbox{\boldmath$p$}^2 (\mbox{\boldmath$n$} \cdot \mbox{\boldmath$p$})^2 - 
3\nu^2 (\mbox{\boldmath$n$} \cdot \mbox{\boldmath$p$})^4] \nonumber \\
&+& \frac{1}{2q^2} \, [(5+8\nu) \, \mbox{\boldmath$p$}^2 + 3\nu \, 
(\mbox{\boldmath$n$} \cdot \mbox{\boldmath$p$})^2] - \frac{1}{4q^3} \, (1+3\nu) 
\, , \label{eq2.6c}
\end{eqnarray}
\end{mathletters}
in which $q \equiv \vert \mbox{\boldmath$q$} \vert \equiv 
(\mbox{\boldmath$q$}^2)^{1/2}$ and $\mbox{\boldmath$n$} \equiv 
\mbox{\boldmath$q$} / q$. When convenient, we shall also use 
the notation $r$ for the reduced radial separation $q$ (and $R$ for 
the unreduced one $Q$). [As in Eqs.~(\ref{eq2.8})--(\ref{eq2.new}) below.]

The relative-motion Hamiltonian (\ref{eq2.5}) is invariant under time 
translations and space rotations. The associated conserved quantities are the 
reduced center-of-mass (c.m.) energy and angular momentum of the binary system:
\begin{equation}
\widehat{H} \, (\mbox{\boldmath$q$},\mbox{\boldmath$p$}) = \widehat{\cal 
E}^{\rm NR} \equiv \frac{{\cal E}_{\rm c.m.}^{\rm NR}}{\mu}\,, \quad  \quad 
\mbox{\boldmath$q$} \times \mbox{\boldmath$p$} = \mbox{\boldmath$j$} \equiv 
\frac{\mbox{\boldmath${\cal J}$}_{\rm c.m.}}{\mu GM} \, . \label{eq2.7}
\end{equation}
A convenient way of solving the 2PN relative-motion dynamics is to use the 
Hamilton-Jacobi approach. The motion in the plane of the relative trajectory is 
obtained, in polar coordinates
\begin{equation}
q^x = r \cos \varphi \,, \quad \quad q^y = r \sin \varphi \,, \quad \quad q^z = 0 \, , 
\label{eq2.8}
\end{equation}
by separating the time and angular coordinates in the (planar) reduced action
\begin{equation}
\widehat S \equiv \frac{S}{\mu GM} = - \widehat{\cal E}^{\rm NR} \, \widehat t 
+ j\,\varphi + \widehat{S}_r (r, \widehat{\cal E}^{\rm NR} , j) \, . \label{eq2.9}
\end{equation}
The time-independent Hamilton-Jacobi equation $\widehat{H}^{\rm NR} 
(\mbox{\boldmath$q$},\mbox{\boldmath$p$}) = \widehat{\cal E}^{\rm NR}$ with 
$\mbox{\boldmath$p$} = \partial \widehat{S} / \partial \mbox{\boldmath$q$}$ can be 
(iteratively) solved with respect to $(d \widehat{S}_r / dr)^2$ with a result of the form
\begin{equation}
\widehat{S}_r (r, \widehat{\cal E}^{\rm NR} , j) = \int dr \ \sqrt{{\cal R} (r, 
\widehat{\cal E}^{\rm NR} , j)} \, . \label{eq2.10}
\end{equation}
The radial ``effective potential'' ${\cal R} (r, \widehat{\cal E}^{\rm NR} , 
j)$ is a fifth-order polynomial in $1/r \equiv 1/q$ which is explicitly written down in 
equations (3.4) of \cite{DS88}. In this section, we shall only need the 
corresponding (integrated) radial action variable
\begin{equation}
i_r \equiv \frac{I_R}{\mu GM} \equiv \frac{2}{2\pi} \int_{r_{\min}}^{r_{\max}} 
dr \ \sqrt{{\cal R} (r, \widehat{\cal E}^{\rm NR} , j)} \, . \label{eq2.11}
\end{equation}
The function $i_r (\widehat{\cal E}^{\rm NR} , j)$ has been computed, at the 
2PN accuracy, in Ref.~\cite{DS88} (see equation (3.10) there). To clarify some 
issues connected with the fact that the natural scalings in the ``effective 
one-body problem'' (to be considered below) differ from those in the present, 
real two-body problem, let us quote the expression of the unscaled radial 
action variable 
\beq
I_R = \alpha \, i_r =\frac{2}{2 \pi}\,\int_{R_{\rm min}}^{R_{\rm max}} 
dR\,\frac{dS_R(R,{\cal E}^{\rm NR},{\cal J})}{dR}
\,, 
\label{eq2.new}
\eeq
in terms of the unscaled variables 
${\cal E}^{\rm NR} = \mu \, \widehat{\cal E}^{\rm NR}$ and ${\cal J} = \alpha 
\, j$. Here $R = Q = G\,M\,r = G\,M\,q$, and 
we introduced the shorthand notation
\begin{equation}
\alpha \equiv \mu \, GM = G \, m_1 \, m_2 \label{eq2.12}
\end{equation}
for the gravitational two-body coupling constant. We have
\begin{eqnarray}
I_R ({\cal E}^{\rm NR} , {\cal J}) &=& \frac{\alpha \, \mu^{1/2}}{\sqrt{-2 \, 
{\cal E}^{\rm NR}}} \, \left[ 1 + \left( \frac{15}{4} - \frac{\nu}{4} \right) 
\, \frac{{\cal E}^{\rm NR}}{\mu c^2} + \left( \frac{35}{32} + \frac{15}{16} \, 
\nu + \frac{3}{32} \, \nu^2 \right) \left( \frac{{\cal E}^{\rm NR}}{\mu c^2} 
\right)^2 \right] \nonumber \\
&-& {\cal J} + \frac{\alpha^2}{c^2 {\cal J}} \,\left [ 
3 + \left( \frac{15}{2} - 3\nu \right) \, \frac{{\cal E}^{\rm NR}}{ \mu\,c^2} \right ]
+ \left(\frac{35}{4} - \frac{5}{2} \, \nu \right)\,\frac{\alpha^4}{c^4\,{\cal J}^3}\,. 
\label{eq2.13}
\end{eqnarray}
Equation (\ref{eq2.13}) can also be solved with respect to ${\cal E}^{\rm NR} 
\equiv {\cal E}^{\rm R} - Mc^2$ with the (2PN-accurate) result (see equation 
(3.13) of Ref.~\cite{DS88})
\begin{eqnarray}
{\cal E}^{\rm R} ({\cal N} , {\cal J}) &=& M c^2 - \frac{1}{2} \, \frac{\mu 
\alpha^2}{{\cal N}^2} \, \Biggl[ 1 + \frac{\alpha^2}{c^2} \left( \frac{6}{{\cal 
N} {\cal J}} - \frac{1}{4} \, \frac{15 - \nu}{{\cal N}^2} \right) \nonumber \\
&+& \frac{\alpha^4}{c^4} \left( \frac{5}{2} \, \frac{7 - 2\nu}{{\cal N} {\cal 
J}^3} + \frac{27}{{\cal N}^2 {\cal J}^2} - \frac{3}{2} \, \frac{35-4\nu}{{\cal 
N}^3 {\cal J}} + \frac{1}{8} \, \frac{145 - 15\nu + \nu^2}{{\cal N}^4} \right) 
\Biggl] \, , \label{eq2.14}
\end{eqnarray}
where ${\cal N}$ denotes the Delaunay action variable ${\cal N} \equiv I_R + 
{\cal J}$. The notation is chosen so as to evoke the one often used in the 
quantum Coulomb problem. Indeed, the classical action variables $I_R$ and 
${\cal J}$, or their combinations ${\cal N} = I_R + {\cal J}$ and ${\cal J}$, 
are adiabatic invariants which, according to the Bohr-Sommerfeld rules, become 
(approximately) quantized in units of $\hbar$ for the corresponding quantum 
bound states. More precisely ${\cal N} / \hbar$ becomes the ``principal quantum 
number'' and ${\cal J} / \hbar$ the total angular-momentum quantum number. The 
fact that the Newtonian-level non-relativistic energy ${\cal E}^{\rm NR} = - 
\frac{1}{2} \, \mu \alpha^2 / {\cal N}^2 + {\cal O} (c^{-2})$ depends only on 
the combination ${\cal N} = I_R + {\cal J}$ is the famous special degeneracy of 
the Coulomb problem. Note that 1PN (and 2PN) effects lift this degeneracy by 
bringing an extra dependence on ${\cal J}$. There remains, however, the 
degeneracy associated with the spherical symmetry of the problem, which implies 
that the energy does not depend on the ``magnetic quantum number'', i.e. on 
${\cal M} = {\cal J}_z$, but only on the magnitude of the angular momentum 
vector ${\cal J} = \sqrt{\mbox{\boldmath${\cal J}$}^2}$. Though we shall only 
be interested in the classical gravitational two-body problem, it is 
conceptually useful to think in terms of the associated quantum problem. From 
this point of view, the formula (\ref{eq2.14}) describes, when ${\cal N} / 
\hbar$ and ${\cal J} / \hbar$ take (non zero) integer values, all the quantum 
{\it energy levels} as a function of the parameters 
$M = m_1 + m_2$, $\mu = m_1 \, m_2 / (m_1 + 
m_2)$, $\alpha = G \, m_1 \, m_2$ and $\nu = \mu / M$. It is to be noted that 
the function ${\cal E}^{\rm R} ({\cal N} , {\cal J})$ describing the energy 
levels is a coordinate-invariant object.

\section{Second post-newtonian energy levels of the effective one-body problem}
\label{sec3}
The ``energy levels'' (\ref{eq2.14}) summarize, at the 2PN accuracy, the 
dynamics obtained by eliminating the field variables $g_{\mu \nu} (x)$ in the 
total action of a gravitationally interacting binary system
\begin{equation}
S_{\rm tot} [z_1^{\mu} ,z_2^{\mu} ,g_{\mu \nu}] = - \int m_1 \, c \, ds_1 - 
\int m_2 \, c \, ds_2\, + \,S_{\rm field} [g_{\mu \nu} (x)] \, , \label{eq3.1}
\end{equation}
where $ds_1 = \sqrt{-g_{\mu \nu} (z_1^{\lambda}) \, dz_1^{\mu} \, dz_1^{\nu}}$ 
and where $S_{\rm field} [g_{\mu \nu} (x)]$ is the (gauge-fixed) 
Einstein-Hilbert action for the gravitational field. Let $S_{\rm real} 
[z_1^{\mu} ,z_2^{\mu}]$ be the Fokker-type action obtained by (formally) 
integrating out $g_{\mu \nu} (x)$ in (\ref{eq3.1}). [See, e.g., \cite{DS85} for 
more details on Fokker-type actions. As we work here only at the 2PN level, 
and take advantage of the explicit results of Refs.~\cite{DD81b}, \cite{D82}, 
we do not need to enter the subtleties of the elimination of the field degrees 
of freedom, which are probably best treated within the ADM approach. See 
\cite{S85}, \cite{S86}.]

The basic idea of the present work is to, somehow, associate to the ``real'' 
two-body dynamics $S_{\rm real} [z_1^{\mu} ,z_2^{\mu}]$ some ``effective'' 
one-body dynamics in an external spacetime, as described by the action
\begin{equation}
S_{\rm eff} \, [z_0^{\mu}] = - \int m_0 \, c \, ds_0 \, , \label{eq3.2}
\end{equation}
where $ds_0 = \sqrt{-g_{\mu \nu}^{\rm eff} (z_0^{\lambda}) \, dz_0^{\mu} \, 
dz_0^{\nu}}$, with some spherically symmetric static effective metric
\bea
ds_{\rm eff}^2 = g_{\mu \nu}^{\rm eff} (x_{\rm eff}^{\lambda}) \, dx_{\rm 
eff}^{\mu} \, dx_{\rm eff}^{\nu} &=& -A (R_{\rm eff}) \, c^2 \, dt_{\rm eff}^2 
+ B (R_{\rm eff}) \, dR_{\rm eff}^2  \nonumber \\
&+& C (R_{\rm eff}) \, R_{\rm eff}^2 (d\theta_{\rm eff}^2 + \sin^2  
\theta_{\rm eff} \, d\varphi_{\rm eff}^2) \, . \label{eq3.3}
\eea
To ease the notation we shall, henceforth in this section, suppress the 
subscript ``eff'' on the coordinates used in the effective problem. [Later in 
this paper we shall explicitly relate the coordinates $z_0^{\mu}$ of the 
effective particle to the coordinates $z_1^{\mu}$, $z_2^{\nu}$ of the two real 
particles.] The metric functions $A(R)$, $B(R)$, $C(R)$ will be constructed in 
the form of an expansion in $1/R$:
\begin{eqnarray}
&& A(R) = 1 + \frac{a_1}{c^2 R} + \frac{a_2}{c^4 R^2} + \frac{a_3}{c^6 R^3} + 
\cdots \, , \nonumber \\
&& B(R) = 1 + \frac{b_1}{c^2 R} + \frac{b_2}{c^4 R^2} + \cdots \, . 
\label{eq3.4}
\end{eqnarray}
Beware that the variable $R$ in Eqs.~(\ref{eq3.4}) denotes (in this section) 
the {\it effective} radial coordinate, which differs from the real 
ADM separation $Q = R_{\rm ADM} = GM r$ used in the previous 
section (e.g. in the definition of $I_R$).
We indicate in Eq.~(\ref{eq3.4}) the terms that we shall need at the 2PN 
level. The third function $C(R)$ entering the effective metric will be either 
fixed to $C_S (R) \equiv 1$ (in ``Schwarzschild'' coordinates), or to satisfy 
$C_I (R) \equiv B(R)$ (in ``Isotropic'' coordinates).

There are two mass parameters entering the effective problem: (i) the mass 
$m_0$ of the effective particle, and (ii) some mass parameter $M_0$ used to 
scale the coefficients $a_i$, $b_i$ entering the effective metric. For 
instance, we can define $M_0$ by conventionally setting
\begin{equation}
a_1 \equiv -2 \, G \, M_0 \, . \label{eq3.5}
\end{equation}

By analogy to Eq.~(\ref{eq2.14}), we can summarize, in a {\it 
coordinate-invariant manner}, the dynamics of the effective one-body problem 
(\ref{eq3.2})--(\ref{eq3.4}) by considering the ``energy levels'' of the bound 
states of the particle $m_0$ in the metric $g_{\mu \nu}^{\rm eff}$:
\begin{equation}
{\cal E}_0^{\rm R} = m_0 \, c^2 + {\cal E}_0^{\rm NR} = {\cal F} ({\cal N}_0 , 
{\cal J}_0 ; m_0 , a_i , b_i) \, . \label{eq3.6}
\end{equation}
Here, the relativistic effective energy ${\cal E}_0^{\rm R}$ and the effective 
action variables ${\cal N}_0$, ${\cal J}_0$ are unambiguously defined by the 
action (\ref{eq3.2}). Namely, we can separate the effective Hamilton-Jacobi 
equation
\begin{equation}
g_{\rm eff}^{\mu \nu} \, \frac{\partial S_{\rm eff}}{\partial x^{\mu}} \, 
\frac{\partial S_{\rm eff}}{\partial x^{\nu}} + m_0^2 \, c^2 = 0 \, , 
\label{eq3.7}
\end{equation}
by writing  (considering, for simplicity, only motions in the equatorial plane 
$\theta = \frac{\pi}{2}$)
\begin{equation}
S_{\rm eff} = -{\cal E}_0 \, t + {\cal J}_0 \, \varphi + S_R^0 (R, {\cal E}_0 , 
{\cal J}_0) \, . \label{eq3.8}
\end{equation}
To abbreviate the notation we suppress the superscript ``$R$'' on the 
relativistic effective energy ${\cal E}_0$. Inserting Eq.~(\ref{eq3.8}) in Eq.~(\ref{eq3.7}) 
yields
\begin{equation}
- \frac{1}{A(R)} \, \frac{{\cal E}_0^2}{c^2} + \frac{1}{B(R)} \left( \frac{d 
S_R^0}{dR}\right)^2 + \frac{{\cal J}_0^2}{C(R)\,R^2} + m_0^2 \, c^2 = 0 \, , 
\label{eq3.9}
\end{equation}
and therefore
\begin{equation}
S_R^0 (R,{\cal E}_0 , {\cal J}_0) = \int dR \, \sqrt{{\cal R}_0 (R,{\cal E}_0 , 
{\cal J}_0)} \, , \label{eq3.10}
\end{equation}
where
\begin{equation}
{\cal R}_0 (R,{\cal E}_0 , {\cal J}_0) \equiv \frac{B(R)}{A(R)} \, \frac{{\cal 
E}_0^2}{c^2} - B(R) \left( m_0^2 \, c^2 + \frac{{\cal J}_0^2}{C(R)\,R^2} 
\right) \, . \label{eq3.11}
\end{equation}
The effective radial action variable $I_R^0$ is then defined as
\begin{equation}
I_R^0 ({\cal E}_0 , {\cal J}_0) \equiv \frac{2}{2\pi} 
\int_{R_{\min}}^{R_{\max}} dR \, \sqrt{{\cal R}_0 (R,{\cal E}_0 , {\cal J}_0)} 
\, , \label{eq3.12}
\end{equation}
while the effective ``principal'' action variable ${\cal N}_0$ is defined as 
the combination ${\cal N}_0 \equiv I_R^0 + {\cal J}_0$.

To obtain the effective ``energy levels'' ${\cal E}_0 = {\cal F} ({\cal N}_0 , 
{\cal J}_0)$ one needs to compute the definite radial integral (\ref{eq3.12}). 
Ref.~\cite{DS88} (extending some classic work of Sommerfeld, used in the old 
quantum theory) has shown how to compute the PN expansion of the radial 
integral (\ref{eq3.12}) to any order in the $1/R$ expansions (\ref{eq3.4}). At 
the present 2PN order, Ref.~\cite{DS88} gave a general formula (their equation 
(3.9)) which can be straightforwardly applied to our case.

As we said above, the function describing the ``energy levels'', ${\cal E}_0 = 
{\cal F} ({\cal N}_0 , {\cal J}_0)$, is a {\it coordinate-invariant} construct. 
As a check on our calculations, we have computed it (or rather, we have 
computed the radial action $I_R^0 ({\cal E}_0 ,{\cal J}_0)$) in the two 
preferred coordinate gauges for a spherically symmetric metric: the 
``Schwarzschild gauge'' and the ``Isotropic'' one. If $(a_i , b_i)$ denote the 
expansion coefficients (\ref{eq3.4}) in the Schwarzschild gauge $(C_S (R) 
\equiv 1)$, we find (at the 2PN accuracy)
\begin{eqnarray}
I_R^0 ({\cal E}_0 ,{\cal J}_0) &=& \frac{m_0^{3/2}}{\sqrt{-2 \, {\cal E}_0^{\rm 
NR}}} \left[ A + B \, \frac{{\cal E}_0^{\rm NR}}{m_0 \, c^2} + C \left( 
\frac{{\cal E}_0^{\rm NR}}{m_0 \, c^2} \right)^2 \right] - {\cal J}_0 \nonumber 
\\
&+& \frac{m_0^2}{c^2 \, {\cal J}_0} \left[ D + E \, \frac{{\cal E}_0^{\rm 
NR}}{m_0 \, c^2} \right]  + 
\frac{m_0^4}{c^4 \, {\cal J}_0^3} \, F \, , \label{eq3.13}
\end{eqnarray}
where ${\cal E}_0^{\rm NR} \equiv {\cal E}_0 - m_0 \, c^2$, and where
$$
A = - \frac{1}{2} \, a_1\,, \quad  \quad B = b_1 - \frac{7}{8} \, a_1\,, \quad  
\quad C = \frac{b_1}{4} - \frac{19}{64} \, a_1 \, ,
$$
$$
D = \frac{a_1^2}{2} - \frac{a_2}{2} - \frac{a_1 b_1}{4}\,, \quad  \quad E = a_1^2 
- a_2 - \frac{a_1 b_1}{2} - \frac{b_1^2}{8} + \frac{b_2}{2} \, ,
$$
\begin{equation}
F = \frac{1}{64} \, [24 \, a_1^4 - 48 \, a_1^2 \, a_2 + 8 \, a_2^2 + 16 \, 
a_1 \, a_3 - 8 \, a_1^3 \, b_1 + 8 \, a_1 \, a_2 \, b_1 - a_1^2 \, b_1^2 + 4 \, 
a_1^2 \, b_2] \, . \label{eq3.14}
\end{equation}

Denoting by $(\widetilde{a}_i ,\widetilde{b}_i)$ the expansion coefficients 
(\ref{eq3.4}) in the Isotropic gauge $(C_I (R) \equiv B_I (R))$, we find, by 
calculating $I_R^0$ directly in the isotropic gauge, that the coefficients 
$A,B,\ldots,F$ entering Eq.~(\ref{eq3.13}) have the following (slightly 
simpler) expressions in terms of $\widetilde{a}_i$ and $\widetilde{b}_i$:
$$
A = -\frac{1}{2} \, \widetilde{a}_1\,, \quad  \quad B = \widetilde{b}_1 - 
\frac{7}{8} \, \widetilde{a}_1\,, \quad  \quad C = \frac{\widetilde{b}_1}{4} - 
\frac{19}{64} \, \widetilde{a}_1 \, ,
$$
$$
D = \frac{\widetilde{a}_1^2}{2} - \frac{\widetilde{a}_2}{2} - 
\frac{\widetilde{a}_1 \widetilde{b}_1}{2}\,, \quad  \quad E = \widetilde{a}_1^2 - 
\widetilde{a}_2 - \widetilde{a}_1 \widetilde{b}_1 + \widetilde{b}_2 \, ,
$$
\begin{equation}
F = \frac{1}{8} \, [3 \, \widetilde{a}_1^4 - 6 \, \widetilde{a}_1^2 \, 
\widetilde{a}_2 + \widetilde{a}_2^2 + 2 \, \widetilde{a}_1 \, \widetilde{a}_3 - 
4 \, \widetilde{a}_1^3 \, \widetilde{b}_1 + 4 \, \widetilde{a}_1 \, 
\widetilde{a}_2 \, \widetilde{b}_1 + \widetilde{a}_1^2 \, \widetilde{b}_1^2 + 2 
\, \widetilde{a}_1^2 \, \widetilde{b}_2] \, . \label{eq3.15}
\end{equation}
The numerical values of the coefficients $A,B,\ldots,F$ are checked to be 
coordinate-invariant by using the following relation between the $(a_i , b_i)$ 
and the $(\widetilde{a}_i , \widetilde{b}_i)$ (which is easily derived either 
by integrating $dR_I / R_I = \sqrt{B_S (R_S)} \, dR_S / R_S$ or by using the 
algebraic link $R_S = R_I \, \sqrt{B_I (R_I)}$)
$$
\widetilde{a}_1 = a_1\,, \quad  \quad \widetilde{b}_1 = b_1 \, ,
$$
$$
\widetilde{a}_2 = a_2 - \frac{1}{2} \, a_1 \, b_1\,, \quad  \quad \widetilde{b}_2 
= \frac{1}{2} \, b_2 - \frac{1}{8} \, b_1^2 \, ,
$$
\begin{equation}
\widetilde{a}_3 = a_3 - a_2 \, b_1 + \frac{7}{16} \, a_1 \, b_1^2 - \frac{1}{4} 
\, a_1 \, b_2 \, . \label{eq3.16}
\end{equation}
Finally, solving iteratively Eq.~(\ref{eq3.13}) with respect to ${\cal 
E}_0^{\rm NR}$, we find the analog of Eq.~(\ref{eq2.14}), i.e. the explicit 
formula giving the effective ``energy levels''. It is convenient to write it in 
terms of ${\cal N}_0 \equiv I_R^0 + {\cal J}_0$, of the coupling constant
\begin{equation}
\alpha_0 \equiv G \, M_0 \, m_0 \, , \label{eq3.17}
\end{equation}
where $M_0$ is defined by Eq.~(\ref{eq3.5}), and of the $(GM_0)$-rescaled, 
dimensionless expansion coefficients $\widehat{a}_i$ and $\widehat{b}_i$, of 
the Schwarzschild gauge:
\begin{equation}
\widehat{a}_i \equiv a_i / (GM_0)^i\,, \quad  \quad \widehat{b}_i \equiv b_i / 
(GM_0)^i \, , \label{eq3.18}
\end{equation}
with $\widehat{a}_1 \equiv -2$.

We find
\begin{eqnarray}
{\cal E}_0 ({\cal N}_0 , {\cal J}_0) &= & m_0 \, c^2 - \frac{1}{2} \, \frac{m_0 
\, \alpha_0^2}{{\cal N}_0^2} \Biggl[ 1 + \frac{\alpha_0^2}{c^2} \left( 
\frac{C_{3,1}}{{\cal N}_0 {\cal J}_0} + \frac{C_{4,0}}{{\cal N}_0^2} \right) 
\nonumber \\
&+ & \frac{\alpha_0^4}{c^4} \left( \frac{C_{3,3}}{{\cal N}_0 {\cal J}_0^3} + 
\frac{C_{4,2}}{{\cal N}_0^2 {\cal J}_0^2} + \frac{C_{5,1}}{{\cal N}_0^3 {\cal 
J}_0} + \frac{C_{6,0}}{{\cal N}_0^4} \right) \Biggl] \, , \label{eq3.19}
\end{eqnarray}
where the coefficients $C_{p,q}$ (which parametrize the contributions $\propto 
- \frac{1}{2} \, (\alpha_0 / c)^{p+q} \, {\cal N}_0^{-p} \, {\cal J}_0^{-q}$ to 
${\cal E}_0 / m_0 \, c^2$) are given by
\begin{eqnarray}
&&C_{3,1} = 2\widehat{D}\,, \quad \quad \quad \quad \quad  \quad \quad C_{4,0} = 
-\widehat{B}\,, \nonumber \\
&&C_{3,3} = 2\widehat{F}\,, \quad  \quad \quad \quad \quad  \quad 
\quad C_{4,2} = 3\widehat{D}^2\,, \nonumber \\
&&C_{5,1} = -(4\widehat{B}\,\widehat{D} + \widehat{E})\,,
\quad  \quad \,\,C_{6,0} = \frac{1}{4} \left (5\widehat{B}^2 + 
2 \widehat{C} \right )\,. \label{eq3.20}
\end{eqnarray}
Here, the dimensionless quantities $\widehat{B}, \widehat{C}, \widehat{D}, 
\widehat{E}, \widehat{F}$ are the $GM_0$-rescaled versions of the 
coefficients of Eq.~(\ref{eq3.13}), given by replacing the $a_i$'s by 
$\widehat{a}_i$ in Eqs.~(\ref{eq3.14}). For instance, $\widehat{B}= 
\widehat{b}_1 - 7/8\,\widehat{a}_1 = \widehat{b}_1 + 7/4$, etc..

\section{Relating the ``real'' and the ``effective'' energy levels, and 
determining the effective metric}
\label{sec4}
We still have to define the precise rules by which we wish to relate the real 
two-body problem to the effective one-body one. If we think in quantum terms, 
there is a natural correspondence between ${\cal N}$ and ${\cal N}_0$, and 
${\cal J}$ and ${\cal J}_0$, which are quantized in units of $\hbar$. It is 
therefore very natural to require the identification
\begin{equation}
{\cal N} = {\cal N}_0\,, \quad  \quad {\cal J} = {\cal J}_0 \, , \label{eq4.1}
\end{equation}
between the real action variables and the effective ones, and we will do so in 
the following. What is a priori less clear is the relation between the real 
masses and energies, $m_1$, $m_2$, ${\cal E}_{\rm real}^R = (m_1 + m_2) \, c^2 + 
{\cal E}_{\rm real}^{\rm NR}$, and the effective ones, $m_0$, $M_0$, ${\cal E}_0 
= m_0 \, c^2 + {\cal E}_0^{\rm NR}$. The usual non-relativistic definition of an 
effective dynamics associated to the relative motion of a (Galileo-invariant) 
two-body system introduces an effective particle whose position 
$\mbox{\boldmath$q$}_0$ is the relative position, $\mbox{\boldmath$q$}_0 = 
\mbox{\boldmath$q$}_1 - \mbox{\boldmath$q$}_2$, whose inertial mass $m_0^{\rm 
NR}$ is the ``reduced'' mass $\mu \equiv m_1 \, m_2 / (m_1 + m_2)$, and whose 
potential energy is the potential energy of the system, $V_{\rm eff} 
(\mbox{\boldmath$q$}_0) = V_{\rm real} (\mbox{\boldmath$q$}_1 - 
\mbox{\boldmath$q$}_2)$. In the present case of a gravitationally interacting 
two-body system, with $V_{\rm real}^{\rm NR} = - G \, m_1 \, m_2 / \vert 
\mbox{\boldmath$q$}_1 - \mbox{\boldmath$q$}_2 \vert$, this would determine
\begin{equation}
m_0^{\rm NR} = \mu \,,  \quad \hbox{and} \quad \ M_0^{\rm NR} = m_1 + m_2 \equiv M \, , 
\label{eq4.2}
\end{equation}
such that $\alpha_{\rm real} = G \, m_1 \, m_2= \alpha_0 = G \, M_0^{\rm NR} \, 
m_0^{\rm NR}$.
The non-relativistic identifications (\ref{eq4.2}) are, however, paradoxical 
within a relativistic framework, even if they are modified by 
``relativistic corrections'', so that, say, $m_0 = \mu + {\cal O} (c^{-2})$, 
$M_0 = M + {\cal O} (c^{-2})$, because the reference level (and accumulation point for 
${\cal N} , {\cal J} \rightarrow \infty$) of the real relativistic levels 
(\ref{eq2.14}) will be the total rest-mass-energy $Mc^2$, and will therefore be 
completely different from the reference level $m_0 \, c^2 \simeq \mu \, c^2$ of 
the effective relativistic energy levels. This difference in the relativistic 
reference energy level shows that, while it is very natural to require the 
straightforward identifications (\ref{eq4.1}) of the action variables, the 
mapping between ${\cal E}_{\rm real}$ and ${\cal E}_0$ must be more subtle.

One might a priori think that the most natural relativistic generalization of 
the usual non-relativistic rules for defining an effective one-body problem 
consists in requiring that
\begin{equation}
{\cal E}_0 ({\cal N}_0 , {\cal J}_0) = {\cal E}_{\rm real} ({\cal N} , {\cal 
J}) - c_0 \, , \label{eq4.3}
\end{equation}
with a properly chosen constant $c_0 = M \, c^2 - m_0 \, c^2$ taking care of 
the shift in reference level. The rule (\ref{eq4.3}) is equivalent to requiring 
the identification of the ``non-relativistic" Hamiltonians (with subtraction of 
the rest-mass contribution)
\begin{equation}
H_0^{\rm NR} ({\mbox{\boldmath$q$}}' , {\mbox{\boldmath$p$}}') = H_{\rm 
real}^{\rm NR} (\mbox{\boldmath$q$} , \mbox{\boldmath$p$}) \, , \label{eq4.4}
\end{equation}
where the canonical coordinates in each problem must be mapped (because of the 
identification (\ref{eq4.1})) by a {\it canonical transformation},
\begin{equation}
\sum_i p_i \, dq^i = \sum_i p'_i \, dq'^i + dg (q, q') \, , \label{eq4.5}
\end{equation}
with some ``generating function'' $g(q,q')$.

We have explored the naive identification (\ref{eq4.3}), or (\ref{eq4.4}), and 
found that it was unsatisfactory. Indeed, one finds that it is {\it impossible} 
to require simultaneously that: (i) the energy levels coincide modulo an 
overall shift (\ref{eq4.3}), (ii) the effective mass $m_0$ coincides with the 
usual reduced mass $\mu = m_1 \, m_2 / (m_1 + m_2)$, and (iii) the effective 
metric (\ref{eq3.3}) depends only on $m_1$ and $m_2$. [This impossibility comes 
from the fact that the requirement (\ref{eq4.4}) is a very strong constraint 
which imposes more equations than unknowns.] If one insists on imposing the 
naive identification (\ref{eq4.3}) there is a price to pay: one must drop at 
least one of the requirements (ii) or (iii). Various possibilities are 
discussed in the Appendices of this paper. One possibility is to drop the requirement that 
$m_0 = \mu$. As discussed in App.~\ref{appA}, we find that there is a unique choice 
of masses in the effective problem, namely
\begin{equation}
m_0 = \mu \, \xi^{-2}\,, \quad \quad GM_0 = GM \, \xi^3 \, , \label{eq4.6}
\end{equation}
with
\begin{equation}
\xi^2 = \frac{1}{5} \left[ 2 \sqrt{100 + 30\nu + 4\nu^2} - 15 + \nu \right] \, 
, \label{eq4.7}
\end{equation}
which is compatible with the requirements (i) and (iii) above. However, we feel 
that it is quite unnatural to introduce an effective mass $m_0$ which differs 
from $\mu$ even in the non-relativistic limit $c \rightarrow +\infty$. We feel 
also that this possibility is so constrained that it is only available at the 
2PN level and will not be generalizable to higher post-Newtonian orders.

A second (formal) possibility is to introduce some energy dependence, either in 
$m_0$, say
\begin{equation}
\quad m_0 = \mu \left( 1 + \beta_1 \, \frac{{\cal E}_0^{\rm NR}}{\mu\,c^2} + \beta_2 
\left( \frac{{\cal E}_0^{\rm NR}}{\mu\,c^2} \right)^2 + \cdots \right) \, , 
\label{eq4.8}
\end{equation}
or in the effective metric (\ref{eq3.3}). Namely, the various coefficients 
$a_1 , b_1 , a_2 , b_2 , a_3 , \ldots$ in Eq.~(\ref{eq3.4}) can be expanded as
\begin{equation}
a_1 ({\cal E}_0) = a_1^{(0)} + a_1^{(2)} \, \frac{{\cal E}_0^{\rm NR}}{m_0\,c^2} + 
a_1^{(4)} \left( \frac{{\cal E}_0^{\rm NR}}{m_0\,c^2} \right)^2 + \cdots \, , 
\label{eq4.9}
\end{equation}
etc. These possibilities are discussed, for completeness, in App.~\ref{appB}.

Though the trick of introducing an energy dependence in (both) $m_0$ and the 
effective potential has been advocated, and used, in the quasi-potential 
approach of Todorov \cite{todorov}, \cite{MNT}, we feel that it is 
unsatisfactory. Conceptually, it obscures very much the nature of the mapping 
between the two problems, and, technically, it renders very difficult the 
generalization (we are interested in) to the case where radiation damping is 
taken into account (and where the energy is no longer conserved). We find much 
more satisfactory to drop the naive requirement (\ref{eq4.3}), and to replace 
it by the more general requirement that there exist a certain one-to-one 
mapping between the real energy levels and the effective ones, say
\begin{equation}
{\cal E}_0 ({\cal N}_0 , {\cal J}_0) = f \, [{\cal E}_{\rm real} ({\cal N} , 
{\cal J})] \, . \label{eq4.10}
\end{equation}
In explicit, expanded form, the requirement (\ref{eq4.10}) yields a deformed 
version of Eq.~(\ref{eq4.3}):
\begin{equation}
\frac{{\cal E}_0^{\rm NR}}{m_0 \, c^2} = \frac{{\cal E}_{\rm real}^{\rm 
NR}}{\mu \, c^2} \left( 1 + \alpha_1 \, \frac{{\cal E}_{\rm real}^{\rm 
NR}}{\mu\,c^2} + \alpha_2 
\left( \frac{{\cal E}_{\rm real}^{\rm NR}}{\mu\,c^2} \right)^2 
+ \cdots \right) \, . \label{eq4.11}
\end{equation}
Here, we assume that the standard identification (\ref{eq4.3}) holds (together 
with $m_0 = \mu + {\cal O} (c^{-2})$) in the non-relativistic limit $c 
\rightarrow \infty$.

We are going to show that the a priori arbitrary function $f$, i.e. the 
parameters $\alpha_1 , \alpha_2 , \ldots$ can be uniquely selected (at the 2PN 
level) by imposing the following physically natural requirements: (a) the mass 
of the effective test particle coincides with the usual reduced mass,
\begin{equation}
m_0 = \mu \, , \label{eq4.12}
\end{equation}
and, (b) the {\it linearized} (``one-graviton-exchange'') effective metric 
coincides with the linearized Schwarzschild metric with mass $M \equiv m_1 + 
m_2$, i.e.
\begin{equation}
a_1 = -2 GM \,,\quad \quad   b_1 = 2 GM \, . \label{eq4.13}
\end{equation}
Note that the requirement (\ref{eq4.12}) is actually imposed by dimensional 
analysis as soon as one requires $m_0 = \mu + {\cal O} (c^{-2})$. Indeed, as we 
bar any dependence on the energy, it is impossible to write any correction 
terms ${\cal O} (c^{-2})$ in the link between $m_0$ and $\mu$. The requirement 
(\ref{eq4.13}) is very natural when one thinks that the role of the effective 
metric is to reproduce, at all orders in the coupling constant $G$, the 
interaction generated by exchanging gravitons between two masses $m_1$ and 
$m_2$. The ``one-graviton-exchange'' interaction (linear in $G \, m_1 \, m_2$) 
depends only on the (Lorentz-invariant) relative velocity and corresponds to a 
linearized Schwarzschild effective metric in the test-mass limit $\nu 
\rightarrow 0$. As the coefficient $- \frac{1}{2} \, a_1$ is fixed (by 
dimensional analysis, as above) to its non-relativistic value $- \frac{1}{2} \, 
a_1 \, m_0 = G \, M_0 \, m_0 = G \, m_1 \, m_2$, it is very natural not to 
deform the coefficient $b_1$ by $\nu$-dependent corrections.

Let us now prove the consistency of the requirements (\ref{eq4.12}), 
(\ref{eq4.13}) and determine the energy mapping $f$. We can start from the 
result (\ref{eq3.13}), in which one replaces ${\cal E}_0^{\rm NR}$ by the 
expansion (\ref{eq4.11}). This leads again to an expression of the form 
(\ref{eq3.13}), with ${\cal E}_0^{\rm NR}$ replaced by ${\cal E}_{\rm 
real}^{\rm NR}$. One can simplify this expression by working with scaled 
variables:
\begin{eqnarray}
&& \widehat{I}_R^0 \equiv \frac{I_R^0}{\alpha_0} \,, 
\quad \quad \widehat{I}_R^{\rm real}  \equiv  
\frac{I_R^{\rm real}}{\alpha}\equiv i_r\,,
\quad \quad E_0 \equiv \frac{{\cal E}_0^{\rm NR}}{m_0} \,, 
\quad \quad E_{\rm real} \equiv \frac{{\cal E}_{\rm real}^{\rm NR}}{\mu} , \nonumber \\
&& {j}_0 \equiv \frac{{\cal J}_0}{\alpha_0} \,, \quad \quad {j} \equiv 
\frac{\cal J}{\alpha} \, . \label{eq4.14}
\end{eqnarray}
Here $\alpha_0 \equiv GM_0 \, m_0$ and $\alpha \equiv GM \, \mu \equiv G \, m_1 
\, m_2$ as above. We use also the scaled metric coefficients $\widehat{a}_i$ 
and $\widehat{b}_i$ of Eq.~(\ref{eq3.18}). Let us note, in passing, that, very 
generally, the dimensionless quantity $\widehat{\cal E}_0 / c^2 \equiv {\cal 
E}_0 / (m_0 \, c^2) = 1 + c^{-2} \, E_0$ is expressible entirely in terms of 
the dimensionless scaled action variables $\widehat{I}_a^0 / c = I_a^0 / 
(\alpha_0 \, c)$ and of the dimensionless scaled metric coefficients 
$\widehat{a}_i$, $\widehat{b}_i$. [This scaling behavior can be proved very 
easily by scaling from the start the effective action $S_0 = -\int m_0 \, c \, 
ds_0^{\rm eff} = - \alpha_0 \, c \int d\widehat{s}_0^{\rm eff}$ with 
$d\widehat{s}_0^2 \equiv (GM_0)^{-2} \, ds_0^2$, and by using scaled 
coordinates: $\widehat R = R / GM_0$, $\widehat t = t / GM_0$.]

Let us now make use of the assumptions $m_0 = \mu$ and $GM_0 \equiv -\frac{1}{2} 
\, a_1 = GM$ (so that $\alpha_0 = GM_0 \, m_0 = GM \, \mu = \alpha$). But, let 
us not yet assume the second equation (\ref{eq4.13}), i.e. let us assume 
$\widehat{a}_1 \equiv -2$, but let us not yet assume any value for 
$\widehat{b}_1 \equiv  b_1 / GM_0 \equiv b_1 / GM$. Within these assumptions, 
the scaled version of the result (\ref{eq3.13}), with ${\cal E}_0^{\rm NR}$ 
replaced by (\ref{eq4.11}), reads
\begin{eqnarray}
\widehat{I}_R^0 (E_0 (E_{\rm real}), j_0) &=& -j_0 + \frac{1}{\sqrt{-2 E_{\rm 
real}}} \left[ \widehat A + \widehat{B}' \, \frac{E_{\rm real}}{c^2} + 
\widehat{C}' \left( \frac{E_{\rm real}}{c^2} \right)^2 \right] \nonumber \\
&+& \frac{1}{c^2 j_0} \left[ \widehat D + \widehat E \, \frac{E_{\rm 
real}}{c^2} \right] + \frac{1}{c^4 j_0^3} \, \widehat F \, , \label{eq4.15}
\end{eqnarray}
where
\begin{eqnarray}
&& \widehat A = - \frac{1}{2} \, \widehat{a}_1 = 1 \,, \quad \quad \widehat{B}' = \frac{7}{4} 
+ \widehat{b}_1 - \frac{\alpha_1}{2} , \nonumber \\
&& \widehat{C}' = \frac{19}{32} + \frac{\widehat{b}_1}{4} + \frac{\alpha_1}{2}
\left( \widehat{b}_1 + \frac{7}{4} \right) + \frac{3}{8} \, 
\alpha_1^2 - \frac{\alpha_2}{2} \, , \label{eq4.16}
\end{eqnarray}
and where $\widehat D$, $\widehat E$ and $\widehat F$ are obtained from the 
expressions (\ref{eq3.14}) by the replacements $a_i \rightarrow \widehat{a}_i$, 
$b_i \rightarrow \widehat{b}_i$ (with $\widehat{a}_1 = -2$). Finally, 
identifying $[I_R^0 ({\cal E}_0 , {\cal J}_0)]_{{\cal J}_0 = {\cal J}_{\rm 
real}}^{{\cal E}_0 = f ({\cal E}_{\rm real})}$ with $I_R ({\cal E}_{\rm real} , 
{\cal J}_{\rm real} )$, or equivalently $\widehat{I}_R^0 (E_0 (E_{\rm 
real}),j_0)$ with $\widehat{I}_R (E_{\rm real} , j_0)$, yields five equations to be 
satisfied, namely the equations stating that $\widehat{B}'$, $\widehat{C}'$, 
$\widehat D$, $\widehat E$ and $\widehat F$ coincide with the corresponding 
coefficients in Eq.~(\ref{eq2.13}). The explicit form of these equations is
\bea
&& \frac{7}{4} + \widehat{b}_1 - \frac{\alpha_1}{2} = \frac{15}{4} - \frac{\nu}{4} 
\, , \label{eq4.17} \\
&& \frac{19}{32} + \frac{\widehat{b}_1}{4} + \frac{\alpha_1}{2} \left( 
\widehat{b}_1 + \frac{7}{4} \right) + \frac{3}{8} \, \alpha_1^2 - 
\frac{\alpha_2}{2} = \frac{35}{32} + \frac{15}{16} \, \nu + \frac{3}{32} \, 
\nu^2 \, , \label{eq4.18} \\
&& 2 -  \frac{\widehat{a}_2}{2} + \frac{\widehat{b}_1}{2} = 3 \, , \label{eq4.19} \\
&& 4 - \widehat{a}_2 + \widehat{b}_1 - \frac{\widehat{b}_1^2}{8} + 
\frac{\widehat{b}_2}{2} = \frac{15}{2} - 3\nu \, , \label{eq4.20} \\
&&6 - 3\,\widehat{a}_2 + \frac{\widehat{a}_2^2}{8} - \frac{\widehat{a}_3}{2} + 
\widehat{b}_1 - \frac{1}{4}\widehat{a}_2\,\widehat{b}_1 - \frac{\widehat{b}_1^2}{16} 
+ \frac{\widehat{b}_2}{4}
= \frac{35}{4} - \frac{5}{2} \, 
\nu \, . \label{eq4.21}
\eea
Note that the subsystem made of the two equations (\ref{eq4.17}), 
(\ref{eq4.18}) (corresponding to $\widehat{B}'$ and $\widehat{C}'$) contains 
the three unknowns $\widehat{b}_1$, $\alpha_1$, $\alpha_2$, while the three 
equations (\ref{eq4.19})--(\ref{eq4.21}) (corresponding to $\widehat D$, 
$\widehat E$ and $\widehat F$) contains the unknowns $\widehat{b}_1$, 
$\widehat{b}_2$, $\widehat{a}_2$, $\widehat{a}_3$. In this section we shall 
consider only the first (``$BC$'') subsystem, leaving the ``$DEF$'' system to 
the next section.

It is easily seen that the $BC$ subsystem would admit no solution in 
$\widehat{b}_1$ if we were to impose $\alpha_1 = \alpha_2 = 0$. This proves the 
assertion made above that one needs a non trivial energy mapping ${\cal E}_0 = 
f ({\cal E}_{\rm real})$. On the other hand, if we introduce the two free 
parameters $\alpha_1$, $\alpha_2$ the $BC$ subsystem becomes an indeterminate 
system of two equations for three unknowns. As argued above, it is physically 
very natural to impose that the linearized effective metric coincides with the 
linearized Schwarzschild metric, i.e. that
\begin{equation}
\widehat{b}_1 = 2 \, . \label{eq4.22}
\end{equation}
Then the $BC$ system (\ref{eq4.17}), (\ref{eq4.18}) admits the unique solution:
\begin{equation}
\alpha_1 = \frac{\nu}{2} \,, \quad \quad \alpha_2 = 0 \, . \label{eq4.23}
\end{equation}
This solution corresponds to the link
\begin{equation}
\frac{{\cal E}_0^{\rm NR}}{m_0 \, c^2} = \frac{{\cal E}_{\rm real}^{\rm 
NR}}{\mu \, c^2} \left( 1 + \frac{\nu}{2} \, \frac{{\cal E}_{\rm real}^{\rm 
NR}}{\mu \, c^2} \right) \, , \label{eq4.24}
\end{equation}
which is equivalent to
\begin{equation}
\frac{{\cal E}_0}{m_0 \, c^2} \equiv \frac{{\cal E}_{\rm real}^2 - m_1^2 \, c^4 
- m_2^2 \, c^4}{2 \, m_1 \, m_2 \, c^4} \, . \label{eq4.25}
\end{equation}
Remarkably, the map (\ref{eq4.25}) between the real total relativistic energy 
${\cal E}_{\rm real} = M \, c^2 + {\cal E}_{\rm real}^{\rm NR}$, and the 
effective relativistic energy ${\cal E}_0 = m_0 \, c^2 + {\cal E}_0^{\rm NR}$ 
coincides with the one introduced by Br\'ezin, Itzykson and Zinn-Justin 
\cite{BIZ}, which maps very simply the one-body relativistic Balmer formula onto 
the two-body one (in quantum electrodynamics). The same map was also recently 
used by Damour, Iyer and Sathyaprakash~\cite{DIS}. 
There it was emphasized that the function $\varphi 
(s)$ of the Mandelstam invariant $s = {\cal E}_{\rm real}^2$ appearing on the 
R.H.S. of Eq.~(\ref{eq4.25}) is the most natural symmetric function of the 
asymptotic\footnote{We consider here scattering states. By analytic 
continuation in $s$, the function $\varphi (s)$ is naturally expected to play a 
special role in the energetics of two-body bound states.} 4-momenta 
$p_1^{\mu}$, $p_2^{\mu}$ of a two-particle system which reduces, in the 
test-mass limit $m_2 \ll m_1$, to the energy of $m_2$ in the rest-frame of 
$m_1$. Indeed, (setting here $c=1$ for simplicity)
\begin{equation}
\varphi (s) \equiv \frac{s - m_1^2 - m_2^2}{2 \, m_1 \, m_2} = \frac{-(p_1 + 
p_2)^2 - m_1^2 - m_2^2}{2 \, m_1 \, m_2} = - \frac{p_1 \cdot p_2}{m_1 \, m_2} 
\, . \label{eq4.26}
\end{equation}
Finally, we have two a priori independent motivations for using the function 
$\varphi (s)$, i.e. the link (\ref{eq4.25}), to map the real two-body energy onto  
the effective one-body one: (i) the simplicity, and the symmetry, of the 
expression (\ref{eq4.26}) which generalizes the test-mass conserved energy 
${\cal E}_0 / m_0 = -K_{\mu} \, p_0^{\mu} / m_0$ (where $K_{\mu}$ is the 
Killing vector defined by the time direction of the background field) (see 
\cite{DIS}), and (ii) the fact that it corresponds to a linearized effective 
metric coinciding with the linearized Schwarzschild metric. Actually, these two 
facts are not really independent, because (as discussed in \cite{BIZ} and 
\cite{IZ}) they correspond heuristically to saying that the ``effective 
interaction'' is the interaction felt by any of the two particles in the rest 
frame of the other particle.

Summarizing: The rules we shall assume for relating the real two-body problem 
to the effective one-body one are Eqs.~(\ref{eq4.1}) (or equivalently the 
condition (\ref{eq4.5}) that the phase-space coordinates are canonically 
mapped), and Eq.~(\ref{eq4.25}).

\section{The effective one-body metric and the dynamics it defines}
\label{sec5}
Having specified the rules linking the real two-body problem to the effective 
one-body one, we can now proceed to the determination of the effective metric 
(at the 2PN level). We shall work in Schwarzschild coordinates:
\begin{equation}
ds_{\rm eff}^2 = -A(R) \, c^2 \, dt^2 + B(R) \, dR^2 + R^2 (d \theta^2 + \sin^2 
\theta \, d \varphi^2) \, , \label{eq5.1}
\end{equation}
with $A(R)$ and $B(R)$ constructed as expansions of the form (\ref{eq3.4}). It 
will be useful to rewrite also the effective metric in the form
\begin{equation}
\label{eq4.31}
ds_{\rm eff}^2 = -A(R) \, c^2 \, dt^2 + \frac{D(R)}{A(R)} \, dR^2 + R^2 (d 
\theta^2 + \sin^2 \theta \, d \varphi^2) \, , \label{eq5.2}
\end{equation}
in which we factorize, \`a la Schwarzschild, $g_{00}^{-1}$ in front of the 
$dR^2$ term, and consider that, besides $A(R)$, the second function constructed 
as an expansion in $1/R$ is
\begin{equation}
D(R) = A(R) \, B(R) = 1 + \frac{d_1}{c^2 \, R} + \frac{d_2}{c^4 \, R^2} + 
\cdots \, , \label{eq5.3}
\end{equation}
where
\begin{equation}
d_1 = a_1 + b_1 \,,\quad \quad  d_2 = a_2 + a_1\,b_1 + b_2\, . \label{eq5.4}
\end{equation}

To determine the effective metric, i.e. the coefficients $\widehat{a}_i$ and 
$\widehat{b}_i$, or equivalently $\widehat{a}_i$ and $\widehat{d}_i \equiv d_i 
/ (GM)^i$, we insert the known values of $\widehat{b}_1$, $\alpha_1$ and 
$\alpha_2$ (namely $\widehat{b}_1 = 2$, $\alpha_1 = \nu / 2$, $\alpha_2 = 0$) 
into the remaining equations (\ref{eq4.19})--(\ref{eq4.21}) (``$DEF$ system''). 
This yields three equations for the three unknowns $\widehat{a}_2$, 
$\widehat{b}_2$ and $\widehat{a}_3$. The unique solution of this $DEF$ system 
reads
\begin{equation}
\widehat{a}_2 = 0 \,, \quad \quad \widehat{a}_3 = 2\nu \,,\quad \quad  
\widehat{b}_2 = 4 - 6\nu \,. 
\label{eq5.5}
\end{equation}
In other words, our natural assumptions (\ref{eq4.12}), (\ref{eq4.13}) have led 
us uniquely to the simple energy map (\ref{eq4.25}) and to an effective 
one-body metric given by
\bea
A(R) &=& 1 - \frac{2GM}{c^2 \, R} + 2\nu \left( \frac{GM}{c^2 \, R} \right)^3 + 
\cdots \,, \label{eq5.6} \\
B(R) &=& 1 + \frac{2GM}{c^2 \, R} + (4-6\nu) \left( \frac{GM}{c^2 \, R} \right)^2 
+ \cdots \, , \label{eq5.7} \\
D(R) &=& 1 - 6\nu \left( \frac{GM}{c^2 \, R} \right)^2 + \cdots \, . 
\label{eq5.8}
\eea
The simplicity of the final results (\ref{eq5.6})--(\ref{eq5.8}) is striking. 
The effective metric (\ref{eq5.2}) is a simple deformation of the Schwarzschild 
metric ($A_s(R) = 1-2GM/c^2 \, R$, $D_s (R) = 1$) with deformation parameter 
$\nu$. Note also that there are no $\nu$-dependent corrections to $A(R)$ at the 
1PN level, i.e. no $\nu (GM/c^2 \, R)^2$ contribution to $A(R)$. The first 
$\nu$-dependent corrections enter at the 2PN level. Remembering that the 
(2PN) effective metric fully encodes the information contained in the 
complicated 2PN expressions (\ref{eq2.13}) or (\ref{eq2.14}), it is remarkable 
that the metric coefficients (\ref{eq5.6})--(\ref{eq5.8})  be so simple. The 
previous approach of Ref.~\cite{MNT} led to much more complicated expressions 
at the 1PN level (to which it was limited).

In this paper, we propose to trust the physical consequences of the effective 
metric (\ref{eq5.2}), with $A(R)$ given by Eq.~(\ref{eq5.6}) and $D(R)$ 
given by Eq.~(\ref{eq5.8}), even in the region where $R$ is of order of a few times 
$GM/c^2$. Note that even in the extreme case where $\nu = 1/4$ and $R \simeq 2 \, 
GM / c^2$ the $\nu$-dependent additional terms entering the effective metric 
remain relatively small: Indeed, in this case, $\delta_\nu A(R) = 
2\nu (GM/c^2 \, R)^3 = 1/16$ 
and $-\delta_\nu D(R) = 
6\nu \, (GM / c^2 \, R)^2 = 3/8$. We expect, therefore, that it should be 
a fortiori possible to trust the predictions of the effective metric (\ref{eq4.31}) near 
the innermost stable circular orbit, i.e.  around $R 
\simeq 6\,GM / c^2$ (where $\delta_\nu A(R) \simeq 2 \times 10^{-3}$ and 
$-\delta_\nu D(R) \simeq 4 \times 10^{-2}$). Note that this nice feature of having only a small 
deformation of Schwarzschild, even when $\nu = 1/4$, is not shared by the 
``hybrid'' approach of Kidder, Will and Wiseman \cite{KWW}. Indeed, as 
emphasized in Ref.~\cite{DIS}, the $\nu$-deformations considered in the hybrid approach 
are, for some 
coefficients, larger than unity when $\nu = 1/4$. This is related to the fact 
pointed out by Sch\"afer and Wex \cite{WS},\cite{SWmoriond} that, by applying the hybrid 
approach of \cite{KWW} to the Hamiltonian, instead of the equations of motion, 
one gets significantly different predictions.

Let us note also that, if we decide to write the effective metric in the form 
(\ref{eq5.2}), the existence of a simple zero in the function $A(R)$, say 
$A(R_H) = 0$, implies (if $D(R_H) \not= 0$, and $D(R) > 0$ for $R > R_H$) that 
the hypersurface $R = R_H$ is (like in the undeformed Schwarzschild case) a 
regular (Killing) horizon. As usual, one can define Kruskal-like coordinates to 
see explicitly the regular nature of the horizon $R = R_H$ (made of two 
intersecting null hypersurfaces). In our case, one checks easily that the 
function $A_{2{\rm PN}} (R)$ defined by the first three terms on the R.H.S. of 
(\ref{eq5.6}) admits a simple zero\footnote{
We consider here only the zero of $A_{\rm 2PN}(R)$ which is continuously connected 
to the usual horizon $R_H^S = 2GM/c^2$ when $\nu \rightarrow 0$.}
 at some $R_H (\nu)$, when $0 \leq \nu \leq 
\frac{1}{4}$. The position $R_H (\nu)$ of this ``effective horizon'' smoothly, 
and monotonically, evolves with the deformation parameter $\nu$ between $R_H 
(0) = 2 GM / c^2$ and
\begin{equation}
R_H \left( {1}/{4} \right) \simeq 0.9277 \left( {2GM}/{c^2} \right) \, . 
\label{eq5.9}
\end{equation}
This relatively small change of the horizon toward a smaller value, i.e. a 
smaller horizon area (to quote an invariant measure of the location of the 
horizon), suggests that the dynamics of trajectories in the effective metric 
will also be only a small deformation of the standard Schwarzschild case.

\begin{figure}
\centerline{\epsfig{file=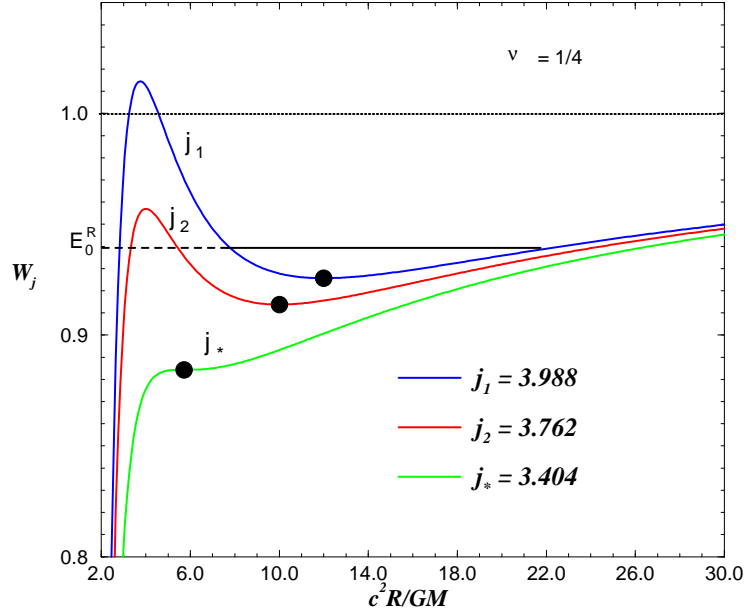,width=0.6\textwidth,angle=-90}}
\caption{\sl The effective radial potential $W_j(R)$ 
(at the 2PN level and for $\nu =1/4$) versus the dimensionless 
radial variable $c^2R/(GM)$ for three 
different values of the dimensionless angular momentum 
$j = c {\cal J}_{\rm real}/(GM\mu)$. Note that the  
effective radial potential tends to one for $R \rightarrow \infty$.
The stable circular orbits are 
located at the minima of the effective potential and are indicated 
by heavy black circles. The innermost stable circular orbit 
corresponds to the critical value $j_*$. In the case of the $j_1$ 
curve the orbit of a particle with  energy $E_0^R = \widehat{\cal E}_0$
is an elliptical rosette.}
\label{fig1}
\end{figure}

One of the main aims of the present work is indeed to study the dynamics (and 
the energetics) in the effective metric (\ref{eq5.2}). In particular, as 
gravitational radiation damping is known to circularize binary orbits, we are 
especially interested in studying the stable circular orbits in the effective 
metric. A convenient tool for doing this is to introduce an effective potential 
\cite{MTW}, \cite{LL}. Note that the Hamilton-Jacobi equation (\ref{eq3.9}) 
yields
\begin{equation}
\left( \frac{{\cal E}_0}{m_0 \, c^2} \right)^2 = W_{{\cal J}_0} (R) + 
\frac{A(R)}{B(R)} \left( \frac{P_R}{m_0 \, c} \right)^2 \geq W_{{\cal J}_0} (R) 
\, , \label{eq5.10}
\end{equation}
where $P_R \equiv \partial S_{\rm eff}/\partial R$ is the effective 
radial momentum, and where 
the ``effective radial potential'' $W_{{\cal J}_0} (R)$ is defined as
\begin{equation}
W_{{\cal J}_0} (R) \equiv A(R) \left[ 1 + \frac{({\cal J}_0 / m_0 \, c)^2}{C(R) 
\, R^2} \right] \, . \label{eq5.11}
\end{equation}
We read also from Eq.~(\ref{eq5.10}) the relativistic effective Hamiltonian 
\bea
H_0^R(R, P_R, P_\varphi) &=& m_0\,c^2\,
\sqrt{A(R)\,\left [ 1 + \frac{P_R^2}{m_0^2\,c^2\,B(R)} + 
\frac{P_\varphi^2}{m_0^2\,c^2\,C(R)\,R^2}\right ]}\,, \nonumber \\
&\equiv& m_0\,c^2\,\sqrt{W_{P_\varphi}(R) + 
\frac{A(R)}{B(R)} \left( \frac{P_R}{m_0 \, c} \right)^2 }\,.
\eea
The coordinate angular frequency along circular orbits is obtained 
by differentiating the Hamiltonian, that is  
\beq
\omega_0 \equiv \left (\frac{d\varphi}{d t}\right )_{\rm circ} =   
{\left (\frac{\partial H_0^R(R, P_R, P_\varphi)}{\partial P_\varphi} \right )}_{P_R = 0}\,,
\eeq  
which gives explicitly (using $P_\varphi = {\cal J}_0$)
\beq
\omega_0 = \frac{{\cal J}_0}{m_0\,C(R)\,R^2}\,\frac{\sqrt{A(R)}}
{\sqrt{1 + \frac{ {\cal J}_0^2}{m_0^2\,c^2\,C(R)\,R^2}}}\,.
\label{omega}
\eeq
Eqs.~(\ref{eq5.11}) and (\ref{omega}) are valid in an arbitrary radial 
coordinate gauge, but we shall use them  
in the Schwarzschild gauge where the metric coefficient $C(R) 
\equiv 1$. Note that $W(R)$ and $\omega_0$ then depend only on the metric coefficient $A(R)$. 
In dimensionless scaled variables $\widehat R \equiv c^2 \, R /(GM)$, $j_0 
\equiv c \, {\cal J}_0 / (GM \, \mu)$, $\widehat \omega_0 \equiv GM \omega_0/c^3$   
(in our case $M_0 = M$ and $m_0 = \mu$), 
the effective potential and the orbital frequency (along circular 
orbits) are quite simple:
\bea
W_{j_0} ( \widehat R ) &=& A( \widehat R ) \left[ 1 + \frac{j_0^2}{\widehat{R}^2} 
\right] \,, \nonumber \\
\widehat \omega_0 &=& \frac{j_0}{\widehat{R}^2}\,\frac{\sqrt{A(\widehat{R})}}
{\sqrt{1 + \frac{ j_0^2}{\widehat{R}^2}}}
\, . \label{eq5.12}
\eea
If we define the 2PN-accurate $A(R)$ by the straightforward truncation of 
Eq.~(\ref{eq5.6}), namely
\begin{equation}
A_{2 {\rm PN}} ( \widehat R ) = 1 - \frac{2}{\widehat R} + 
\frac{2\nu}{\widehat{R}^3} \, , \label{eq5.13}
\end{equation}
$W_{j_0}$ is a fifth-order polynomial in $u \equiv 1 / \widehat R \equiv GM/(c^2 
R)$. As the analytical study of the extrema of $W_{j_0}$ is rather 
complicated, we have used a numerical approach. When $\nu$ varies between $0$ 
and $1/4$, $W_{j_0}$ evolves into a smoothly deformed version of the standard 
Schwarzschild effective potential. To illustrate this fact, we plot in 
Fig.~\ref{fig1} $W_{j_0} ( \widehat R )$ for $\nu = \frac{1}{4}$ and for 
various values of the dimensionless angular momentum $j_0$. Note that the 
latter quantity coincides (in view of our rules) with the corresponding real 
two-body dimensionless angular momentum $j$:
\begin{equation}
j_0 \equiv \frac{c \, {\cal J}_0}{G \, M_0 \, m_0} = \frac{c \, {\cal J}_{\rm 
real}}{G \, M \, \mu} \equiv j \, . \label{eq5.14}
\end{equation}
[Note that our definition of the $j$'s differs by a factor $c$ from the one 
used in the previous section.]

As usual, because of the inequality (\ref{eq5.10}), when $j$ and $\widehat{\cal 
E}_0 \equiv {\cal E}_0 / (m_0 \, c^2)$ are fixed, the trajectory of a particle 
following a geodesic in the effective metric (\ref{eq5.2}) can be qualitatively 
read on Fig.~\ref{fig1}. For instance, in the case illustrated for the $j_1$ 
curve ($E_0^R \equiv \widehat{\cal E}_0$ line), the orbit will be an elliptical rosette, 
with the radial variable oscillating between a minimum and a maximum 
(solid line in Fig.~\ref{fig1}). The 
stable circular orbits are located at the minima of the effective potential 
(the maxima being unstable circular orbits). The innermost stable circular 
orbit (ISCO) corresponds to the critical value $j_*$ of the angular momentum 
where the maximum and the minimum of the effective potential fuse together to 
form an horizontal inflection point:
\begin{equation}
\frac{\partial \, W_{j_*}}{\partial \widehat{R}_*} = 0 = \frac{\partial^2 \, 
W_{j_*}}{\partial \widehat{R}_*^2} \, . \label{eq5.15}
\end{equation}
Let us, for comparison with our deformed case, recall the standard results for 
circular orbits in a Schwarzschild spacetime \cite{MTW}, \cite{LL}. With the 
notation $u \equiv GM_0 / c^2 \, R$ (for a Schwarzschild metric of mass $M_0$), 
the location, orbital frequency\footnote{Here, as well as in 
Eqs.~(\ref{eq5.22}) and (\ref{eq5.28}) below, $\omega$ denotes the angular 
frequency $d\varphi / dt$ on a circular orbit (in the equatorial plane).}, and 
energy of circular orbits are given, when $j$ varies, by
\begin{equation}
u = \frac{1}{6} \left[ 1 - \sqrt{1 - \frac{12}{j^2}} \,\right] \, , 
\label{eq5.16}
\end{equation}
\begin{equation}
\widehat{\omega}_S \equiv \frac{GM_0}{c^3} \, \omega = u^{3/2} \, , 
\label{eq5.17}
\end{equation}
\begin{equation}
\widehat{\cal E}_S \equiv \left( \frac{{\cal E}_0}{m_0 \, c^2} \right)^S = j 
(1-2u) \, u^{1/2} \, . \label{eq5.18}
\end{equation}
The ISCO corresponds to the critical values
\begin{equation}
j_*^S = \sqrt{12} \,, \quad \quad u_*^S = \frac{1}{6} \,, 
\quad \quad \widehat{\omega}_*^S = \frac{1}{6 \sqrt 6} \,, \quad \quad 
\widehat{\cal E}_*^S = \sqrt{\frac{8}{9}} \, . 
\label{eq5.19}
\end{equation}
In the deformed Schwarzschild case defined by Eq.~(\ref{eq5.13}), the ISCO for 
the extreme case $\nu = \frac{1}{4}$ is numerically found to correspond to the 
values
\bea
 && j_*^{2 {\rm PN}} \equiv \left( \frac{c \, {\cal J}_{\rm real}}{GM\mu} 
\right)_{\rm ISCO} = 3.404 = 0.983 \, j_*^S \, , \label{eq5.20} \\
&& u_{0*}^{2 {\rm PN}} \equiv \left( \frac{GM}{c^2 \, R} \right)_{\rm ISCO} = 0.1749 
= 1.049\, u_*^S \, , \label{eq5.21} \\
&& \widehat{\omega}_{0*}^{2 {\rm PN}} \equiv \left (\frac{GM\,\omega_0}{c^3}\right )_{\rm  
ISCO} = 0.07230 = 1.063 \, \widehat{\omega}_*^S \, , \label{eq5.22}\\
&&\widehat{\cal E}_{0*}^{2 {\rm PN}} \equiv \left( \frac{{\cal E}_0}{\mu \, c^2} 
\right)_{\rm ISCO} = 0.94040  = 0.99744 \, \widehat{\cal E}_*^S \, . \label{eq5.23}
\eea
Note that the Schwarzschild-coordinate radius of the effective ISCO is (when 
$\nu = 1/4$) $R^{\rm ISCO} = 5.718 \, GM / c^2$, i.e. lower than the standard 
Schwarzschild value $6\,GM/c^2$ corresponding to the total mass $M = m_1 + m_2$. 
This is consistent with the fact that the effective horizon was drawn in below 
$2 GM / c^2$ when $\nu$ was turned on. Note, however, that the three quantities 
$u_0^{2 {\rm PN}}$, $\omega_0^{2 {\rm PN}}$ and ${\cal E}_0^{2 {\rm PN}}$ 
entering equations (\ref{eq5.21})--(\ref{eq5.23}) are mathematical quantities 
defined in the {\it effective} problem, and not physical quantities defined in 
the real problem (hence the subscript $0$ added as a warning). [By contrast, 
$j^{2 {\rm PN}}$, Eq.~(\ref{eq5.20}) is directly related to the real, two-body 
angular momentum.] For physical (and astrophysical) purposes, we need to 
transform the information contained in Eqs.~(\ref{eq5.21})--(\ref{eq5.23}) into 
numbers concerning physical quantities defined in the real, two-body problem. 
For the energy, this is achieved (by definition) by using Eq.~(\ref{eq4.25}) to 
compute the real, two-body total energy ${\cal E}_{\rm real}$. Explicitly, the 
solution of Eq.~(\ref{eq4.25}) is (see also \cite{DIS})
\begin{equation}
{\cal E}_{\rm real} = M \, c^2 \ \sqrt{1+2\nu \left( \frac{{\cal E}_0 - m_0 \, 
c^2}{m_0 \, c^2} \right)} \, . \label{eq5.24}
\end{equation}
We need also to transform the effective orbital frequency $\omega_0$. This is 
easily done as follows. We know that the Hamiltonians of the real and effective 
problems are related by a mapping
\begin{equation}
H_{\rm real} \, (I_a^{\rm real}) = h (H_0 (I_a^0)) \, , \label{eq5.25}
\end{equation}
where $a = R,\theta ,\varphi$ (for the 3-dimensional problem), 
and where the function $h$ (the inverse of the 
function $f$ of Eq.~(\ref{eq4.10})) is, in our case, explicitly defined by 
Eq.~(\ref{eq5.24}). On the other hand, we know that the action variables are 
identically mapped onto each other: $I_a^0 = I_a^{\rm real}$ (canonical 
transformation). The frequency of the motion of any separated degree of freedom 
is given by the general formulas $\omega_a^0 = \partial \, 
H_0(\mbox{\boldmath$I$}^0) / \partial \,I_a^0$, $\omega_a^{\rm real} = \partial \, 
H_{\rm real}(\mbox{\boldmath$I$}^{\rm real}) / \partial \, I_a^{\rm real}$, 
where the Hamiltonians are considered as functions of the canonically 
conjugate action-angle variables $(I_a, \theta_a)$ (remembering that for such 
integrable systems, the Hamiltonian does not depend on the $\theta$'s).
Therefore the frequencies of the real problem are all obtained from the 
frequencies of the effective one by a common, energy-dependent factor
\begin{equation}
\frac{\omega_a^{\rm real}}{\omega_a^0} = \frac{dt_0}{dt^{\rm real}} = \frac{d 
H_{\rm real}}{d H_0} = \frac{\partial \, h(H_0)}{\partial \, H_0} \, . 
\label{eq5.26}
\end{equation}
In our case this ``blue shift''\footnote{For bound states, $\omega^{\rm real} > 
\omega^0$.} factor reads
\begin{equation}
\frac{\omega_a^{\rm real}}{\omega_a^0} = \frac{dt_0}{dt^{\rm real}} = 
\frac{1}{\sqrt{1 + 2\nu \, ({\cal E}_0 - m_0 \, c^2) / m_0 \, c^2}}  \, . 
\label{eq5.27}
\end{equation}
\begin{figure}
\begin{center}
\begin{tabular}{cc}   
\hspace{-1.2cm} 
\epsfig{file=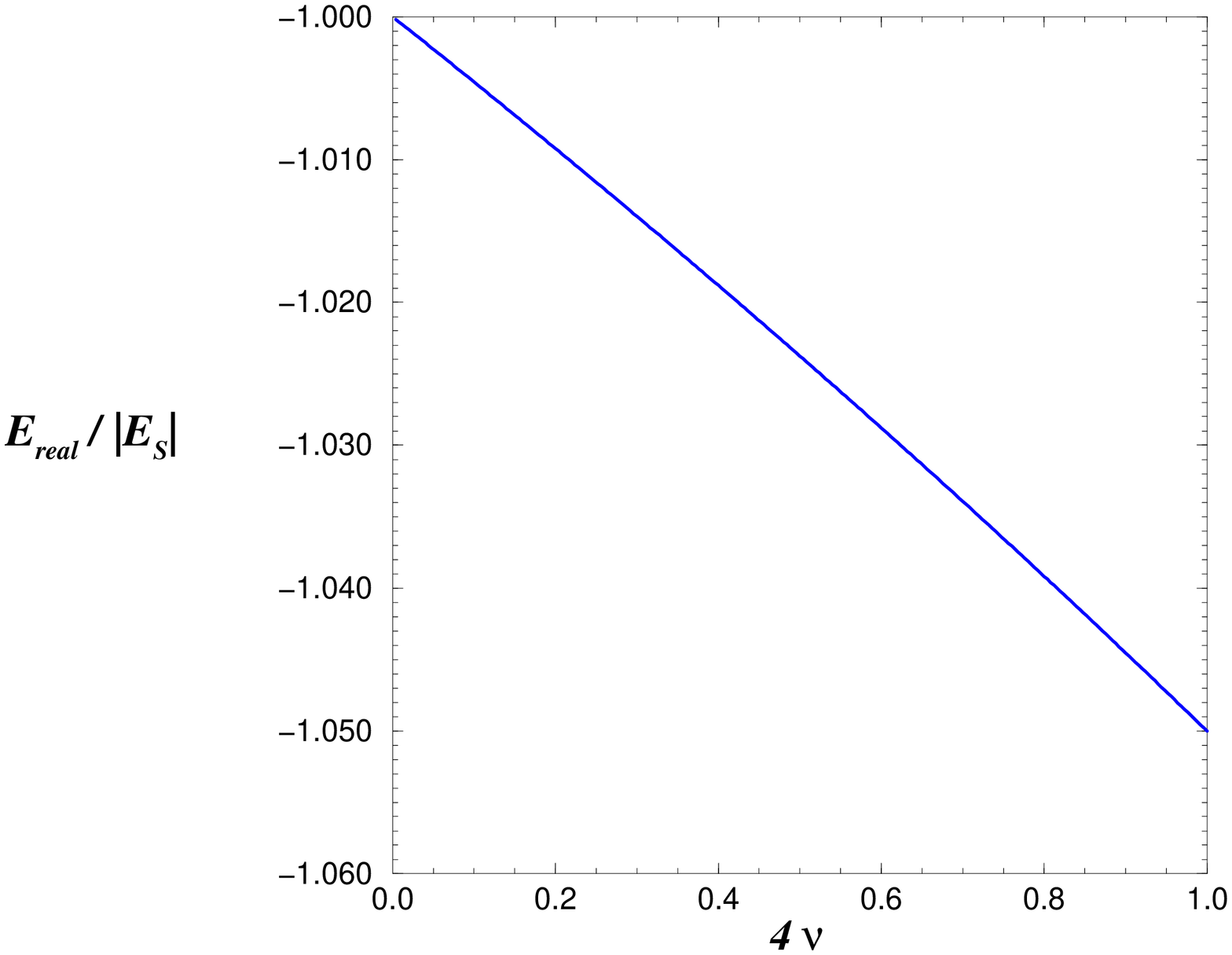,width=0.5\textwidth,height = 0.5\textwidth,angle=-90} & 
\hspace{-1cm}
\epsfig{file=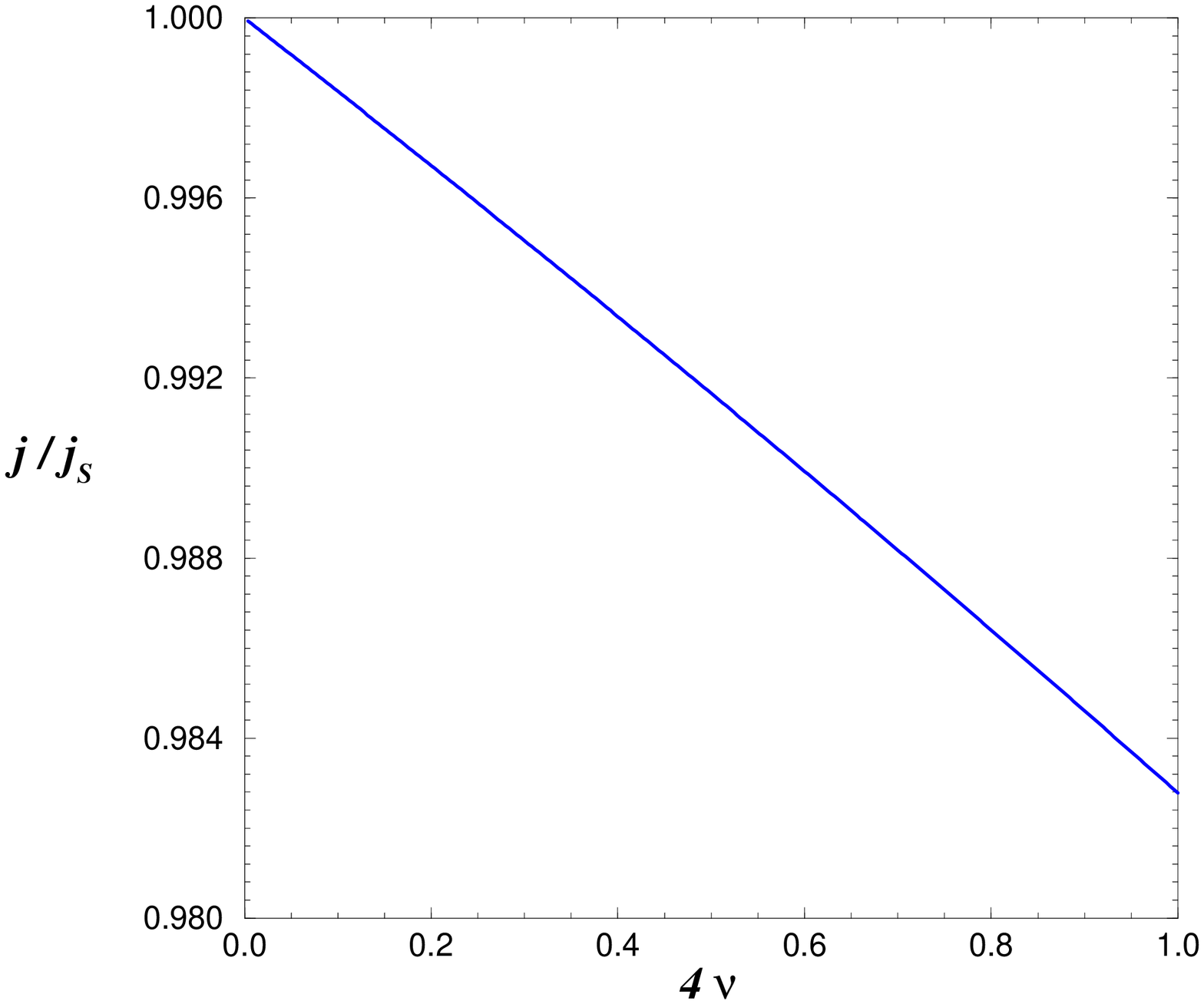,width=0.5\textwidth,height = 0.5\textwidth,angle=-90}
\end{tabular}
\caption{\sl Variation with $\nu$ (at the 2PN level) of the ISCO values 
of the real non-relativistic 
energy $E_{\rm real} \equiv \widehat{\cal E}_{\rm real}^{\rm NR} \equiv 
({\cal E}_{\rm real} -M\,c^2)/\mu c^2$ (on the left) and 
of the real angular momentum  ${j} \equiv c{\cal J}_{\rm real}/GM\mu$ 
(on the right), divided by the corresponding Schwarzschild values 
$|E_{\rm S}| \equiv 
|\widehat{\cal E}_{\rm S}^{\rm NR}| = 1 - \sqrt{8/9} \simeq 0.05719$ and ${j}_{\rm S} = 
\sqrt{12}$, respectively.}
\label{fig2}
\end{center}
\end{figure}
As indicated in Eqs.~(\ref{eq5.26}) and (\ref{eq5.27}) the same 
energy-dependent ``blue shift'' factor maps the effective and the real times 
(along corresponding orbits). 
Note that we have here a simple generalization of the spatial canonical 
transformation $(d \mbox{\boldmath$p$} \wedge d \mbox{\boldmath$q$} = d 
\mbox{\boldmath$p$}_0 \wedge d \mbox{\boldmath$q$}_0)$ to the time domain $(dH 
\wedge dt = dH_0 \wedge dt_0)$.

\begin{figure}
\centerline{\epsfig{file=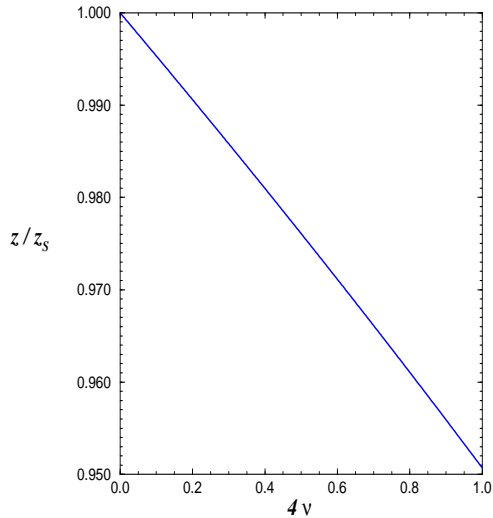,width=0.5\textwidth,
height = 0.5\textwidth,angle=-90}}
\caption{\sl ISCO values (at the 2PN level) 
of the quantity $z =(GM \omega_{\rm real}/c^3)^{-2/3}$, 
divided by the Schwarzschild value 
$z_S = 6$, versus $\nu$.}
\label{fig3}
\end{figure}

Applying the transformations (\ref{eq5.24}) and (\ref{eq5.26}), we obtain the 
physical quantities\footnote{In Eq.~(\ref{eq5.28}) $\omega_{\rm real} = d {\cal 
E}_{\rm real} / d {\cal J}_{\rm real}$ is again the angular frequency on a 
circular orbit. It should not be confused with the radial (periastron to periastron)
frequency $\omega_R$ 
for non-circular, rosette orbits.} predicted by our effective 2PN metric, 
still in the extreme case $\nu = 1/4$,
\bea
&& \widehat{\omega}_{{\rm real} *}^{2 {\rm PN}} = \left (\frac{GM}{c^3} \,\omega_{\rm 
real}\right )_{\rm ISCO} = 1.079 \, \widehat{\omega}_*^S = 0.07340\, , \label{eq5.28}\\
&& \left( \frac{{\cal E}_{\rm real}^{2 {\rm PN}} - M c^2}{\mu \, c^2} \right)_{\rm 
ISCO} = 1.050\,(\widehat{\cal E}_*^S - 1) =  - 0.06005\,. \label{eq5.29}
\eea
We represent in Figs.~\ref{fig2} and ~\ref{fig3}  
the variation with $\nu$ of the ISCO values of 
the real non-relativistic energy, $E_{\rm real} 
\equiv \widehat{\cal E}_{\rm real}^{\rm NR} \equiv 
({\cal E}_{\rm real} -M\,c^2)/\mu c^2$, 
the real angular momentum, ${j} \equiv c{\cal J}_{\rm real}/GM\mu$, 
and of the quantity
\begin{equation}
z \equiv \left( \frac{GM}{c^3} \, \omega_{\rm real} \right)^{-2/3} \, , 
\label{eq5.30}
\end{equation}
which is an invariant measure of the radial 
position of the orbit, and which 
coincides with the scaled Schwarzschild radius $\widehat R = c^2 R/(GM)$ in 
the test-mass limit $\nu \rightarrow 0$. One checks that our ISCO values 
respect the ``black hole limit'' ${\cal J}_{\rm real} < G\,{\cal E}^2_{\rm real}/c^5$, 
so that the system does not need to radiate a lot of gravitational waves in the final 
coalescence before being able to settle down as a black hole.

Let us now briefly compare our predictions with previous ones in the 
literature. The first attempt to address the question of the ISCO for binary 
systems of comparable masses was made by Clark and Eardley \cite{CE}. They 
worked only at the 1PN level, and predicted that the ISCO should be 
significantly more tightly bound than in the Schwarzschild case (with $M_0 = M 
= m_1 + m_2$): ${\cal E}_{\rm CE}^{\rm NR} / \mu \, c^2 \simeq -0.1$ 
when $\nu = 1/4$, compared to ${\cal E}_{\rm Schwarz}^{\rm NR} / m_0 \, c^2 = 
\sqrt{8/9} - 1 \simeq -0.0572$. Blackburn and Detweiler \cite{BD92} used an 
initial value formalism (which is only a rough approximation, even in the 
test-mass limit) to predict an extremely tight ISCO when $\nu = 1/4 : {\cal 
E}_{\rm BD}^{\rm NR} / \mu c^2 \simeq -0.7$. Kidder, Will and Wiseman 
\cite{KWW} were the first to try to use the full 2PN information contained in 
the Damour-Deruelle equations of motion (\ref{eq1.1}) to estimate analytically 
the change of the ISCO brought by turning on a finite mass ratio $\nu$. They 
introduced an ``hybrid'' approach in which one re-sums exactly the 
``Schwarzschild'' ($\nu$-independent) terms in the equations of motion, and 
treats the $\nu$-dependent terms as additional corrections. In contrast
with our present 2PN-effective approach (and also with the less reliable 
previous studies \cite{CE}, \cite{BD92}), they predict\footnote{
We use here the values read on the figures 3 and 4 of Ref.~\cite{KWW}:
for ${\cal E}^{\rm NR}$ and $(m f) = 0.00963$, which refer to a static ISCO 
without radiation damping.} that, when $\nu$ 
increases, the ISCO becomes markedly less tightly bound: e.g. ${\cal E}_{\rm 
KWW}^{\rm NR} / \mu c^2 \simeq -0.0377$ when $\nu = 1/4$. If their trend were 
real, this would imply that, except for the very stiff equations of state of 
nuclear matter (leading to large neutron star radii), the final plunge 
triggered when the ISCO is reached by an inspiraling 
$(1.4 M_\odot + 1.4 M_\odot)$ neutron star binary would probably take place 
before tidal disruption. 
However, both the robustness and the consistency of the hybrid approach of 
\cite{KWW} have been questioned. Wex and Sch\"afer \cite{WS} showed that the 
predictions of the hybrid approach were not ``robust'' in that they could be 
significantly modified by applying this approach to the Hamiltonian, rather 
than to the equations of motion. Sch\"afer and Wex \cite{SWmoriond} further 
showed that the predictions of the hybrid approach were not robust under a change 
of coordinate system. Moreover, 
Ref.~\cite{DIS} has questioned the consistency 
of the hybrid approach by pointing out that the formal ``$\nu$-corrections'' 
represent, in several cases, a very large (larger than 100\%) modification of 
the corresponding $\nu$-independent terms. This unreliability of the hybrid 
approach casts a doubt on the ISCO estimates of Ref.~\cite{Lombardi} which 
are based on hybrid orbital terms, and which use only 1PN accuracy in most terms.

Damour, Iyer and Sathyaprakash \cite{DIS} have introduced (at the 2PN level) another 
analytical approach to the determination of the ISCO, based on the Pad\'e approximants of 
some invariant energy function (closely related with the energy transformation 
(\ref{eq4.25})). Their trend is consistent with the one found in the present 
paper, namely a more tightly bound ISCO: for $\nu = 1/4$, the Pad\'e approximant 
approach predicts ${\cal E}_{\rm DIS}^{\rm NR} / \mu c^2 \simeq -0.0653$. 
\begin{figure}
\begin{center}
\begin{tabular}{cc}   
\hspace{-1.2cm} 
\epsfig{file=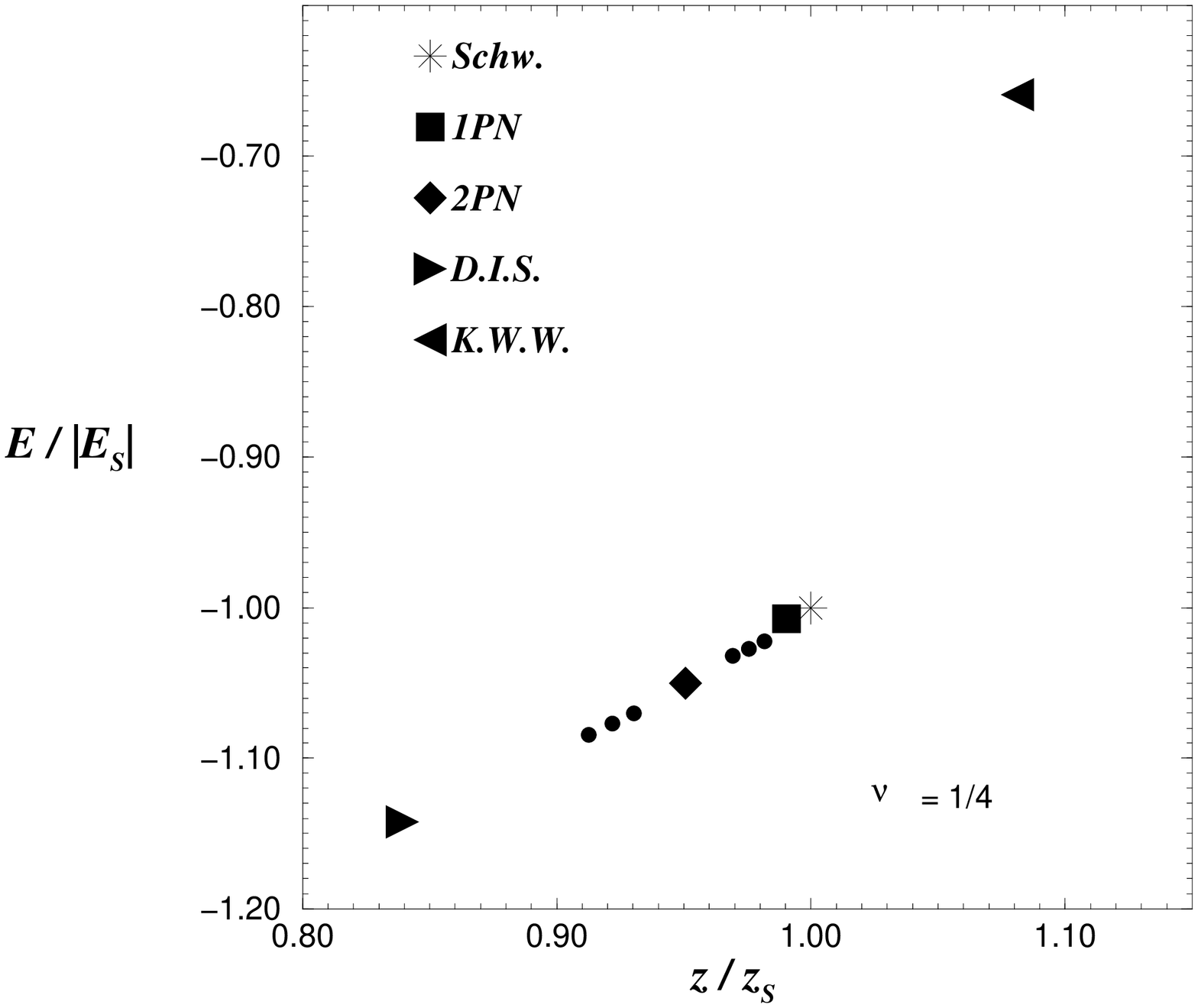,width=0.5\textwidth,height = 0.5\textwidth,angle=-90} & 
\hspace{-1cm}
\epsfig{file=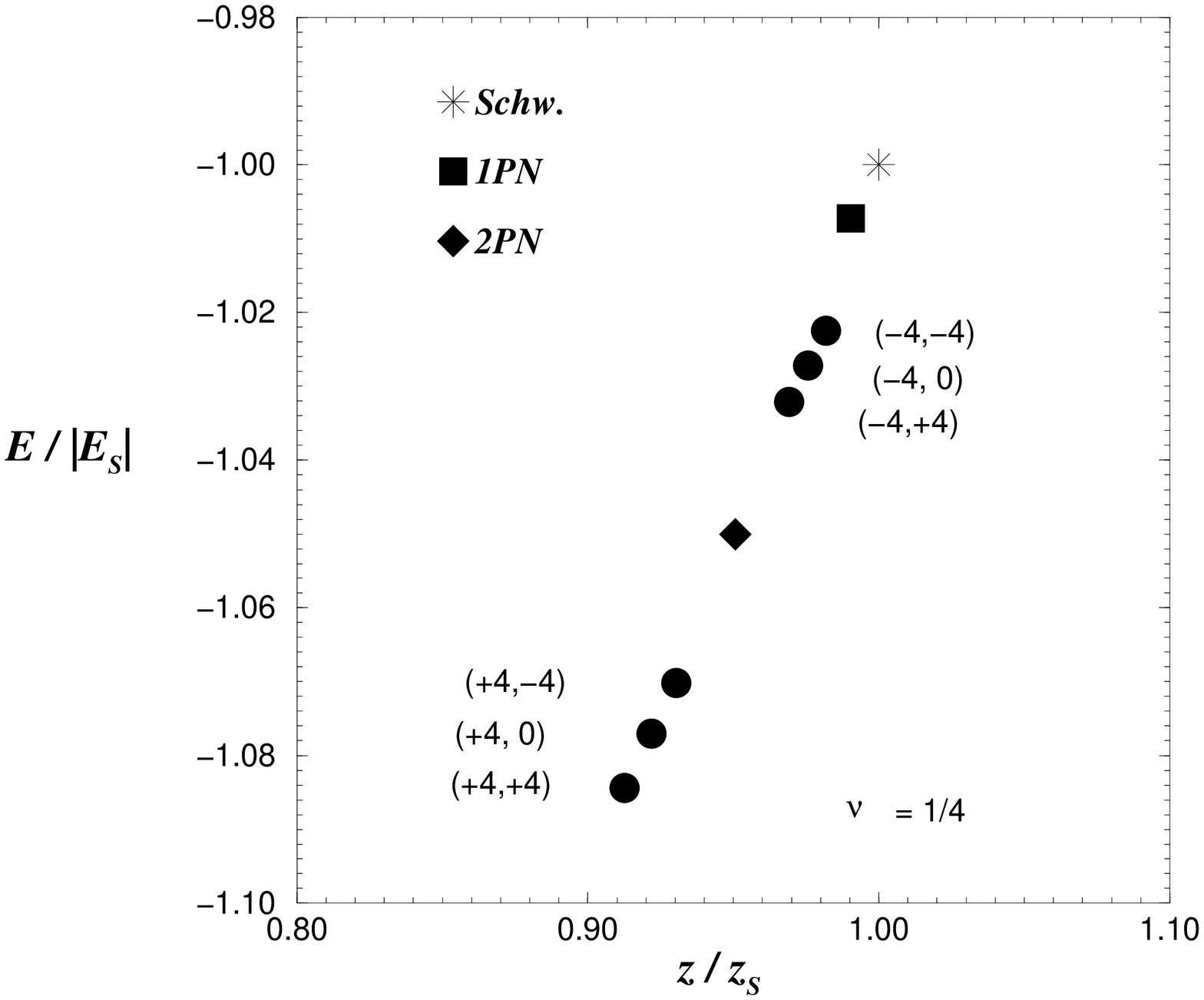,width=0.5\textwidth,height = 0.5\textwidth,angle=-90}
\end{tabular}
\caption{\sl ISCO values (for $\nu = 1/4$) 
of the real non-relativistic energy 
$E \equiv \widehat{\cal E}_{\rm real}^{\rm NR}$,  
divided by the corresponding Schwarzschild value $E_S \equiv 
\widehat{\cal E}_{\rm S}^{\rm NR}$, 
versus $z/z_S$. On the left we have compared 
our predictions at the 1PN level $({\blacksquare})$  and 
2PN level $(\blacklozenge)$ with the results obtained 
in~[21] $(\blacktriangleright)$ and~[22] $(\blacktriangleleft)$. 
The $(\ast)$ indicates the Schwarzschild predictions. 
The right panel is a magnification of the part of the 
left one in which we analyze the robustness of our method by exhibiting 
the points $(\bullet)$ 
obtained by introducing in the effective metric reasonable 
3PN and 4PN contributions: $(a'_4 , a'_5) = (\pm 4 , -4)$, $(\pm 
4 , 0)$ and $(\pm 4 , +4)$ in the notation of Eq.~(\ref{eq5.31}).}
\label{fig4}
\end{center}
\end{figure}

Numerical methods have recently been used to try to locate the ISCO for binary 
neutron stars \cite{Wilson}, \cite{Baumgarte}. However, we do not 
think that the {\it truncation} of Einstein's field equations (to a conformally 
flat spatial metric) used in these works is a good approximation 
for close orbits. Indeed, at the 2PN approximation, some numerically 
significant terms in the interaction potential come from the transverse-traceless 
part of the metric \cite{OOKH74}, \cite{D82}, \cite{DS85}. Moreover, 
the (unrealistic) assumption used in these works that the stars 
are corotating has probably also a significant effect on the location 
of the ISCO by adding both spin-orbit and spin-spin interaction terms.

This large scatter in the predictions for the location of the ISCO for 
comparable masses poses the question of the ``robustness'' of our new, 
effective-action approach. The main problem can be formulated as follows. 
Assuming that the effective-action approach (for the time-symmetric part of the 
dynamics) makes sense at higher post-Newtonian levels, the ``exact'' effective 
function $A(R)$ will read
\begin{table}
\begin{center}
\begin{tabular}{lcccc}
{\sl Method}   & {\sl ${\cal E}^{\rm NR}_{\rm real}/M c^2$} & $z$ & {\sl $\widehat{\omega}_
{\rm real}$} & {\sl $f_\odot$ (kHz)} \\ \hline
``Schwarzschild'' & $-0.01430$ & $6$  & $0.06804$& $2.199$\\
{\sl Eff. action 1PN} & $-0.01440$ & $5.942$ & $ 0.06904$ & $2.231$\\
{\sl Eff. action 2PN} & $-0.01501$ & $5.704$ & $ 0.07340$ & $2.372$ \\
{\sl Eff. action $(a'_4 , a'_5) = (-4,-4)$} & $-0.01462$ & $5.891$ &$ 0.06994$ & $2.260$ \\
{\sl Eff. action $(a'_4 , a'_5) = (-4, 0)$} & $-0.01469$ & $5.854$ & $0.07061$ & $2.267$ \\
{\sl Eff. action $(a'_4 , a'_5) =(-4,+4)$} & $-0.01476$ & $5.815$ &$ 0.07131$ & $2.304$ \\
{\sl Eff. action $(a'_4 , a'_5) =(+4,-4)$} & $-0.01530$ &$ 5.583$ & $0.07582$ & $2.450$ \\
{\sl Eff. action $(a'_4 , a'_5) =(+4, 0)$} & $-0.01540$ &$ 5.531$ &$ 0.07688$ & $2.484$ \\
{\sl Eff. action $(a'_4 , a'_5) =(+4,+4)$} &$ -0.01551$ & $5.475$ &$ 0.07806$ & $2.522$ \\
{\sl D.I.S. [21]} & $-0.01633$ & $5.036$ & $0.08850$ & $2.860$ \\
{\sl K.W.W. [22]} & $-0.00943$ &$ 6.49$ & $0.0605$ & $1.96$
\end{tabular}
\caption{\sl Summary of the ISCO values used in  Fig.~4 ($\nu =1/4$). 
Note that we give here ${\cal E}^{\rm NR}_{\rm real}/M c^2$, that is the ratio  
between the energy that can be radiated in gravitational waves before the final 
plunge and  the total mass-energy initially available. 
The first row refers to the naive estimate defined by a test particle 
of mass $\mu$ in a Schwarzschild spacetime of mass $M$.
We show also in the last column the solar-mass-scaled orbital frequency $f_\odot$ 
defined by $f_{\rm real} = \omega_{\rm real}/(2 \pi) \equiv f_\odot \,(M_\odot/M)$.}
\label{tab1}
\end{center}
\end{table}
\begin{equation}
A(R) = 1 - 2 \left( \frac{GM}{c^2 \, R} \right) + 2\nu \left( \frac{GM}{c^2 \, 
R} \right)^3 + \nu a'_4 \left( \frac{GM}{c^2 \, R} \right)^4 + \nu a'_5 \left( 
\frac{GM}{c^2 \, R} \right)^5 + \cdots\,. \label{eq5.31}
\end{equation}
The question is then to know how sensitive is the location of the ISCO to the 
values of the (still unknown) coefficients $a'_4 , a'_5 , \ldots$. One 
should have some a priori idea of the reasonable range of values of $a'_4 , a'_5 
, \ldots$. A rationale for deciding upon the reasonable values of $a'_4$ is the 
following. At the 2PN level, it is formally equivalent to use (with $u \equiv GM/c^2 
\, R$) $A_{2 {\rm PN}} = 1 - 2u + 2\nu \, u^3$ or the factorized form $A'_{2 
{\rm PN}} = (1 - 2u) (1+2\nu \, u^3)$. However, $A'_{2 {\rm PN}} = A_{2 {\rm 
PN}} - 4\nu \, u^4$ which corresponds to $a'_4 = -4$. This suggest that $-4 \leq 
a'_4 \leq +4$ is a reasonable range. We shall also consider $-4 \leq a'_5 \leq 
+4$ as a plausible range. Note that both choices correspond to having 
coefficients of $u^n$ which vary between $-1$ and $+1$ when $\nu = 1/4$. The 
robustness of our effective-action predictions against the introduction of 
$a'_4$ and $a'_5$ is illustrated in Fig.~\ref{fig4}. The numerical values used in 
Fig.~\ref{fig4} are exhibited in Tab.~\ref{tab1}.

Fig.~\ref{fig4} plots the ratio $E / \vert E_S \vert$ where $E \equiv 
{\cal E}_{\rm real}^{\rm NR} / \mu c^2 \equiv ({\cal E}_{\rm real} - M\,c^2)/\mu c^2$
at the ISCO (for $\nu = 1/4$) and $E_S = 
\sqrt{(8/9)} - 1 \simeq -0.05719$ is the corresponding ``Schwarzschild'' value, versus $z/z_S$ 
where $z$ is defined in Eq.~(\ref{eq5.30}), and where $z_S = 6$. This figure
compares the predictions of Ref.~\cite{KWW}, of Ref.~\cite{DIS} and of our new, 
effective-action prediction (at the 2PN level). We have also added what would 
be the prediction of the effective-action approach at the 1PN level. Note that, 
at the 1PN level, the function $A(R)$, Eq.~(\ref{eq5.6}), exactly coincides 
with the Schwarzschild one, but that the energy mapping (\ref{eq4.24}) 
introduces a slight deviation from the test-mass limit. Fig.~\ref{fig4} exhibits 
also the points obtained when considering $(a'_4 , a'_5) = (\pm 4 , -4)$, $(\pm 
4 , 0)$ and $(\pm 4 , +4)$. We see on this figure that the main prediction 
of the present approach (a prediction already clear from the fact that the 2PN 
contribution to $A(R)$ is fractionally small), namely that the ISCO is only 
slightly more bound than in the test-mass limit, is robust under the addition of 
higher PN contributions. The sensitivity to $a^\prime_4$ 
of the binding energy is only at the $\sim 3 \%$ level (for 
$a^\prime_4 = \pm 4$), while its sensitivity to the 4PN-coefficient 
$a^\prime_5$ is further reduced to the $\sim 0.6 \%$ level (for $a^\prime_5 = 
\pm 4)$. Still, it would be important to determine the 
3PN coefficient $a'_4$ to refine the determination of the ISCO quantities.

\section{Explicit mapping between the real problem and the effective one}
\label{sec6}
The basic idea of the effective one-body approach is to map the complicated and 
badly-convergent PN-expansion of the dynamics of a two-body system onto a 
simpler auxiliary one-body problem. We have shown in the previous sections that 
by imposing some simple, coordinate-invariant requirements, we could uniquely 
determine that the one-body dynamics was defined (at the 2PN level) by geodesic 
motion in a certain deformed Schwarzschild spacetime. The latter dynamics can be 
solved exactly by means of quadratures (e.g. by using the Hamilton-Jacobi 
method, see Eqs.~(\ref{eq3.7})--(\ref{eq3.12})). Note that this exact solution 
defines a particular re-summation of the original 2PN-expanded dynamics. The 
hope (that we tried to subtantiate in Sec.~\ref{sec5}) is that this re-summation 
captures, with sufficient approximation, the crucial non-perturbative aspects of 
the two-body dynamics, such as the existence of an ISCO.

As all the current work about the equations of motion, and/or the 
gravitational-wave radiation, of binary systems is done in some specific 
coordinate systems (harmonic or ADM), we need to complete the 
(coordinate-invariant) work done in the previous sections by explicitly 
constructing the transformation which maps the variables entering the effective 
problem onto those of the real one. We have already mentioned that the 
transformation between harmonic and ADM coordinates has been explicitly worked 
out in Refs.~\cite{DS85} and \cite{DS88}. Here, we shall explicitly relate the 
ADM phase-space variables $\mbox{\boldmath$Q$} = \mbox{\boldmath$q$}_1  - 
\mbox{\boldmath$q$}_2$ and $\mbox{\boldmath$P$} = \partial \, S / \partial \, 
\mbox{\boldmath$Q$}$ of the {\it relative motion} (as defined in Sec.~\ref{sec2} 
above) to the coordinate and momenta of the effective problem. More precisely, 
we shall construct the map
\begin{equation}
q'^i = {\cal Q}^i (q^j , p_j) \,,\quad \quad  p'_i = {\cal P}_i (q^j , p_j) \, , 
\label{eq6.1}
\end{equation}
transforming the {\it reduced} ADM relative position and momenta $(q^i , p_i)$, 
defined in Eq.~(\ref{eq2.4}), into the corresponding {\it reduced 
cartesian-like} position and momenta $(q'^i , p'_i)$ canonically defined by the 
(Schwarzschild-gauge) effective action (\ref{eq3.2}). In other words,
\begin{equation}
q'^i = \frac{Q'^i}{GM} \,,\quad \quad  p'_i = \frac{P'_i}{\mu} \, , \label{eq6.2}
\end{equation}
with $Q'^1 = R \, \sin \theta \, \cos \varphi$, $Q'^2 = R \, \sin \theta \, \sin 
\varphi$, $Q'^3 = R \, \cos \theta$, and $P'_i = \partial \, S_{\rm eff}/ 
\partial \, Q'^i$. Here, the ``effective'' coordinates $R,\theta ,\varphi$ are 
those of Eq.~(\ref{eq5.1}) (in Schwarzschild gauge) and $S_{\rm eff} = - \int 
\mu c \, ds_{\rm eff}$. The corresponding effective Hamiltonian (with respect to 
the coordinate time $t$ of the effective problem) is easily found by solving 
$g_{\rm eff}^{\mu \nu} (Q') \, P'_{\mu} \, P'_{\nu} + m_0^2 \, c^2 = 0$ in terms 
of the energy ${\cal E}_0 = - P'_0$. Transforming the usual polar-coordinate 
result (equivalent to Eq.~(\ref{eq5.10})) into cartesian coordinates leads to
\begin{equation}
H_{\rm eff} (\mbox{\boldmath$Q$}' , \mbox{\boldmath$P$}') = \mu \, c^2 \, 
\sqrt{A(Q') \left[ 1 + \frac{(\mbox{\boldmath$n$}' \cdot 
\mbox{\boldmath$P$}')^2}{\mu^2 \, c^2 \, B(Q')} + \frac{(\mbox{\boldmath$n$}' 
\times \mbox{\boldmath$P$}')^2}{\mu^2 \, c^2} \right]} \, , \label{eq6.3}
\end{equation}
where $Q' \equiv \sqrt{\delta_{ij} \, Q'^i \, Q'^j} = R$, where 
$\mbox{\boldmath$n$}'^i = Q'^i / Q'$ is the unit vector in the radial direction,  
and where the scalar and vector products are performed as in Euclidean space. 
When scaling the effective coordinates as in (\ref{eq6.2}), we need to scale 
correspondingly the time variable, the Hamiltonian and the action of the 
effective problem:
\begin{equation}
\widehat t \equiv \frac{t}{GM} \,, \quad \quad 
\widehat{H}_{\rm eff} \equiv \frac{H_{\rm eff}}{\mu}\,, 
\quad \quad \widehat{S}_{\rm eff} \equiv \frac{S_{\rm eff}}{\mu GM} \, . \label{eq6.4}
\end{equation}
Note that the effective Hamiltonian (\ref{eq6.3}) contains the rest-mass contribution. The 
scaled version of (\ref{eq6.3}) simplifies to
\begin{equation}
\widehat{H}_{\rm eff} (\mbox{\boldmath$q$}' , \mbox{\boldmath$p$}') = c^2 \, 
\sqrt{A(q') \left[ 1 + \frac{\mbox{\boldmath$p$}'^2}{c^2} + 
\frac{(\mbox{\boldmath$n$}' \cdot \mbox{\boldmath$p$}')^2}{c^2} \left( 
\frac{1}{B(q')} - 1 \right) \right]} \, , \label{eq6.5}
\end{equation}
where $q' \equiv \sqrt{\delta_{ij} \, q'^i \, q'^j} = R / GM$ and $n'^i \equiv 
q'^i / q'$. As was mentioned above the identification of the action variables in 
the real and effective problems guarantees that the two problems are mapped by a 
canonical transformation, i.e. a transformation such that Eq.~(\ref{eq4.5}) is 
satisfied. It will be more convenient to replace the generating function 
$g(q,q')$ of Eq.~(\ref{eq4.5}) by the new generating function $\widetilde G 
(q,p') = g(q,q') + p'_i\,q'^i$ such that
\begin{equation}
p_i \, dq^i + q'^i \, dp'_i = d \, \widetilde G (q,p') \, . \label{eq6.6}
\end{equation}
We can further separate $\widetilde G (q,p')$ into $\widetilde{G}_{\rm id} 
(q,p') \equiv q^i \, p'_i$, which generates the identity transformation, and an 
additional (perturbative) contribution $G(q,p')$:
\begin{equation}
\widetilde G (q,p') = q^i \, p'_i + G (q,p') \,,\quad \quad   G (q,p') = \frac{1}{c^2} \, 
G_{1 {\rm PN}} (q,p') + \frac{1}{c^4} \, G_{2 {\rm PN}} (q,p') \, . 
\label{eq6.7}
\end{equation}
Eqs.~(\ref{eq6.6}), (\ref{eq6.7}) yield the link
\begin{equation}
q'^i = q^i + \frac{\partial \, G(q,p')}{\partial \, p'_i} \,, \quad \quad  p'_i = p_i - 
\frac{\partial \, G(q,p')}{\partial \, q^i} \, . \label{eq6.8}
\end{equation}
Note that Eqs.~(\ref{eq6.8}) are exact and determine $q'$ and $p$ in function of 
$q$ and $p'$. We have, however, written them in a form appropriate for 
determining, by successive {\it iteration}, $q'$ and $p'$ in function of $q$ and 
$p$. If needed (e.g. for applications of the present work to the direct 
numerical calculation of the effective dynamics in the original $q,p$ 
coordinates), it is numerically fast to iterate Eqs.~(\ref{eq6.8}) to get 
Eqs.~(\ref{eq6.1}). For our present purpose we need an explicit analytical 
approximation of Eqs.~(\ref{eq6.1}) at the 2PN level. Remembering that $G$ 
starts at order $1/c^2$, one easily finds that
\begin{eqnarray}
&&q'^i = q^i + \frac{\partial \, G(q,p)}{\partial \, p_i} - \frac{\partial \, 
G(q,p)}{\partial \, q^j} \ \frac{\partial^2 \, G(q,p)}{\partial \, p_j \, 
\partial \, p_i} + {\cal O} \left( \frac{1}{c^6} \right) \, , \nonumber \\
&&p'_i = p_i - \frac{\partial \, G(q,p)}{\partial \,q^i} + \frac{\partial \, 
G(q,p)}{\partial \, q^j} \ \frac{\partial^2 \, G(q,p)}{\partial \, p_j \, 
\partial \, q^i} + {\cal O} \left( \frac{1}{c^6} \right) \, . \label{eq6.9}
\end{eqnarray}
In the terms linear in $G(q,p)$ one needs to use the full (1PN $+$ 2PN) 
expression of $G(q,p)$, while in the quadratic terms it is enough to use $G_{1 
{\rm PN}} / c^2$.

To determine the generating function $G(q,p)$ we need to write the equation 
stating that, under the canonical transformation (\ref{eq6.8}), the effective 
Hamiltonian $H_{\rm eff} (q',p')$ is mapped into a function of $q$ and $p$ which 
is linked to the real (relativistic) Hamiltonian $H_{\rm real}^R (q,p)$ by our 
rule (\ref{eq4.25}). If we write this link in terms of the reduced effective 
Hamiltonian (\ref{eq6.5}), and of the reduced, non-relativistic real Hamiltonian 
$\widehat{H}_{\rm real}^{\rm NR} \equiv (H_{\rm real}^R - M c^2) / \mu$ (the 
same as $\widehat H$ appearing in Eqs.~(\ref{eq2.5}), (\ref{eq2.6}) above), it 
reads
\begin{equation}
1 + \frac{\widehat{H}_{\rm real}^{\rm NR} (q,p)}{c^2} \left( 1 + \frac{\nu}{2} \ 
\frac{\widehat{H}_{\rm real}^{\rm NR} (q,p)}{c^2} \right) = \frac{1}{c^2}\,\widehat{H}_{\rm 
eff} [q' (q,p) , p' (q,p) ] \, . \label{eq6.10}
\end{equation}
Actually, we found more convenient to work with the square of 
Eq.~(\ref{eq6.10}), so as to get rid of the square root in $\widehat{H}_{\rm 
eff}$, Eq.~(\ref{eq6.5}). Hence, writing (half) the square of Eq.~(\ref{eq6.10}),  
and Taylor-expanding $\widehat{H}_{\rm 
eff} [q' (q,p) , p' (q,p) ]$ using Eqs.~(\ref{eq6.7})--(\ref{eq6.9}), 
we get  at order $1/c^4$, the following partial differential equation 
for $G_{1 {\rm PN}} (q,p)$
\beq
\frac{\partial \, \widehat{H}_{\rm Newt}}{\partial \, q^i} \,\frac{\partial \, 
G_{1{\rm PN}}}{\partial \, p_i} - \frac{\partial \, 
\widehat{H}_{\rm Newt}}{\partial \, 
p_i} \ \frac{\partial \, G_{1 {\rm PN}}}{\partial \, q^i} =
\frac{\nu}{2}\,\mbox{\boldmath$p$}^4 - (1 + \nu)\,
\frac{\mbox{\boldmath$p$}^2}{q} + \left ( 1 - \frac{\nu}{2} \right )
\,\frac{(\mbox{\boldmath$n$} \cdot \mbox{\boldmath$p$})^2}{q}  
+ \left ( 1 + \frac{\nu}{2} \right )\,\frac{1}{q^2}\,,
\label{eqn1}
\eeq
where we have denoted the Newtonian Hamiltonian 
as $\widehat{H}_{\rm Newt}  \equiv \widehat{H}_{0} = \mbox{\boldmath$p$}^2/2 
- 1/q$ (see Eq.~(\ref{eq2.6a})). At order $1/c^6$, 
a more complex calculation gives the partial differential equation for $G_{2 {\rm 
PN}} (q,p)$
\vskip 0.2truecm 
\bea
&& \frac{\partial \, \widehat{H}_{\rm Newt}}{\partial \, q^i} \,\frac{\partial \, 
G_{2{\rm PN}}}{\partial \, p_i} - \frac{\partial \, 
\widehat{H}_{\rm Newt}}{\partial \, 
p_i} \ \frac{\partial \, G_{2 {\rm PN}}}{\partial \, q^i} = 
\frac{\nu}{2}\,\widehat{H}_{0}^3 + (1 + \nu)\,\widehat{H}_{0}\,\widehat{H}_{2}
+ \widehat{H}_{4}
- (2 + 3\,\nu)\,
\frac{(\mbox{\boldmath$n$} \cdot \mbox{\boldmath$p$})^2}{q^2} \nonumber \\
&-& \frac{\nu}{q^3} + 
\frac{\partial \,{\cal R}}{\partial \, q^i} \,\frac{\partial \, 
G_{1{\rm PN}}}{\partial \, p_i} - \frac{\partial \,{\cal R}}{\partial \, 
p_i} \ \frac{\partial \, G_{1 {\rm PN}}}{\partial \, q^i}
+ \frac{\partial \,G_{1{\rm PN}}}{\partial \, q^j} \,\frac{\partial^2 \, 
G_{1{\rm PN}}}{\partial \, p_j\,\partial \,p_i}\,
\frac{\partial \, \widehat{H}_{\rm Newt}}{\partial \, q^i}
- \frac{\partial \,G_{1{\rm PN}}}{\partial \, q^j} \,\frac{\partial^2 \, 
G_{1{\rm PN}}}{\partial \, p_j\,\partial \,q^i}\,
\frac{\partial \, \widehat{H}_{\rm Newt}}{\partial \, p_i} \nonumber \\
&-&\frac{1}{2}\,\frac{\partial \,G_{1{\rm PN}}}{\partial \, p_i} \,\frac{\partial \, 
G_{1{\rm PN}}}{\partial \, p_j}\,
\frac{\partial^2 \, \widehat{H}_{\rm Newt}}{\partial \, q^i
\,\partial \, q^j}
- \frac{1}{2}\,\frac{\partial \,G_{1{\rm PN}}}{\partial \, q^i} \,\frac{\partial \, 
G_{1{\rm PN}}}{\partial \, q^j}\,
\frac{\partial^2 \, \widehat{H}_{\rm Newt}}{\partial \, p_i
\,\partial \, p_j}\,,
\label{eqn2}
\eea
\vskip 0.2truecm 
where $\widehat{H}_{2}$ and $\widehat{H}_{4}$ are given by 
Eqs.~(\ref{eq2.6b}),~(\ref{eq2.6c}), while 
\beq
{\cal R} = \frac{1}{q}\,((\mbox{\boldmath$n$} \cdot \mbox{\boldmath$p$})^2 
 + \mbox{\boldmath$p$}^2)\,.
\eeq
The partial differential equations (\ref{eqn1}) and ~(\ref{eqn2})
have the general form
\begin{equation}
\frac{\partial \, \widehat{H}_{\rm Newt}}{\partial \, q^i} \ \frac{\partial \, 
G_n}{\partial \, p_i} - \frac{\partial \, \widehat{H}_{\rm Newt}}{\partial \, 
p_i} \ \frac{\partial \, G_n}{\partial \, q^i} = \frac{q^i}{q^3} \ 
\frac{\partial \, G_n}{\partial \, p_i} - p_i \, \frac{\partial \, G_n}{\partial 
\, q_i} = K_n (q,p) \, , \label{eq6.11}
\end{equation}
where, at each PN order $n = 1$PN or 2PN, the R.H.S. is a known source term
 $K_n(q,p)$. Note that the L.H.S. of Eq.~(\ref{eq6.11}) is the Poisson bracket 
$\{ \widehat{H}_{\rm Newt}, G_n \}$, or, equivalently, minus the time derivative 
of $G_n$ along the Newtonian motion. 
It is easily checked that the solution of Eq.~(\ref{eq6.11}) is unique modulo the 
addition of terms generating a constant time shift or a spatial rotation. 
[Indeed, the homogeneous scalar solutions of Eq.~(\ref{eq6.11}) must correspond 
to the scalar constants of motion of the Keplerian motion: $\widehat{H}_{\rm 
Newt}(\mbox{\boldmath$q$}, \mbox{\boldmath$p$})$ 
and $(\mbox{\boldmath$q$} \times \mbox{\boldmath$p$})^2$.] If we 
require (as we can) that $G(q,p)$ changes sign when $\mbox{\boldmath$q$}$ or 
(separately) $\mbox{\boldmath$p$}$ change sign, the generating function is 
uniquely fixed. In particular, at 1PN level,  
by looking at the structure of the source terms, i.e. the R.H.S. of Eq.~(\ref{eqn1}),
we can prove in advance that $G_{1 {\rm PN}}$ must be of the form
\begin{equation}
G_{1 {\rm PN}} (\mbox{\boldmath$q$} , \mbox{\boldmath$p$}) = 
(\mbox{\boldmath$q$} \cdot \mbox{\boldmath$p$}) \left[ \alpha_1 \, 
\mbox{\boldmath$p$}^2 + \frac{\beta_1}{q} \right] \,. \label{eq6.12}
\end{equation}
Inserting Eq.~(\ref{eq6.12})  in the equation 
to be satisfied (\ref{eqn1}) gives a system of four equations for the two 
unknown coefficients $\alpha_1$ and $\beta_1$. Two of these equations 
give directly the values $\alpha_1$ and $\beta_1$, 
\beq
\label{sol}
\alpha_1 = - \frac{\nu}{2} \,, \quad  \quad \quad \quad \beta_1 = 1 + \frac{\nu}{2} \,,
\eeq
while the two redundant equations, 
\beq
\alpha_1 - \beta_1 = -1 - \nu\,, \quad \quad 
2\,\alpha_1 + \beta_1 = 1 - \frac{\nu}{2}\,,
\eeq
are identically satisfied by the solution (\ref{sol}). 

Using these 1PN-results we can go further and evaluate the 2PN-source 
term $K_2(q,p)$ in Eq.~(\ref{eq6.11}):
\bea
K_2(q,p) &=& -\frac{\nu}{8}\,(1 + 3\,\nu)\,\mbox{\boldmath$p$}^6 + 
\frac{\nu}{8}\,(-1 + 8\,\nu)\,\frac{\mbox{\boldmath$p$}^4}{q} - 
\frac{\nu}{4}\,(9 + \nu)\,\frac{ (\mbox{\boldmath$n$} \cdot \mbox{\boldmath$p$})^2\,
\mbox{\boldmath$p$}^2}{q} + \frac{3}{8}\,\nu\,(8 + 3\,\nu)\,
\frac{(\mbox{\boldmath$n$}\cdot\mbox{\boldmath$p$})^4}{q} \nonumber \\
&+& \frac{1}{8}\,(-2 + 16\,\nu - 7\,\nu^2)\,\frac{\mbox{\boldmath$p$}^2}{q^2} + 
\frac{1}{8}\,(4 + 3\,\nu^2)\,\frac{(\mbox{\boldmath$n$}\cdot \mbox{\boldmath$p$})^2}{q^2} 
+ \frac{1}{4}\,(1 - 7\,\nu + \nu^2)\,\frac{1}{q^3}\,.
\label{eqn3}
\eea
By looking at the structures in Eq.~(\ref{eqn3}) we deduce that 
the most general form of  $G_{2 {\rm PN}}$ is
\begin{equation}
G_{2 {\rm PN}} (\mbox{\boldmath$q$} , \mbox{\boldmath$p$}) = 
(\mbox{\boldmath$q$} \cdot \mbox{\boldmath$p$}) \left[ \alpha_2 \, 
\mbox{\boldmath$p$}^4 + \frac{1}{q} (\beta_2 \, \mbox{\boldmath$p$}^2 + \gamma_2 
(\mbox{\boldmath$n$} \cdot \mbox{\boldmath$p$})^2) + \frac{\delta_2}{q^2} 
\right] \, . \label{eq6.13}
\end{equation}
Inserting the Ansatz (\ref{eq6.13}), and the 1PN-results, 
in Eq.~(\ref{eqn2}), we get again more equations than unknowns:
\bea
&& -\alpha_2 + \frac{\nu}{8} + \frac{3}{8}\,\nu^2 = 0 \,,
\quad \quad \quad \quad \quad  
\quad \quad \alpha_2 - \beta_2 + \frac{\nu}{8} - \nu^2 =0 \,,\nonumber \\
&& 4\,\alpha_2 + \beta_2 - 3\,\gamma_2 + \frac{9}{4}\,\nu + 
\frac{\nu^2}{4}=0 \,, \quad \quad  
3\,\gamma_2 -3\,\nu - \frac{9}{8}\,\nu^2 =0 \,,\nonumber \\
&& \frac{1}{4} + \beta_2 - \delta_2 - 2\,\nu + \frac{7}{8}\,\nu^2 =0
\,,\quad \quad \,\,\,
- \frac{1}{2} + 2\,\beta_2 + 2\,\delta_2 + 3\,\gamma_2 
-\frac{3}{8}\,\nu^2 = 0\,,\nonumber \\
&&-\frac{1}{4} + \delta_2 + \frac{7}{4}\,\nu - \frac{\nu^2}{4} = 0\,.
\eea
As it should 
(in view of the work of the previous sections) one finds that all the 
redundant equations can be satisfied. The final, unique solutions for the 
coefficients $\alpha_2, \beta_2, \gamma_2$ and $\delta_2$ are:
\begin{eqnarray}
&&\alpha_2 = \frac{\nu + 3\nu^2}{8} \,, \quad \quad \beta_2 = \frac{2\nu - 5\nu^2}{8} \, 
, \nonumber \\
&&\gamma_2 = \frac{8\nu + 3\nu^2}{8} \,, \quad \quad  \delta_2 = \frac{1-7\nu + \nu^2}{4} 
\, . \label{eq6.14}
\end{eqnarray}
Finally, we give the explicit form of the canonical 
transformation between the coordinates $(q,p)$ and $(q^\prime,p^\prime)$ 
at the 2PN level (see Eq.~(\ref{eq6.9})): 
\bea
\label{can1}
q^{\prime\,i} - q^i &=& \frac{1}{c^2}\,\left [ 
\left (1 + \frac{\nu}{2} \right )\,\frac{q^i}{q} - \frac{\nu}{2}\,q^i\,
\mbox{\boldmath$p$}^2  - \nu\,p^i\,(\mbox{\boldmath$q$}\cdot\mbox{\boldmath$p$})
\right ] \nonumber \\
&+& \frac{1}{c^4}\, \left [ \nu\,\left (1 + \frac{\nu}{8} \right)\,
\frac{q^i\,(\mbox{\boldmath$q$}\cdot\mbox{\boldmath$p$})^2}{q^3} + 
\frac{\nu}{4}\,\left ( 5 - \frac{\nu}{2} \right )\,
\frac{q^i\,\mbox{\boldmath$p$}^2}{q} + \frac{3}{2}\,\nu\,\left ( 1 - \frac{\nu}{2} 
\right )\frac{p^i\,(\mbox{\boldmath$q$}\cdot \mbox{\boldmath$p$})}{q} \right .
\nonumber \\
&+&\left .  \frac{1}{4}\left ( 1 - 7\,\nu + \nu^2 \right)\,\frac{q^i}{q^2} + 
\frac{\nu}{8}\,(1 - \nu)\,q^i\,\mbox{\boldmath$p$}^4 + 
\frac{\nu}{2}\,(1 + \nu)\,p^i\,\mbox{\boldmath$p$}^2\,(\mbox{\boldmath$q$}\cdot 
\mbox{\boldmath$p$}) \right ]\,, \\
\label{can2}
p^{\prime}_i - p_i &=& \frac{1}{c^2}\,\left [ 
-\left ( 1 + \frac{\nu}{2} \right )\,\frac{p_i}{q} + 
\frac{\nu}{2}\,p_i\,\mbox{\boldmath$p$}^2 + \left ( 1 + \frac{\nu}{2} \right )\,
\frac{q_i\,(\mbox{\boldmath$q$}\cdot \mbox{\boldmath$p$})}{q^3} \right ]\nonumber\\ 
&+& \frac{1}{c^4}\,\left [
\frac{\nu}{8}\,\left ( -1 + 3\,\nu\right )\,p_i\,\mbox{\boldmath$p$}^4 + 
\frac{1}{4}\,(3 + 11\,\nu) \frac{p_i}{q^2} 
- \frac{3}{4}\,\nu\,\left ( 3 + \frac{\nu}{2} \right )\,
\frac{p_i\,\mbox{\boldmath$p$}^2}{q} \right . \nonumber \\
&+& \left .  \frac{1}{4}\,(-2 -18\,\nu + \nu^2)\,\frac{q_i\,
(\mbox{\boldmath$q$}\cdot \mbox{\boldmath$p$})}{q^4} + 
\frac{\nu}{8}\,(10 - \nu)\,\frac{q_i\,
(\mbox{\boldmath$q$}\cdot \mbox{\boldmath$p$})\,\mbox{\boldmath$p$}^2}{q^3} 
\right . \nonumber \\
&-& \left . \frac{\nu}{8}\,(16 + 5\,\nu) \, \frac{p_i\,
(\mbox{\boldmath$q$}\cdot \mbox{\boldmath$p$})^2}{q^3} 
+ \frac{3}{8}\,\nu\,(8 + 3\,\nu)\,\frac{q_i\,(\mbox{\boldmath$q$}\cdot 
\mbox{\boldmath$p$})^3}{q^5}\right ]\,.
\eea
Note that the $\nu \rightarrow 0$ limit of Eq.~(\ref{can1}) gives 
$q^{\prime i} = (1 + 1/(2c^2q))^2\,q^i$ which is (as it should) the 
relation between ``Schwarzschild'' ($q^\prime$) and ``Isotropic'' ($q$) 
quasi-cartesian coordinates in a Schwarzschild spacetime. [In this 
case, ADM = Isotropic]. 
As a check on Eqs.~(\ref{can1}), (\ref{can2}) we have verified that (at the 2PN level) 
$ \mbox{\boldmath$q$}^\prime \times \mbox{\boldmath$p$}^\prime$ 
coincides with $ \mbox{\boldmath$q$} \times \mbox{\boldmath$p$}$. 
[They should coincide exactly, when solving exactly Eqs.~(\ref{eq6.8}) with 
any (spherically symmetric) generating function $G(q,p)$.] 
Let us quote, for completeness,  
the partial derivatives of the generating function 
$G = c^{-2}\,G_{\rm 1PN} + c^{-4}\,G_{\rm 2PN}$, 
that must be used to solve by successive iterations the exact 
equations (\ref{eq6.8}) and determine $q^\prime$ and $p^\prime$ in 
terms of $q$ and $p$:
\bea
\frac{\partial G_{\rm 1PN}(q,p)}{\partial q^i} &=& 
-\frac{\nu}{2}\,p_i\,\mbox{\boldmath$p$}^2 + 
\left ( 1 + \frac{\nu}{2} \right )\,\frac{p_i}{q} - 
\left ( 1 + \frac{\nu}{2} \right )\,\frac{q_i\,
(\mbox{\boldmath$q$}\cdot \mbox{\boldmath$p$})}{q^3}\,,\\
\frac{\partial G_{\rm 1PN}(q,p)}{\partial p_i} &=&
-\frac{\nu}{2}\,q^i\,\mbox{\boldmath$p$}^2 + 
\left (1 + \frac{\nu}{2} \right )\,\frac{q^i}{q} - 
\nu\,p^i\,(\mbox{\boldmath$q$}\cdot \mbox{\boldmath$p$})\,, 
\eea
\bea
\frac{\partial G_{\rm 2PN}(q,p)}{\partial q^i} &=& \frac{1}{8}\,\nu\,(1 + 3\,\nu)\,
p_i\,\mbox{\boldmath$p$}^4 + \frac{\nu}{8}\,(2 - 5\,\nu)\,
\frac{p_i\,\mbox{\boldmath$p$}^2}{q} + 
\frac{3}{8}\,\nu\,(8 + 3\,\nu)\,\frac{p_i\,
(\mbox{\boldmath$q$}\cdot\mbox{\boldmath$p$})^2}{q^3} \nonumber \\
&-& \frac{3}{8}\,\nu\,(8 + 3\,\nu)\,\frac{q_i\,
(\mbox{\boldmath$q$}\cdot\mbox{\boldmath$p$})^3}{q^5} + 
\frac{1}{4}\,(1 - 7\,\nu + \nu^2)\,\frac{p_i}{q^2} - 
\frac{\nu}{8}\,(2 - 5\,\nu)\,\frac{q_i\, 
(\mbox{\boldmath$q$}\cdot\mbox{\boldmath$p$})\,
\mbox{\boldmath$p$}^2}{q^3} \nonumber \\
&-& \frac{1}{2}\,(1-7\,\nu+\nu^2)\,\frac{q_i\,
(\mbox{\boldmath$q$}\cdot\mbox{\boldmath$p$})}{q^4}
\,, \\
\frac{\partial G_{\rm 2PN}(q,p)}{\partial p_i} &=& 
\frac{1}{8}\,\nu\,(1 + 3\,\nu)\,
q^i\,\mbox{\boldmath$p$}^4 + 
\frac{\nu}{8}\,(2 - 5\,\nu)\,\frac{q^i\,\mbox{\boldmath$p$}^2}{q} 
+ \frac{3}{8}\,\nu\,(8+3\,\nu)\,
\frac{q^i\,(\mbox{\boldmath$q$}\cdot\mbox{\boldmath$p$})^2}{q^3} \nonumber \\
&+& \frac{1}{4}\,(1 - 7\,\nu + \nu^2)\,\frac{q^i}{q^2} + 
\frac{\nu}{2}\,(1 + 3\,\nu)\,p^i\,\mbox{\boldmath$p$}^2\,
(\mbox{\boldmath$q$}\cdot\mbox{\boldmath$p$}) 
+\frac{\nu}{4}\,(2 - 5\,\nu)\,\frac{p^i\,
(\mbox{\boldmath$q$}\cdot\mbox{\boldmath$p$})}{q}\,. 
\eea

\section{Inclusion of radiation reaction effects and transition between inspiral 
and plunge}
\label{sec7}
In the preceding sections we have limited our attention to the conservative 
(time-symmetric) part of the dynamics of a two-body system, i.e. the one 
defined, at the 2PN level, by neglecting $\mbox{\boldmath$A$}_a^{\rm reac}$ in 
Eq.~(\ref{eq1.1}). We expect that the separation of the dynamics in a 
conservative part plus a reactive part, makes sense also at higher PN orders 
(though it probably gets blurred at some high PN level). However, there exists, 
at present, no algorithm defining precisely this separation. Anyway we shall 
content ourselves here to working at the 2.5PN level where this separation is 
well-defined, as shown in Eq.~(\ref{eq1.1}). When dealing with the relative 
motion we find it convenient to continue using an Hamiltonian formalism. 
Sch\"afer \cite{S85}, \cite{S86}, \cite{SGauss} has shown how to treat radiation reaction 
effects within the ADM canonical formalism. His result (at the 2.5PN level) is 
that it is enough to use as Hamiltonian for the dynamics of two masses a {\it 
time-dependent} Hamiltonian obtained by adding to the conservative 2PN 
Hamiltonian $H_{2 {\rm PN}} (\mbox{\boldmath$q$}_1 , \mbox{\boldmath$q$}_2 , 
\mbox{\boldmath$p$}_1 , \mbox{\boldmath$p$}_2)$ the following ``reactive'' 
Hamiltonian
\begin{equation}
H_{\rm reac} (\mbox{\boldmath$q$}_1 , \mbox{\boldmath$q$}_2 , 
\mbox{\boldmath$p$}_1 , \mbox{\boldmath$p$}_2 ; t) = - h_{ij}^{TT {\rm reac}} 
(t) \left[ \frac{p_1^i \, p_1^j}{2m_1} + \frac{p_2^i \, p_2^j}{2m_2} - 
\frac{1}{2} \, G \, m_1 \, m_2 \, \frac{(q_1^i - q_2^i)(q_1^j - q_2^j)}{\vert 
\mbox{\boldmath$q$}_1 - \mbox{\boldmath$q$}_2 \vert^3} \right] \, , 
\label{eq7.1}
\end{equation}
where
\begin{equation}
h_{ij}^{TT {\rm reac}} (t) = - \frac{4}{5} \ \frac{G}{c^5} \ \frac{d^3 \, Q_{ij} 
(t)}{dt^3} \, , \label{eq7.2}
\end{equation}
$Q_{ij}$ denoting the quadrupole moment of the two-body system
\begin{equation}
Q_{ij} (t) = \sum_{a=1,2} m_a \, \left( q_a^i \, q_a^j - \frac{1}{3} \ 
\mbox{\boldmath$q$}_a^2 \, \delta^{ij} \right) \, . \label{eq7.3}
\end{equation}
Note that $h_{ij}^{TT {\rm reac}}$ in Eq.~(\ref{eq7.1}) should be treated as a 
given, time-dependent external field, considered as being independent of the 
canonical variables $\mbox{\boldmath$q$}_a$, $\mbox{\boldmath$p$}_a$. In other 
words, when writing the canonical equations of motion $\dot q = \partial \, 
H_{\rm tot} / \partial \, p$, $\dot p = -\partial \, H_{\rm tot} / \partial \, 
q$, one should consider only the explicit $q-p$ dependence appearing in the 
square bracket on the R.H.S. of Eq.~(\ref{eq7.1}). After differentiation with 
respect to $q$ and $p$ one can insert the explicit phase-space expression of the 
third time derivative of $Q_{ij} (t)$ (obtained, with sufficient precision, by 
using the Newtonian-level dynamics, i.e. by computing a repeated Poisson bracket 
of $Q_{ij} (q,p)$ with $H_{\rm Newton} (q,p)$).

Finally, we propose to graft radiation-reaction effects onto the 
non-perturbatively re-summed conservative dynamics defined by our 
effective-action approach in the following way. The total Hamiltonian for the 
relative motion $Q$, $P$ in ADM coordinates is
\begin{equation}
H_{\rm tot} (Q,P;t) = H_{\rm real}^{\rm improved} (Q,P) + H^{\rm reac} (Q,P;t) 
\, , \label{eq7.4}
\end{equation}
where the ``improved 2PN'' Hamiltonian is that defined by solving 
Eq.~(\ref{eq4.25}) for ${\cal E}_{\rm real} = H_{\rm real}^R$, i.e.
\begin{equation}
\frac{H_{\rm real}^{\rm improved} (Q,P)}{M \, c^2} = \sqrt{1 + 2\nu \left( 
\frac{H_{\rm eff} (Q'(Q,P),P'(Q,P))}{\mu \, c^2} - 1 \right)} \, , \label{eq7.5}
\end{equation}
on the R.H.S. of which one must transform, by the canonical transformation 
discussed in Sec.~\ref{sec6}, the (exact) effective Hamiltonian defined by 
Eq.~(\ref{eq6.3}). In the latter, we propose to use our current best estimates 
of the effective metric coefficients $A(Q')$, $B(Q')$, namely
\begin{eqnarray}
&&A(Q') \equiv 1 - \frac{2GM}{c^2 \, Q'} + 2\nu \left( \frac{GM}{c^2 \, Q'} 
\right)^3 \, , \nonumber \\
&&B(Q') \equiv A^{-1} (Q') \left[ 1 - 6\nu \left( \frac{GM}{c^2 \, Q'} \right)^2 
\right] \, . \label{eq7.6}
\end{eqnarray}
On the other hand the ``reactive'' contribution to the total Hamiltonian 
(\ref{eq7.4}) is the center of mass reduction $(\mbox{\boldmath$p$}_1 = - 
\mbox{\boldmath$p$}_2 = \mbox{\boldmath$P$} , \mbox{\boldmath$Q$} = 
\mbox{\boldmath$q$}_1 - \mbox{\boldmath$q$}_2)$ of Eq.~(\ref{eq7.1}).

\begin{figure}
\vspace{-3cm}
\begin{center}
\begin{tabular}{cc}   
\hspace{-1.2cm} 
\epsfig{file=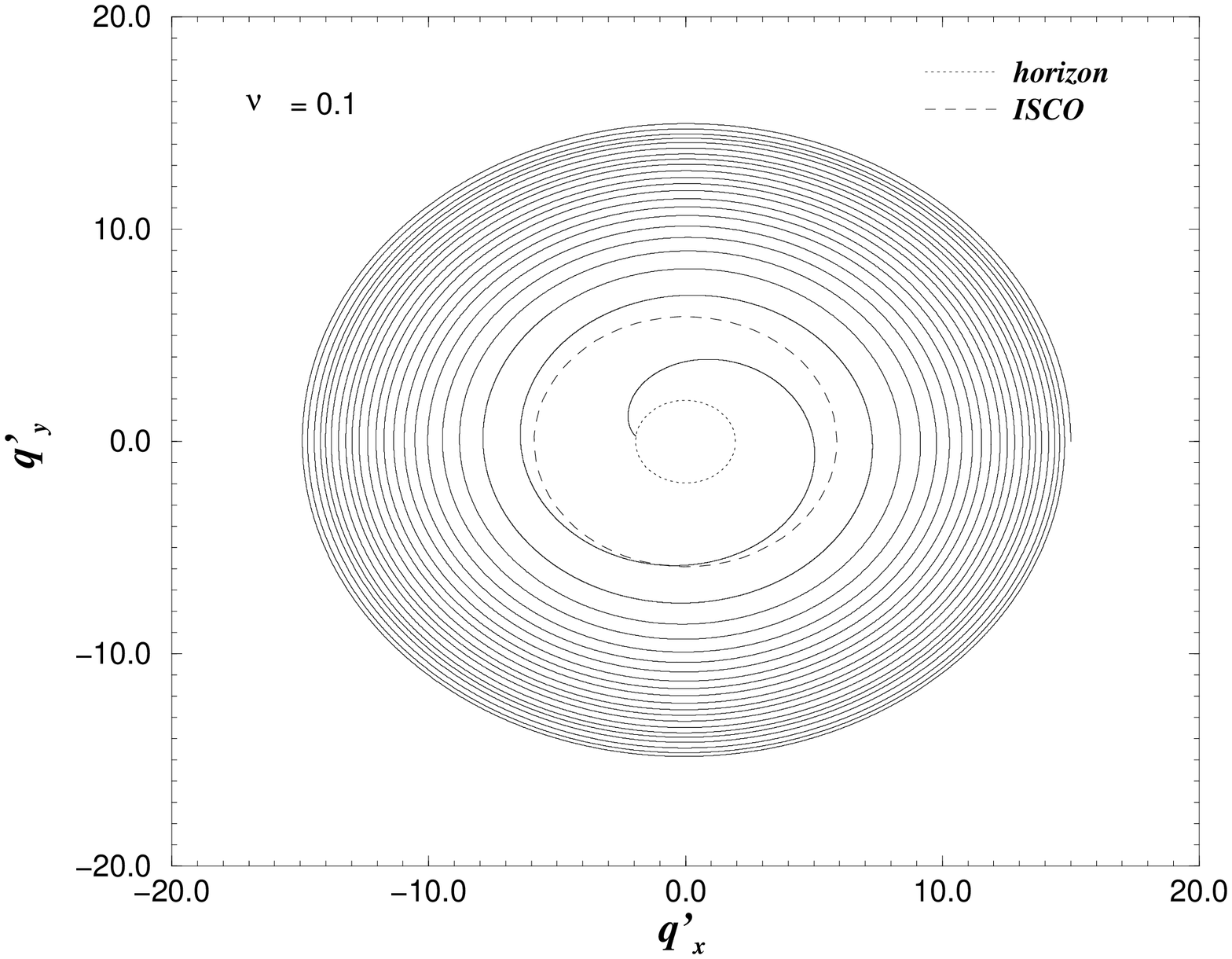,width=0.5\textwidth,height = 0.5\textwidth,angle=-90} & 
\hspace{-1cm}
\epsfig{file=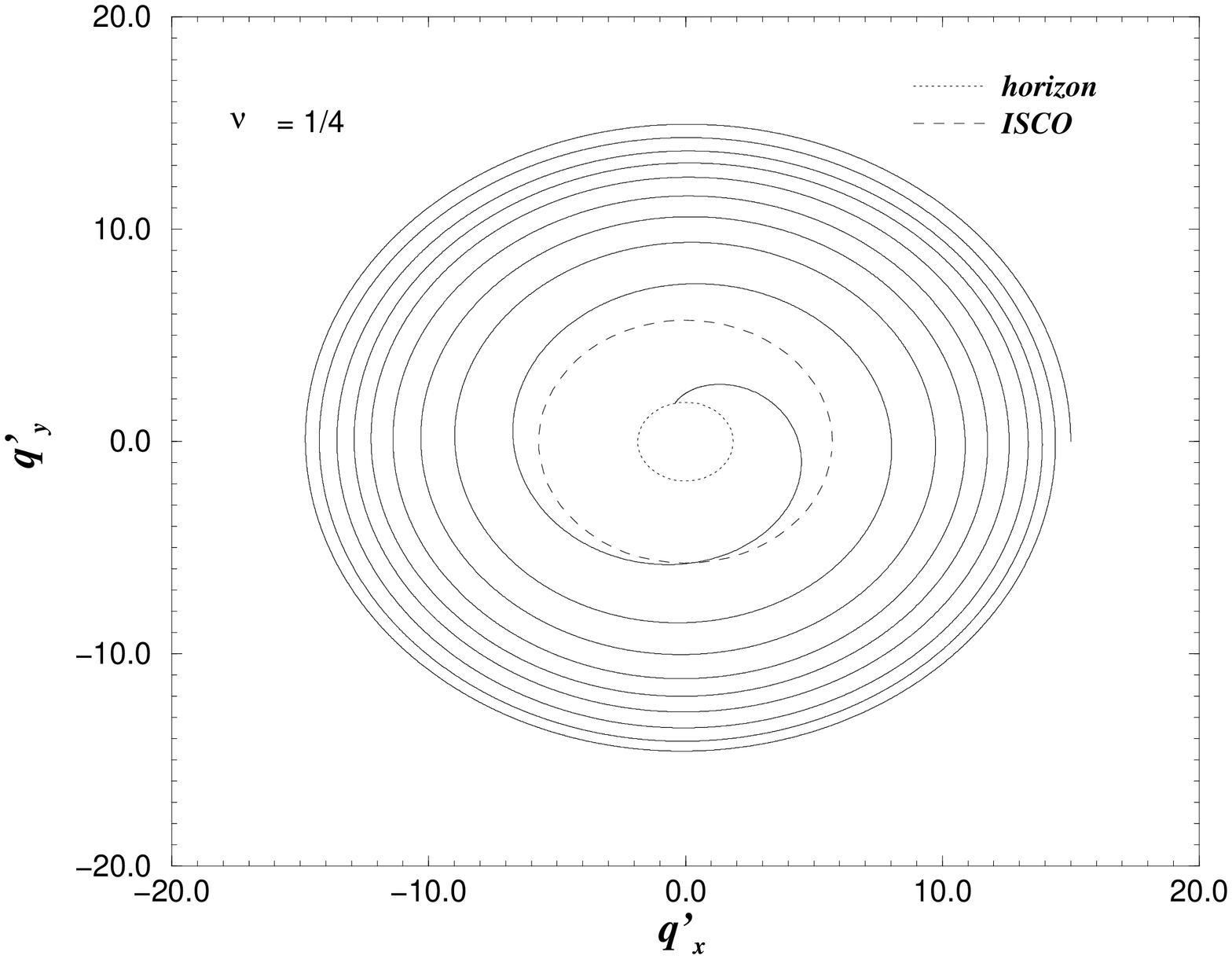,width=0.5\textwidth,height = 0.5\textwidth,angle=-90}
\end{tabular}
\caption{\sl Inspiraling circular orbits in $(q^\prime, p^\prime)$ coordinates 
including radiation reaction effects for $\nu = 0.1$ (left panel) and 
$\nu = 1/4$ (right panel). The location of the ISCO and of the horizon 
are indicated.}
\label{fig5}
\end{center}
\end{figure}

In terms of reduced variables $(q=Q/GM , p=P/\mu)$ and of the non-relativistic 
reduced Hamiltonian, $\widehat{H}_{\rm real}^{\rm NR} \equiv (H_{\rm real}^R - M 
c^2) / \mu$, our proposal reads
\begin{equation}
\widehat{H}_{\rm tot}^{\rm NR} (q,p;t) = \widehat{H}_{\rm real}^{\rm NR \, 
improved} (q,p) + \widehat{H}^{\rm reac} (q,p;t) \, , \label{eq7.7}
\end{equation}
with
\begin{equation}
\widehat{H}_{\rm real}^{\rm NR \, improved} (q,p) \equiv \frac{c^2}{\nu}\, \left[ 
\sqrt{1+2\nu \, \left [\frac{1}{c^2}\,\widehat{H}_{\rm eff} (q'(q,p),p'(q,p))-1
\right ]} - 1 \right] \, , 
\label{eq7.8}
\end{equation}
where $\widehat{H}_{\rm eff} (q',p')$ is defined by inserting (\ref{eq7.6}) into 
Eq.~(\ref{eq6.5}), and with
\begin{equation}
\widehat{H}^{\rm reac} (q,p;t) = - h_{ij}^{TT {\rm reac}} (t) \left[ \frac{1}{2} 
\, p^i \, p^j - \frac{1}{2} \, \frac{q^i \, q^j}{q^3} \right] \, , \label{eq7.9}
\end{equation}
\begin{equation}
h_{ij}^{TT {\rm reac}} (t) = -\frac{4}{5\,c^5} \, \frac{\nu}{q^2} \left[ -4 (p^i \, 
n^j + p^j \, n^i) + 6 \, n^i \, n^j (\mbox{\boldmath$n$} \cdot 
\mbox{\boldmath$p$}) + \frac{2}{3} \, (\mbox{\boldmath$n$} \cdot 
\mbox{\boldmath$p$}) \, \delta^{ij} \right] \, , \label{eq7.10}
\end{equation}
where $n^i \equiv q^i/q$. 
As explained above, the quantity $h_{ij}^{TT {\rm reac}} (t)$ should not be 
differentiated with respect to $q$ and $p$ when writing the equations of motion
\begin{eqnarray}
&&{\dot q}^i = \frac{\partial \, \widehat{H}_{\rm real}^{\rm NR \, improved} 
(q,p)}{\partial \, p_i} + \frac{\partial \, \widehat{H}^{\rm reac} 
(q,p;h_{ij}^{TT {\rm reac}} (t))}{\partial \, p_i} \, , \nonumber \\
&&{\dot p}_i = -\frac{\partial \, \widehat{H}_{\rm real}^{\rm NR \, improved} 
(q,p)}{\partial \, q^i} - \frac{\partial \, \widehat{H}^{\rm reac} 
(q,p;h_{ij}^{TT {\rm reac}} (t))}{\partial \, q^i} \, . \label{eq7.11}
\end{eqnarray}
When inserting, after differentiation, Eq.~(\ref{eq7.10}), the equations of 
motion (\ref{eq7.11}) become an explicit, autonomous (time-independent) 
evolution equation in phase space: $\dot{\mbox{\boldmath$x$}} = 
f(\mbox{\boldmath$x$})$ where $\mbox{\boldmath$x$} = (q^i , p_i)$. From the 
study in Sec.~\ref{sec5} above of the circular orbits defined by the exact, 
non-perturbative Hamiltonian $H_{\rm eff}$, we expect that the combined dynamics 
(\ref{eq7.11}) will exhibit a transition from inspiral to plunge when $q = \vert 
\mbox{\boldmath$q$} \vert$ (which decreases under radiation damping) reaches the 
image in the $q-p$ phase space of the ISCO, studied above in $q'$, $p'$ 
coordinates. We have in mind here quasi-circular, inspiraling orbits 
(circularized by radiation reaction), though, evidently, our approach can be 
used to study all possible orbits. We further expect that, when $\nu \ll 1$ the 
inspiral will be very slow (the reaction Hamiltonian being proportional to 
$\nu$, see Eq.~(\ref{eq7.10})) and therefore the transition to plunge will be quite 
sharp, and well located at the ISCO. When $\nu = 1/4$ the radiation reaction 
effects are numerically smallish, but not parametrically small at the ISCO, and 
the transition to plunge cannot be expected to be very sharp. These expected 
behaviors are illustrated in Fig.~\ref{fig5}.

For simplicity, we have computed the orbits exhibited in these figures in 
$\mbox{\boldmath$q$}'$ space, neglecting the (formally 3.5PN) effect of the 
$(q,p) \rightarrow (q',p')$ transformation on the reactive part of the equations 
of motion. [Thanks to the canonical invariance of the Hamilton equations of 
motion, the crucial conservative part of the evolution in $q',p'$ space is 
simply obtained from the Hamiltonian $\widehat{H}_{\rm real}^{\rm NR \, improved} (q',p')$ 
defined by keeping the variables $q'$ and $p'$ on the R.H.S. of Eq.~(\ref{eq7.8}).]

Let us finally mention another possibility for incorporating radiation reaction 
effects directly in the effective one-body dynamics. In the $q-p$ coordinates 
the (2.5PN) reaction Hamiltonian (\ref{eq7.1}) can be simply seen as due to 
perturbing the Euclidean metric $g_{ij}^0 = \delta_{ij}$ appearing in the lowest 
order Newtonian Hamiltonian $(q_{ab}^i \equiv q_a^i - q_b^i)$
\begin{equation}
H_{\rm Newtonian} (q_a , p_a) = \sum_a \frac{g_0^{ij} \, p_{ai} \, p_{aj}}{2 
m_a} - \sum_{a<b} \frac{G \, m_a \, m_b}{(g_{ij}^0 \, q_{ab}^i \, 
q_{ab}^j)^{1/2}} \, , \label{eq7.12}
\end{equation}
by taking into account the near zone radiative field:
\begin{equation}
g_{ij} \simeq g_{ij}^0 + h_{ij}^{TT {\rm reac}} (t) \,,\quad \quad  g^{ij} \simeq g_0^{ij} 
- h^{ij\,TT}_ {\rm reac} (t) \, . \label{eq7.13}
\end{equation}
By mapping back (through our $(qp) \leftrightarrow (q'p')$ link) the metric 
perturbation $h_{ij}^{TT {\rm reac}}$ onto the effective problem, one might try 
to incorporate reaction effects by defining a suitable ``reactive'' perturbation 
of our effective metric:
\begin{equation}
g_{\mu \nu} (q') = g_{\mu \nu}^{\rm eff} (q') + \delta^{\rm reac} \, g_{\mu 
\nu}^{\rm eff} (q') \, . \label{eq7.14}
\end{equation}
This approach might be useful for trying to go beyond the 2.5PN level discussed 
here and to define a ``re-summed'' version of reaction effects. Alternatively, if 
one has at one's disposal a more complete PN-expanded reactive force expressed 
in the original $q$ coordinates~\cite{rad}, one can, following 
the strategy proposed in Eq.~(\ref{eq7.4}), graft this improved (perturbative) 
reactive force onto the non-perturbatively improved conservative force defined 
by mapping back our effective dynamics onto the $q$ coordinates.

\section{Conclusions}
\label{sec8}
We have introduced a novel approach to studying the late dynamical evolution of 
a coalescing binary system of compact objects. This approach is based on mapping 
(by a canonical transformation) the dynamics of the relative motion of a 
two-body system, with comparable masses $m_1 , m_2$, onto the dynamics of one 
particle of mass $\mu = m_1 \, m_2 / (m_1 + m_2)$ moving in some effective 
metric $ds_{\rm eff}$. When neglecting radiation reaction, the mapping rules 
between the two problems are best interpreted in quantum terms (mapping between 
the discrete energy spectrum of bound states). They involve a physically natural 
transformation of the energy axis between the two problems, stating essentially 
that the effective energy of the effective particle is the energy of particle 1 
in the rest-frame of particle 2 (or reciprocally), 
see Eq.~(\ref{eq4.26}). The usefulness of this energy 
mapping was previously emphasized both in quantum two-body problems \cite{BIZ}, 
and in classical ones \cite{DIS}.

Starting from the currently most accurate knowledge of two-body dynamics 
\cite{DD81a}, \cite{D82}, we have shown that, when neglecting radiation reaction, 
our rules uniquely determine the effective metric $g_{\mu \nu}^{\rm eff} (q')$ 
in which the effective particle moves. This metric is a simple deformation of a 
Schwarzschild metric of mass $M = m_1 + m_2$, with deformation parameter $\nu = 
\mu / M$. Our suggestion is then to {\it define} (as is done in quantum two-body 
problems \cite{BIZ}, \cite{todorov}) a particular non-perturbative re-summation 
of the usual, badly convergent, post-Newtonian-expanded dynamics by considering 
the dynamics defined by the effective metric as exact. This definition leads, in 
particular, to specific predictions for the characteristics of the innermost 
stable circular orbit (ISCO) for comparable-mass systems. In agreement with some 
previous predictions (notably one based on Pad\'e approximants \cite{DIS}), but 
in disagreement with the predictions of the ``hybrid'' approach of 
Ref.~\cite{KWW}, we predict an ISCO which is more tightly bound than the usual 
test-mass-in-Schwarzschild one. The invariant physical characteristics of our 
predicted ISCO are given in Eqs.~(\ref{eq5.28}) and ~(\ref{eq5.29}), 
see also Tab.~\ref{tab1}. Note in particular that 
the binding energy at the ISCO is robustly predicted to be 
${\cal E}^{\rm NR}_{\rm real} \simeq - 1.5 \% M c^2$ (for equal-mass 
systems; $\nu = 1/4$), while the   
orbital frequency at the ISCO is numerically predicted to be 
(again for $\nu = 1/4$)
\begin{equation}
f^{\rm ISCO} = 2372\,{\rm Hertz} \left( \frac{M_\odot}{M} \right) \, . 
\label{eq8.1}
\end{equation}
Note that this corresponds to $\sim 847\,{\rm Hertz}$ 
for $(1.4 M_\odot , 1.4 M_\odot)$ neutron 
star systems.

We have argued, by studying the effects of higher (time-symmetric) 
post-Newtonian contributions, that our predictions for the characteristics of 
the ISCO are rather robust (especially when compared to the scatter of previous 
predictions). See Fig.~\ref{fig4} and Tab.~\ref{tab1}. 
We note, however, that the knowledge of 
the 3PN dynamics (currently in progress \cite{JS}, 
\cite{Betal}) would significantly reduce the present 
(2PN-based) uncertainty on the knowledge of the effective metric.

The coordinate separation, in effective Schwarzschild coordinates, corresponding 
to the ISCO is $Q' = R \simeq 5.72\,GM/c^2$, i.e. $ \sim 23.6\,{\rm km}$ for a 
$(1.4 M_\odot , 1.4 M_\odot)$ neutron star system (from our canonical transformation (\ref{eq6.8}), 
this corresponds to an ADM-coordinate relative separation of 
$ Q \simeq 4.79\,GM / c^2$). This value is near the sum of the nominal radii of (isolated) neutron 
stars for most nuclear equations of state \cite{AB77}. This suggests that the 
inspiral phase of coalescing neutron star systems might terminate into tidal 
disruption (or at least tidally-dominated dynamics) without going through a 
well-defined plunge phase. Fully relativistic 3D numerical simulations are 
needed to investigate this question. We note that a positive aspect of having 
(as predicted here) a rather low ISCO is that the end of the inspiral phase 
might well be very sensitive to the nuclear equation of state, so that 
LIGO/VIRGO observations might teach us something new about dense nuclear matter.

Finally, we have proposed two ways of adding radiation reaction effects to our 
effective one-body dynamics. The most straightforward one consists in directly 
combining radiation effects determined in the real two-body problem with the 
non-perturbative conservative dynamics (which, in particular, features a dynamical 
instability at our ISCO) obtained by mapping the effective dynamics onto some 
standard (ADM or harmonic) two-body coordinate system: see Eq.~(\ref{eq7.7}). A 
more subtle approach, which needs to be further developed, would consist in 
adding radiation reaction effects at the level of the effective metric itself, 
see Eq.~(\ref{eq7.14}). We have illustrated in Fig.~\ref{fig5} the transition from 
inspiral to plunge implied by (an approximation to) Eq.~(\ref{eq7.7}). In 
principle, this transition, and in particular the frequency at the ISCO, will be 
observable in gravitational wave observations of systems containing black holes.

We hope that the approach presented here will also be of value for supplementing 
numerical relativity investigations. Indeed, our main (hopeful) claim is that 
the effective one-body dynamics is a ``good'' non-perturbative re-summation of 
the standard post-Newtonian-expanded results. Therefore, it gives a simple way 
of boosting up the accuracy of many PN-expanded results. 
[We leave to future work a more systematic analysis of the extension of our 
approach to higher post-Newtonian orders.]
Effectively, this 
extends the validity of the post-Newtonian expansions in a new way (e.g., 
different from Pad\'e approximants\footnote{It should be, however, possible to 
combine the effective one-body approach with Pad\'e approximants, thereby 
defining an even better scheme.}). In particular, our results could be used to 
define initial conditions for two-body systems very near, or even at the ISCO, 
thereby cutting down significantly the numerical work needed to evolve fully 
relativistic 3D binary-system simulations.

As a final remark, let us note that many extensions of the approach presented 
here are possible. In particular, the addition of the (classical) {\it spin} 
degrees of freedom to the effective one-body problem (in the 
effective metric and/or in the effective particle) suggests itself as an 
interesting issue (with possibly important physical consequences).
\acknowledgments
We thank Gerhard Sch\"afer for useful comments.
\appendix      
\section{}
\label{appA}
In this appendix we determine, at the 2PN level and 
in the Schwarzschild gauge, the effective metric 
\beq
\label{eqa1}
ds_{\rm eff}^2 = -A(R) \, c^2 \, dt^2 
+ B (R) \, dR^2  + R^2 (d\theta^2 + \sin^2 \theta\, d\varphi^2) \,, 
\eeq
\beq
A(R) = 1 + \frac{a_1}{c^2 R} + \frac{a_2}{c^4 R^2} + \frac{a_3}{c^6 R^3} \,, 
\quad \quad 
B(R) = 1 + \frac{b_1}{c^2 R} + \frac{b_2}{c^4 R^2}\,,
\eeq
when requiring simultaneously that: a) the energy levels of the 
``effective'' and ``real'' problems coincide  modulo an 
overall shift, i.e. ${\cal E}_0({\cal N}_0, {\cal J}_0) = 
{\cal E}_{\rm real}({\cal N}, {\cal J}) -c_0$, with $ c_0 = M\,c^2 - m_0\,c^2$, 
${\cal J}_0 = {\cal J}$
and $ {\cal N}_0 = {\cal N}$ and b) the effective metric depends only on 
$m_1$ and $m_2$. In this case, as anticipated in Sec.~\ref{sec4}, we will see 
that it not possible to satisfy the condition $m_0 = \mu$. 

The radial action $I^0_R({\cal E}_0, {\cal J}_0)$ of the 
``effective'' description is
\bea
I_R^0 ({\cal E}_0,{\cal J}_0) &=& \frac{\alpha_0\,m_0^{1/2}}{\sqrt{-2 \, {\cal E}_0^{\rm 
NR}}} \left[ \widehat{A} + \widehat{B} \, \frac{{\cal E}_0^{\rm NR}}{m_0 \, c^2} + 
\widehat{C} \left( \frac{{\cal E}_0^{\rm NR}}{m_0 \, c^2} \right)^2 \right] - 
{\cal J}_0 \nonumber \\
&+& \frac{\alpha_0^2}{c^2 \, {\cal J}_0} \left[ \widehat{D} + 
\widehat{E}\,\frac{{\cal E}_0^{\rm NR}}{m_0 \, c^2} \right]  + 
\frac{\alpha_0^4}{c^4 \, {\cal J}_0^3} \, \widehat{F} \,,
\label{eqa2} 
\eea
where ${\cal E}_0^{\rm NR} \equiv {\cal E}_0 - m_0 \, c^2$, 
$\alpha_0 \equiv GM_0 m_0$, 
$$
\widehat{A} = - \frac{1}{2} \, \widehat{a}_1\,, \quad  \quad 
\widehat{B} = \widehat{b}_1 - \frac{7}{8} \, \widehat{a}_1\,, \quad  
\quad \widehat{C} = \frac{\widehat{b}_1}{4} - \frac{19}{64} \, 
\widehat{a}_1 \, ,
$$
$$
\widehat{D} = \frac{\widehat{a}_1^2}{2} - \frac{\widehat{a}_2}{2} - 
\frac{\widehat{a}_1 \widehat{b}_1}{4}\,, \quad  \quad 
\widehat{E} = \widehat{a}_1^2 
- \widehat{a}_2 - \frac{\widehat{a}_1 \widehat{b}_1}{2} - 
\frac{\widehat{b}_1^2}{8} + \frac{\widehat{b}_2}{2} \, ,
$$
\begin{equation}
\widehat{F} =  \frac{1}{64} \, [24 \, \widehat{a}_1^4 - 
48 \, \widehat{a}_1^2 \, \widehat{a}_2 + 8 \, \widehat{a}_2^2 + 16 \, 
\widehat{a}_1 \, \widehat{a}_3 - 8 \, \widehat{a}_1^3 \, 
\widehat{b}_1 + 8 \, \widehat{a}_1 \, \widehat{a}_2 \, \widehat{b}_1 - 
\widehat{a}_1^2 \, \widehat{b}_1^2 + 4 \, 
\widehat{a}_1^2 \, \widehat{b}_2] \,,
\end{equation}
and we have introduced the dimensionless coefficients:
\beq
\widehat{a}_i = \frac{a_i}{(GM_0)^i} \,, \quad \quad 
\widehat{b}_i = \frac{b_i}{(GM_0)^i}\,.
\eeq
We define the mass $M_0$ used to scale the coefficients 
$a_i$ and $b_i$ by requiring $\widehat{a}_1 \equiv -2$ 
(i.e. $a_1 \equiv -2GM_0$). Identifying Eq.~(\ref{eqa2}) 
with the radial action $I^0_R({\cal E}^{\rm NR}, {\cal J})$ of the ``real'' 
problem, i.e.  
\bea
I_R ({\cal E}^{\rm NR} , {\cal J}) &=& \frac{\alpha \, \mu^{1/2}}{\sqrt{-2 \, 
{\cal E}^{\rm NR}}} \, \left[ 1 + \left( \frac{15}{4} - \frac{\nu}{4} \right) 
\, \frac{{\cal E}^{\rm NR}}{\mu c^2} + \left( \frac{35}{32} + \frac{15}{16} \, 
\nu + \frac{3}{32} \, \nu^2 \right) \left( \frac{{\cal E}^{\rm NR}}{\mu c^2} 
\right)^2 \right] \nonumber \\
&-& {\cal J} + \frac{\alpha^2}{c^2 {\cal J}} \,\left [ 
3 + \left( \frac{15}{2} - 3\nu \right) \, \frac{{\cal E}^{\rm NR}}{ \mu\,c^2} \right ]
+ \left(\frac{35}{4} - \frac{5}{2} \, \nu \right)\,\frac{\alpha^4}{c^4\,{\cal J}^3}\,,
\label{eqa3}
\eea
where $\alpha \equiv GM \mu$ and 
${\cal E}^{\rm NR} \equiv {\cal E}_{\rm real} - M \, c^2$,  
yields six equations to be satisfied. The requirement a) above implies 
the simple identification of the variables entering 
Eqs.~(\ref{eqa2}) and (\ref{eqa3}): ${\cal E}_0^{\rm NR} = {\cal E}^{\rm NR}$,
${\cal J}_0 = {\cal J}$, $I^0_R = I_R$.   
The explicit form of the equations stating that 
$\widehat{A}\,m_0^{1/2}\,\alpha_0$ (0PN level), 
$\widehat{B}\,m_0^{-1/2}\,\alpha_0$, 
$\widehat{D}\,\alpha_0^2$ (1PN level) and $\widehat{C}\,m_0^{-3/2}\,
\alpha_0$, $\widehat{E}\,{\alpha_0^2}/{m_0}$ and $\widehat{F}\,\alpha_0^4$ (2PN level) 
in Eq.~(\ref{eqa2}) coincide with the analogous coefficients in Eq.~(\ref{eqa3}) 
yields: 
\bea
\label{eqa4}
&&m_0^{1/2}\,\alpha_0 = \mu^{1/2}\,\alpha \,,\\
\label{eqa5}
&& \left (\widehat{b}_1 + \frac{7}{4}\right )\,m_0^{-1/2}\,
\alpha_0 = 
\frac{1}{4}\,(15 - \nu)\,\mu^{-1/2}\,\alpha\,,\\
\label{eqa6}
&& \left (4 - \widehat{a}_2 + \widehat{b}_1\right )\,\alpha_0^2 = 
6\,\alpha^2\,, \\
\label{eqa7}
&& \left (\frac{19}{32} + 
\frac{\widehat{b}_1}{4} \right )\,m_0^{-3/2}\,
\alpha_0 = 
\left ( \frac{35}{32} + \frac{15}{16}\,\nu + \frac{3}{32}\,\nu^2 \right )
\,\mu^{-3/2}\,\alpha\,,\\
\label{eqa8}
&& \left ( 4 -\widehat{a}_2 + \widehat{b}_1 - \frac{\widehat{b}_1^2}{8} 
+ \frac{\widehat{b}_2}{2} \right )\,\frac{\alpha_0^2}{m_0} = 
\left ( \frac{15}{2} - 3\nu\right )\,\frac{\alpha^2}{\mu}\,, \\
\label{eqa9}
&& \widehat{F}\,\alpha_0^4 = \left ( \frac{35}{4} - \frac{5}{2}\,\nu 
\right)\,\alpha^4\,.
\eea
It is to be noted that if we impose $m_0 = \mu$ and $GM_0 = GM$ 
(so that $\alpha_0 = \alpha$) 
we get an incompatibility at the 2PN level. Indeed, Eq.~(\ref{eqa4}) is satisfied and 
we can solve Eqs.~(\ref{eqa5}),(\ref{eqa6}) 
in terms of the 1PN-coefficients  
$\widehat{b}_1$ and $\widehat{a}_2$, but then the 
2PN-equation (\ref{eqa7}), which contains only  
$\widehat{b}_1$, is not satisfied. [This problem 
is due to the fact that we have more equations 
than unknowns.] Hence, we are obliged to relax 
the constraint $m_0 = \mu$. Let us introduce 
the parameter $\xi$, defined by $m_0 \equiv \mu\,\xi^{-2}$. 
Eq.~(\ref{eqa4}) then gives   
$GM_0 = GM\,\xi^3$. Note that we are crucially using here the fact that the 
Newton-order energy levels ${\cal E}^{\rm NR} = -m_0 \alpha_0^2/(2{\cal N}_0)
+ {\cal O}(c^{-2})$ do not depend separately on $m_0$ and 
$\alpha_0 = G\,M_0\,m_0$, but only on the combination $m_0\,\alpha_0^2 = 
G^2\,M_0^2\,m_0^3$. Solving the 1PN-level Eqs.~(\ref{eqa5}), (\ref{eqa6}) 
we then get:
\beq
\label{eqa10}
\widehat{b}_1 = \frac{1}{4\,\xi^2}\,(15 - 7\xi^2 -\nu)\,, \quad \quad 
\widehat{a}_2 = \frac{1}{4\,\xi^2}\,(-9 + 9\xi^2 -\nu)\,,
\eeq
while the 2PN-level Eq.~(\ref{eqa7}) gives a quadratic 
equation in $\xi^2$ which fixes uniquely its value 
(as well as that of the positive parameter $\xi$), namely
\beq
\xi^2 = \frac{\mu}{m_0} = \frac{1}{5}\,
\left [ -15 + \nu + 2\sqrt{2}\sqrt{50 + 15 \nu + 2\nu^2}\, \right]\,.
\eeq
Finally, the remaining 2PN equations (\ref{eqa8}) and (\ref{eqa9}) determine 
the coefficients of the effective metric at the 2PN level 
\bea
\widehat{b}_2 &=& \frac{1}{64\,\xi^2}\,\left (1185 - 978\,\xi^2 
+ 49\,\xi^4 - 414\,\nu + 14\,\xi^2\,\nu + \nu^2 \right )\,,\\ 
\widehat{a}_3 &=& \frac{1}{64\,\xi^4}\,
(-289 + 402\,\xi^2 -113\, \xi^4 + 158\,\nu + 50 \xi^2\,\nu - \nu^2)\,.
\label{eqa11}
\eea
The complexity of the results (\ref{eqa10})--(\ref{eqa11}), compared  
to the simplicity of our preferred solution (\ref{eq5.6})--(\ref{eq5.8}), 
convinced us that the requirement a) above should be relaxed. Also, it seems fishy
to have an effective mass $m_0$ which differs from $\mu$ even in the non-relativistic 
limit $c \rightarrow \infty$. Finally, it is not evident that 
this method can be generalized  to higher post-Newtonian orders 
(where more redundant equations will have to be satisfied).

\section{}
\label{appB}
In this appendix we describe an alternative, more formal method 
to map the ``effective'' one-body problem onto the ``real'' two-body one.
We work in the Schwarzschild gauge. 
Here we require simultaneously that: a) the energy levels of the 
``effective'' and ``real'' descriptions coincide  modulo an 
overall shift, i.e. ${\cal E}_0({\cal N}_0, {\cal J}_0) = 
{\cal E}_{\rm real}({\cal N}, {\cal J}) -c_0$, with $ c_0 = M\,c^2 - m_0\,c^2$, 
${\cal J}_0 = {\cal J}$
and $ {\cal N}_0 = {\cal N}$ and b) the effective mass $m_0$ is equal to the  
reduced mass $\mu = m_1 \, m_2 / (m_1 + m_2)$.
Introducing the dimensionless quantities:
\begin{eqnarray}
&& \widehat{I}_R^0 \equiv \frac{I_R^0}{\alpha_0} \,, 
\quad \quad \widehat{I}_R^{\rm real}  \equiv  
\frac{I_R^{\rm real}}{\alpha}\,,
\quad \quad E_0 \equiv \frac{{\cal E}_0^{\rm NR}}{m_0} \,, 
\quad \quad E_{\rm real} \equiv \frac{{\cal E}_{\rm real}^{\rm NR}}{\mu} , \nonumber \\
&& {j}_0 \equiv \frac{{\cal J}_0}{\alpha_0} \,, \quad \quad {j} \equiv 
\frac{\cal J}{\alpha} \,,
\label{eqb1}
\end{eqnarray}
where $\alpha_0 \equiv GM_0 \, m_0$ and $\alpha \equiv GM \, \mu \equiv G \, m_1 
\, m_2$,  we can re-write the radial action for the ``effective'' problem, 
Eq.~(\ref{eq3.13}), in the form: 
\beq
\widehat{I}_R^0 (E_0,j_0) = \frac{1}{\sqrt{-2 \, E_0}} 
\left[ \widehat{A} + \widehat{B} \, \frac{E_0}{c^2} + 
\widehat{C} \left( \frac{E_0}{c^2} \right)^2 \right] - 
{j}_0 
+ \frac{1}{c^2 \, {j}_0} \left[ \widehat{D} + 
\widehat{E}\,\frac{E_0}{c^2} \right]  + 
\frac{1}{c^4 \, {j}_0^3} \, \widehat{F} \,, 
\label{eqb2}
\eeq
where 
$$
\widehat{A} = - \frac{1}{2} \, \widehat{a}_1\,, \quad  \quad 
\widehat{B} = \widehat{b}_1 - \frac{7}{8} \, \widehat{a}_1\,, \quad  
\quad \widehat{C} = \frac{\widehat{b}_1}{4} - \frac{19}{64} \, 
\widehat{a}_1 \, ,
$$
$$
\widehat{D} = \frac{\widehat{a}_1^2}{2} - \frac{\widehat{a}_2}{2} - 
\frac{\widehat{a}_1 \widehat{b}_1}{4}\,, \quad  \quad 
\widehat{E} = \widehat{a}_1^2 
- \widehat{a}_2 - \frac{\widehat{a}_1 \widehat{b}_1}{2} - 
\frac{\widehat{b}_1^2}{8} + \frac{\widehat{b}_2}{2} \, ,
$$
\begin{equation}
\widehat{F} =  \frac{1}{64} \, [24 \, \widehat{a}_1^4 - 
48 \, \widehat{a}_1^2 \, \widehat{a}_2 + 8 \, \widehat{a}_2^2 + 16 \, 
\widehat{a}_1 \, \widehat{a}_3 - 8 \, \widehat{a}_1^3 \, 
\widehat{b}_1 + 8 \, \widehat{a}_1 \, \widehat{a}_2 \, \widehat{b}_1 - 
\widehat{a}_1^2 \, \widehat{b}_1^2 + 4 \, 
\widehat{a}_1^2 \, \widehat{b}_2] \,,
\label{eqb3}
\end{equation}
and where we have used, as above, the scaled metric coefficients 
\beq
\widehat{a}_i = \frac{a_i}{(GM_0)^i} \,, \quad \quad 
\widehat{b}_i = \frac{b_i}{(GM_0)^i}\,.
\eeq
Identifying $\widehat{I}_R^0 (E_0,j_0)$ with the analogous 
expression for the ``real'' problem, 
\bea
\widehat{I}_R ({E_{\rm real}} , {j}) 
&=& \frac{1}{\sqrt{-2 \, 
{E_{\rm real}}}} \, \left[ 1 + \left( \frac{15}{4} - \frac{\nu}{4} \right) 
\, \frac{E_{\rm real}}{c^2} + \left( \frac{35}{32} + \frac{15}{16} \, 
\nu + \frac{3}{32} \, \nu^2 \right) \left( \frac{E_{\rm real}}{c^2} 
\right)^2 \right] \nonumber \\
&-& {j} + \frac{1}{c^2 j} \,\left [ 
3 + \left( \frac{15}{2} - 3\nu \right) \, \frac{E_{\rm real}}{c^2} \right ]
+ \left(\frac{35}{4} - \frac{5}{2} \, \nu \right)\,\frac{1}{c^4\,{j}^3}\,,
\label{eqb4}
\eea
and imposing $ E_0 = E_{\rm real}$, $m_0 = \mu$, $\alpha_0 = \alpha$,  
we get more equations to be satisfied than unknowns, 
\bea
\label{eqb5}
&& -\frac{1}{2}\,\widehat{a}_1 = 1\,,\\
\label{eqb6}
&& \widehat{b}_1 - \frac{7}{8}\,\widehat{a}_1  = 
\frac{1}{4}\,(15 - \nu)\,,\\
\label{eqb7}
&& \widehat{a}_1^2 - \widehat{a}_2 - 
\frac{\widehat{a}_1\,\widehat{b}_1}{2} = 
6\,, \\
\label{eqb8}
&& -\frac{19}{64}\,\widehat{a}_1 + 
\frac{\widehat{b}_1}{4}  = 
 \frac{35}{32} + \frac{15}{16}\,\nu + \frac{3}{32}\,\nu^2 
\,,\\
\label{eqb9}
&&  \widehat{a}_1^2 -\widehat{a}_2 - \frac{\widehat{a}_1\,
\widehat{b}_1}{2} - \frac{\widehat{b}_1^2}{8} 
+ \frac{\widehat{b}_2}{2}  = 
 \frac{15}{2} - 3\nu\,, \\
\label{eqb10}
&& \widehat{F} =  \frac{35}{4} - \frac{5}{2}\,\nu\,.
\eea
Note that Eqs.~(\ref{eqb6}) and ~(\ref{eqb8}) depend 
only on $\widehat{a}_1$ and $\widehat{b}_1$, and cannot both 
be satisfied. To solve this incompatibility we  
consider here the possibility that the various coefficients  
that appear in the effective metric depend on the energy.
Namely, at the 2PN level we consider the following 
expansions
\bea
\label{eqb11}
\widehat{a}_1({E}_0) &=& \widehat{a}_{1}^{(0)} + 
\widehat{a}_{1}^{(2)}\,\left (\frac{E_0}{c^2}\right ) +
\widehat{a}_{1}^{(4)}\,\left (\frac{E_0}{c^2}\right )^2  \,, \\
\label{eqb12}
\widehat{a}_2({E}_0) &=& \widehat{a}_{2}^{(0)} + 
\widehat{a}_{2}^{(2)}\,\left (\frac{{E}_0}{c^2}\right )\,, \\
\widehat{a}_3({E}_0) &=& \widehat{a}_{3}^{(0)}\,,
\eea
and 
\beq
\widehat{b}_1({E}_0) = \widehat{b}_{1}^{(0)} + 
\widehat{b}_{1}^{(2)}\,\left (\frac{{E}_0}{c^2}\right )\,,\quad \quad 
\widehat{b}_2({E}_0) = \widehat{b}_{2}^{(0)}\,.
\eeq
The introduction of an energy dependence in the coefficients 
$\widehat{a}_i$, $\widehat{b}_i$ reshuffles the $c^{-2}$ expansion 
of Eq.~(\ref{eqb2}) and modifies the equations (\ref{eqb5})--(\ref{eqb10}) 
to be satisfied. It is easy to see that the flexibility introduced 
by the new coefficients $\widehat{a}_i^{(2n)}$, $\widehat{b}_i^{(2n)}$
allows one to solve in many ways 
the constraints to be satisfied.
The simplest solution is obtained by requiring that the energy-dependence 
enters only in $\widehat{a}_1(E_0)$ and only at the 2PN level:
\beq
\widehat{a}_{1}^{(2)} = 0\,, \quad \quad \widehat{a}_{2}^{(2)} = 0 \,, \quad \quad 
\widehat{b}_{1}^{(2)} = 0\,,
\eeq
because in this case only Eq.~(\ref{eqb8}) gets modified. Indeed, 
it is straightforward to derive the new equation replacing (\ref{eqb8}): 
\beq
-\frac{19}{64}\,\widehat{a}_1^{(0)} + 
\frac{\widehat{b}_1^{(0)}}{4} - \frac{\widehat{a}_1^{(4)}}{2} = 
\frac{35}{32} + \frac{15}{16}\,\nu + \frac{3}{32}\,\nu^2 
\,.
\label{eqb13}
\eeq
Hence, from Eqs.~(\ref{eqb5})--(\ref{eqb7}) 
we obtain the effective metric coefficients at the 1PN level:
\beq
\widehat{a}_{1}^{(0)} = -2\,, \quad \quad 
\widehat{a}_{2}^{(0)} = -\frac{\nu}{4}\,,\quad \quad 
\widehat{b}_{1}^{(0)} = \frac{1}{4}\,(8 - \nu)\,,
\eeq
while the 2PN-equations (\ref{eqb13}) and (\ref{eqb10}),~(\ref{eqb11}) give:
\beq
\widehat{a}_{1}^{(4)} = - \frac{\nu}{16}\,(32 + 3\nu)\,, \quad \quad 
\widehat{a}_{3}^{(0)} = \frac{\nu}{64}\,(208- \nu)\,, \quad \quad 
\widehat{b}_{2}^{(0)} = \frac{1}{64}\,(256-400\,\nu+\nu^2)\,.
\eeq
Again this solution is more complex than our preferred solution 
(\ref{eq5.6})--(\ref{eq5.8}). Moreover we think that the assumption of an energy 
dependence in the effective metric introduces a conceptual obscurity 
in the entire approach: Indeed, one should introduce two separate 
(effective) energies: the energy parameter $E_0^{(0)}$ appearing 
explicitly in $g_{\mu \nu}^{\rm eff}$, and the conserved energy $E_0^{(1)}$ 
of some individual geodesic motion in $g_{\mu \nu}^{\rm eff}(E_0^{(0)})$. 
They can only be identified, a posteriori, for each specified 
geodesic motion. This makes it also quite difficult to incorporate 
radiation reaction effects. 

Finally, one can require that the effective metric does not depend 
on the energy, but that the effective mass $m_0$ depends on $E_0$. 
One then finds the solution 
\beq
m_0(E_0) = \mu\,\left [ 1 + \frac{\nu}{48}\,(32 + 3\,\nu)\,\left (\frac{E_0}{c^2}
\right )^2 \right ]\,,
\eeq
with a corresponding effective metric defined by 
the energy-independent part $\widehat{a}_i^{(0)}$, $\widehat{b}_i^{(0)}$ 
of the solution 
above. The objections of complexity and conceptual obscurity 
raised above also 
apply to this energy-dependent effective-mass solution.


\begin{references}
\bibitem{nbtod} While both the ``relativistic reduced mass'' $m_w \equiv m_1 \, 
m_2 / w$ and the ``energy of the effective particle'' $E_w \equiv \sqrt{m_w^2 + 
b^2 (w^2)}$, introduced in Ref.~\cite{todorov}, are rather complicated functions 
of the total energy $w = \sqrt s$, their ratio $E_w / m_w$ simplifies to the 
function $\epsilon \equiv (s - m_1^2 - m_2^2) / (2 \, m_1 \, m_2)$, implicit in 
Ref.~\cite{BIZ}, which we use in our approach.
\end{references}
\end{document}